\documentclass[onecolumn,11pt]{article} 

\usepackage[margin=1.1in]{geometry}
  
 \usepackage{setspace}
 \usepackage{float} 
\usepackage{graphicx}
\usepackage{amsmath}
\usepackage{amssymb}
\usepackage{dsfont}
\usepackage[english]{babel}
\usepackage[latin1]{inputenc}
\usepackage[T1]{fontenc}

\font\dsrom=dsrom10 scaled 1200
\def \indic{\textrm{\dsrom{1}}}

\usepackage[authoryear, round]{natbib}

 \usepackage[
                  breaklinks = true,
                 colorlinks = true,
                 linkcolor = red,
                 urlcolor  = black, 
                 citecolor = blue,
                 anchorcolor = green,
                 ]{hyperref}

\newtheorem{prop}{Proposition}[section] 
\graphicspath{{Figures/}}
\newcommand{\bt}{\mathbf{t}}
\newcommand{\bg}{\mathbf{g}}

\newcommand{\bx}{\mathbf{x}}
\newcommand{\bX}{\mathbf{X}}
\newcommand{\by}{\mathbf{y}}
\newcommand{\bY}{\mathbf{Y}}
\newcommand{\bz}{\mathbf{z}}

\newcommand{\bsx}{\boldsymbol{x}}

\newcommand{\cL}{\mathcal{L}}

\newcommand{\cD}{\mathcal{D}}

\newcommand{\cJ}{\mathcal{J}}

\newcommand{\bsbeta}{\boldsymbol{\beta}}

\newcommand{\bstheta}{\boldsymbol{\theta}}

\newcommand{\bsPsi}{\boldsymbol{\Psi}}

\newcommand{\bsxi}{\boldsymbol{\xi}}

\newcommand{\Identity}{\textbf{I}}

\newcommand{\E}{\mathbb{E}}

\newcommand{\R}{\mathbb{R}}
	
\newcommand{\N}{\mathcal{N}}

\newcommand{\ICL}{\text{ICL}}

\date{}
\begin{document}

\title{Piecewise regression mixture for simultaneous functional data clustering and optimal segmentation} 
\author{Faicel Chamroukhi$^{1,2}$}
 \maketitle
\begin{center}
$^1$Aix Marseille Universit\'e, CNRS, ENSAM, LSIS, UMR 7296, 13397 Marseille, France\\
$^2$Universit\'e de Toulon, CNRS, LSIS, UMR 7296, 83957 La Garde, France\\
\href{mailto:chamroukhi@univ-tln.fr}{chamroukhi@univ-tln.fr}
\end{center}

\begin{abstract}
This paper introduces a novel mixture model-based approach for simultaneous clustering and optimal segmentation 
of functional data which are curves presenting regime changes. The proposed model consists in a finite mixture of piecewise polynomial regression models. Each piecewise polynomial regression model is associated with a cluster, and within each cluster, each piecewise polynomial component is associated with a regime (i.e., a segment). We derive two approaches for learning the model parameters. The former is an  estimation approach and consists in maximizing the observed-data likelihood via a dedicated expectation-maximization (EM) algorithm. A fuzzy partition of the curves in $K$ clusters is then obtained at convergence by maximizing the posterior  cluster probabilities. The latter however is a classification approach and optimizes a specific classification likelihood criterion through a dedicated classification expectation-maximization (CEM) algorithm. The optimal curve segmentation is performed by using dynamic programming.
In the classification approach, both the curve clustering and the optimal segmentation are performed simultaneously as the CEM learning proceeds. 
We show that the classification approach is the probabilistic version that generalizes the deterministic $K$-means-like  algorithm proposed in \cite{hebrailEtal:2010}. The proposed approach is evaluated using simulated curves and real-world curves. Comparisons with alternatives including regression mixture models and the $K$-means like algorithm for piecewise regression demonstrate the effectiveness of the proposed approach.  
\end{abstract}

%

\section{Introduction}

Probabilistic modeling approaches are  known by their well-established theoretical background and the associated efficient estimation tools in many problems such as regression, classification or clustering.  In several situations, such models have interpretation as to generalize deterministic algorithms. 
In particular, in model-based clustering \citep{mclachlanFiniteMixtureModels, mclachlan_basford88,banfield_and_raftery_93, Fraley2002_model-basedclustering, celeuxetgovaert92_CEM}, for example, the $K$-means clustering algorithm is well-known to be a particular case of the expectation-maximization (EM) algorithm \citep{dlr, mclachlanEM} for a Gaussian mixture model (GMM). 
Indeed, $K$-means is equivalent to a GMM with the same mixing proportions and identical isotropic covariance matrices when the data are assigned in a hard way after the E-step rather in a soft way, that is the classification EM (CEM) algorithm \citep{celeuxetgovaert92_CEM, celeuxetgovaert_mixture_classif_93}. 
Most of these statistical analyses in model-based clustering are multivariate as they involve reduced dimensional vectors as observations (inputs). However, in many application domains, these observations are functions (e.g., curves) and the statistical methods for analyzing such data are functional as they belong to the functional data analysis (FDA) approaches \citep{ramsayandsilvermanFDA2005}. FDA is therefore the paradigm of data analysis where the basic unit of information is a function rather than a finite dimensional vector.  
The flexibility, easy interpretation and efficiency of mixture model-based approaches for classification, clustering, segmentation, etc., in multivariate analysis, has lead to a growing investigation for adapting them to the framework of FDA, in particular for curve analysis as in \citep{Gaffney99trajectoryclustering, liuANDyangFunctionalDataClustering, Gui_FMDA,ShiW08, XiongY04, chamroukhi_et_al_neurocomputing2010, same_chamroukhi_Adac, chamroukhi_fmda_neucomp2013}.

In this paper we consider the problem of model-based functional data clustering and segmentation. The considered  data are heterogeneous  curves which may also present regime changes. The observed curves are univariate and are values of functions, available at  discretized input time points. 
This type of curves can be found in several application domains, including diagnosis application \citep{chamroukhi_et_al_neurocomputing2010, chamroukhi_ijcnn_2011}, bioinformatics \citep{Gui_FMDA,picardetal2007}, electrical engineering \citep{hebrailEtal:2010}, etc.

\subsection{Problem statement}
\begin{figure}[H]
\centering 
\includegraphics[width=4.5cm]{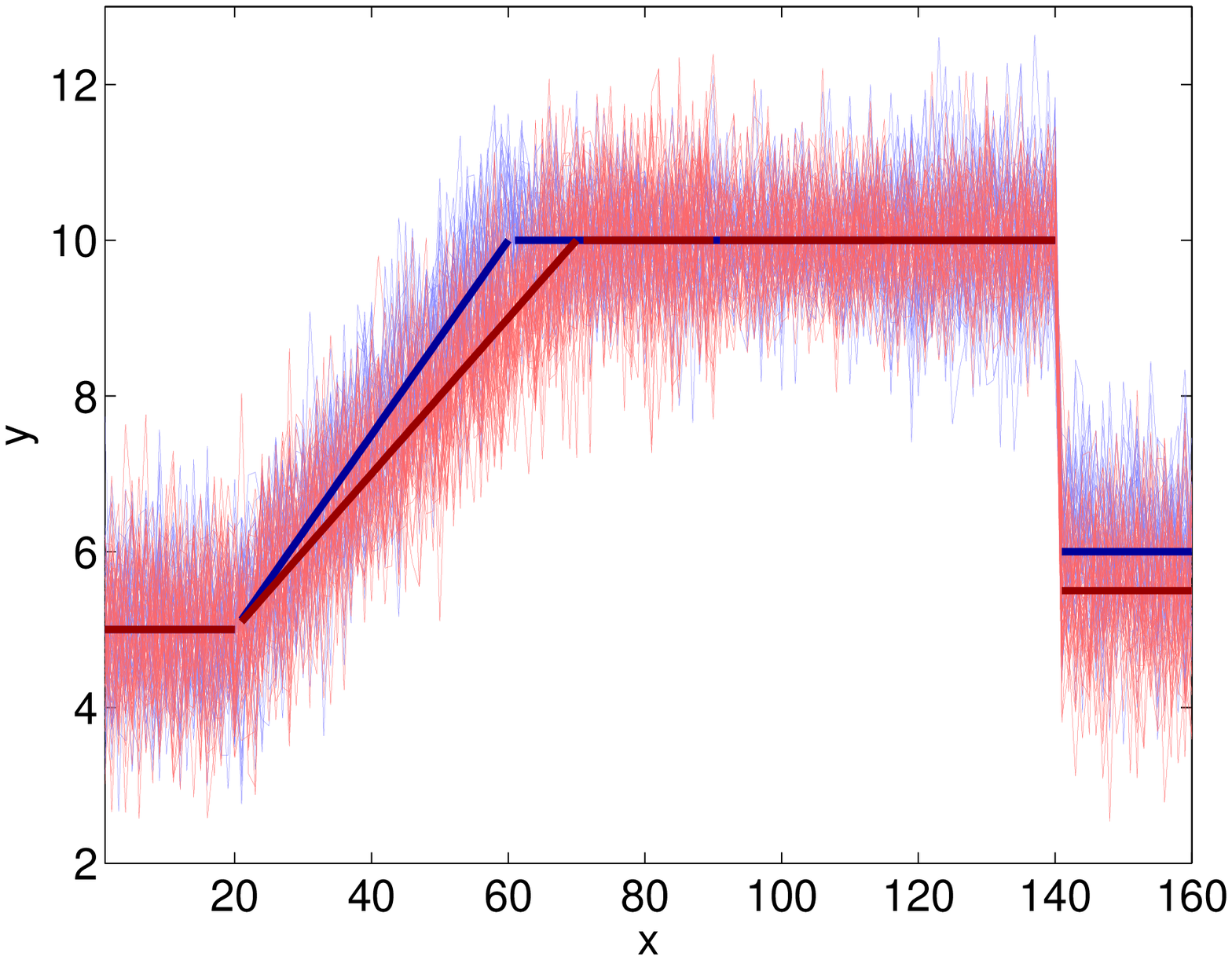} 
\includegraphics[width=4.5cm]{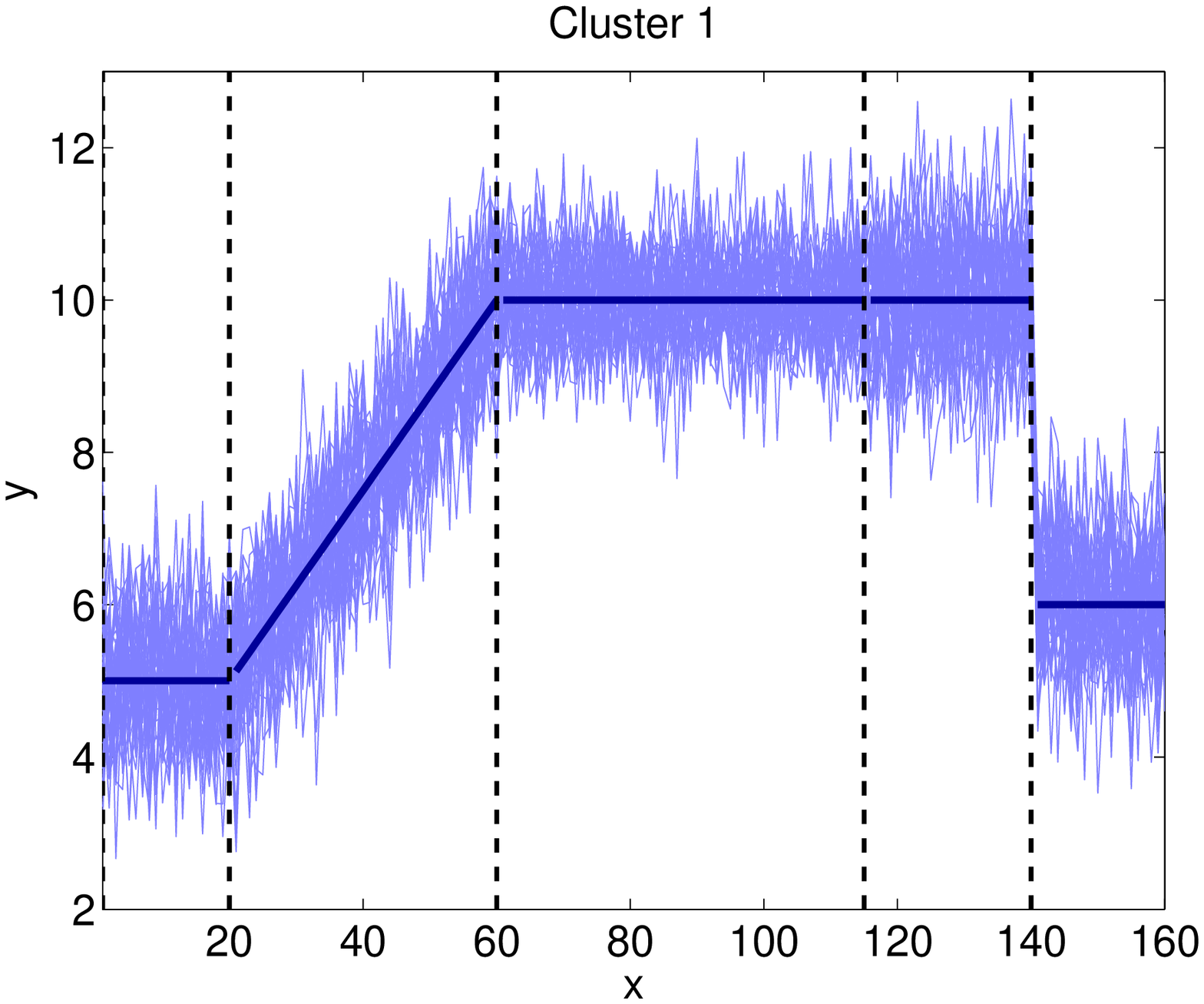}
\includegraphics[width=4.5cm]{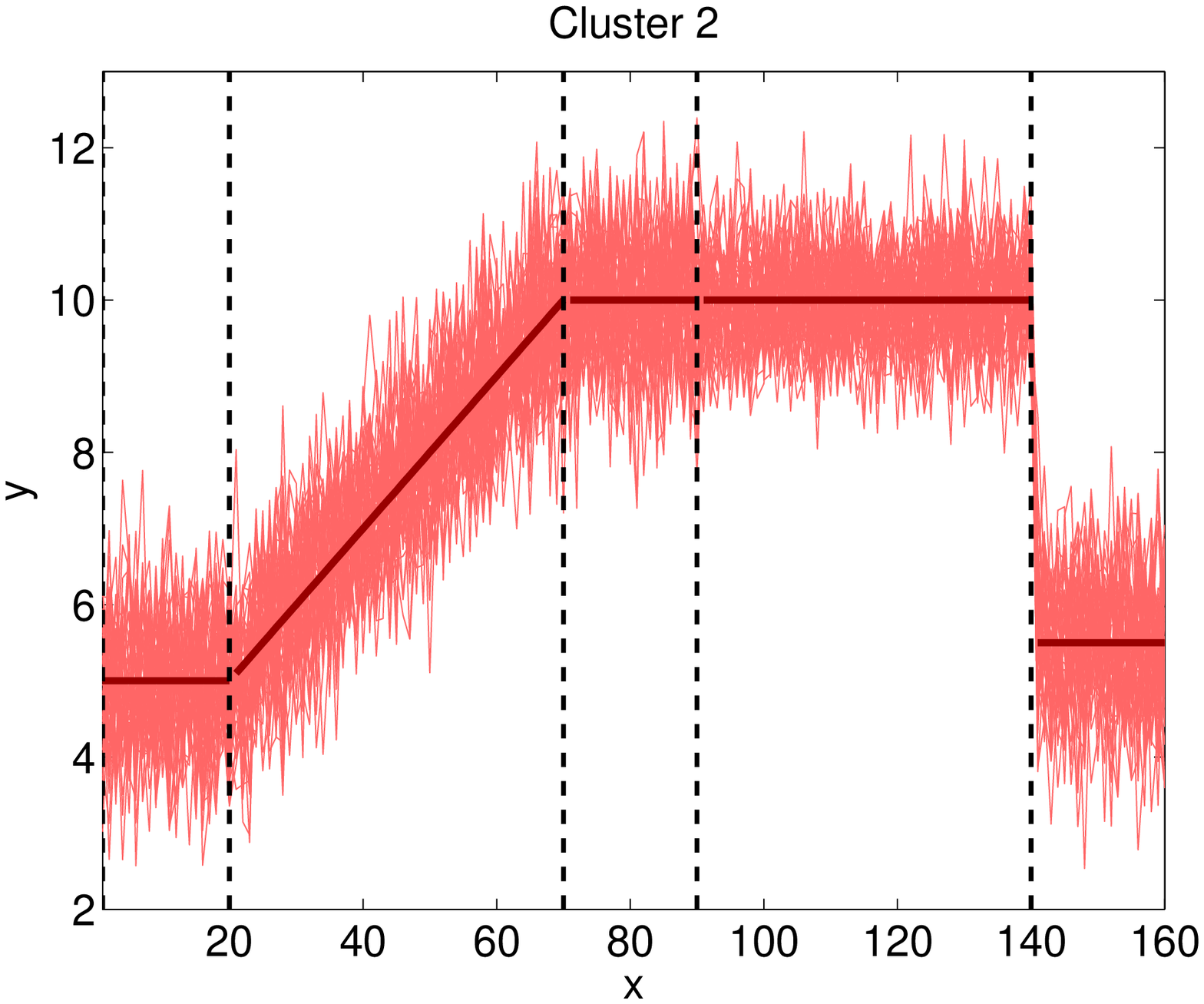}
\caption{\label{fig: example of simulated curves} A two-class data set of  simulated curves, each cluster is composed of five noisy constant/linear regimes. The clusters are colored according to the true partition, and the dashed lines represent the true segmentation of each cluster.} 
 \end{figure}

Let $\bY=(\by_1,\ldots,\by_n)$ be a set of $n$  independent curves where each curve $\by_i$ consists of $m$  measurements (observations) $\by_i = (y_{i1},\ldots,y_{im})$ regularly observed at the (time) points $(x_{i1},\ldots,x_{im})$ with $x_{i1}<\ldots<x_{im}$.  
Finally, let $(z_1,\ldots,z_n)$ be the unknown cluster labels of the curves, with $z_i \in \{1,\ldots,K\}$, $K$ being the number of clusters.
Figure \ref{fig: example of simulated curves} shows an example from a two-class situation of simulated curves which are mixed at random and each cluster contains five regimes. 
The aim is to perform curve clustering. As it can be seen, each cluster is itself very structured as it is a succession of non-overlapping segments, which we call regimes. 
 %
 %
 %
Each regime has it is own characteristics and is active for a certain time range. As it can be seen on each of these two clusters, the change of the characteristics of the regimes may correspond to a change in the mean, in the variance, or in the linearity, etc. Thus, in order to precisely infer the hidden structure of the data, it is central that the clustering method take into account the structure of the data which are composed of several regimes, instead of treating them as simple vectors in $\R^m$. This can be achieved by integrating a segmentation procedure to capture the various regime changes. This is the regime change problem. 


In such a  context, basic regression models (e.g., linear, polynomial) are not suitable. The problem of regime changes has been considered as a multiple regime change point detection problem namely by using Bayesian approaches as in \cite{Fearnhead2006} by using MCMC sampling, and   \cite{Fearnhead2007} with sequential MCMC for online change point detection. However, these approaches only concern inference from a single curve, and do not concern curve clustering as they only perform single curve segmentation.
An alternative approach in this curve clustering context may consist in using cubic splines to model each class of curves \cite{garetjamesJASA2003} but this requires the setting of knots a priori. 
Generative models have been developed by \cite{Gaffney99trajectoryclustering, gaffneyANDsmythNIPS2004} which consist in clustering curves with a mixture of regression models or random effect models. In \cite{liuANDyangFunctionalDataClustering}, the authors proposed a clustering approach based on random effect regression splines where the curves are represented by B-spline functions. However, the first approach does not address the problem of regime changes and the second one requires the setting of the spline knots to learn the model. 
Another approach based on splines is the one of clustering sparsely sampled curves \cite{garetjamesJASA2003}. All these generative approaches use the EM algorithm to estimate the model parameters. 
Recently, in \citep{HugueneyEtAl:ESANN2009, hebrailEtal:2010}, the authors proposed a distance-based approach based on a piecewise regression  model. It allows for fitting several constant (or polynomial) models to the curves and performs simultaneous curve clustering and optimal segmentation using a $K$-means-like algorithm \citep{HugueneyEtAl:ESANN2009, hebrailEtal:2010}.
The $K$-means-like algorithm simultaneously performs curve clustering and optimal segmentation using dynamic programming. It minimizes a distance function in the curve space as the learning proceeds. The curves segmentation is carried out using dynamic programming procedure.
 
The main focus of this paper is to provide a well-established latent data model to simultaneously perform curve clustering and optimal segmentation. We propose a probabilistic generative model for curve clustering and optimal curve segmentation. It combines both a mixture model as to achieve the clustering, and a polynomial piecewise regression model to optimally segment each set (cluster) of homogeneous curves into a finite number of segments using dynamic programming. We show that the proposed probabilistic model generalizes the recently proposed distance-based approach, that is the $K$-means-like algorithm of \cite{hebrailEtal:2010}. More specifically, the proposed model is a mixture of piecewise regression models. 
%
We provide two algorithms for learning the model parameters. The first one is a dedicated EM algorithm to find a fuzzy partition of the data and an optimal segmentation by maximizing the observed-data log-likelihood. The second algorithm consists in maximizing a specific classification likelihood criterion by using a dedicated CEM algorithm in which the curves are partitioned and optimally segmented simultaneously as the learning proceeds. In this CEM-based classification approach, the curves are partitioned in a hard way in contrast to the fuzzy classification approach. For the two algorithms, the optimal curve segmentation is performed by using dynamic programming.

This paper is organized as follows. We first briefly recall the two main approaches for model-based clustering,  
 and its extension to curve clustering. Then, Section \ref{sec: Related work} provides a brief account of related work on model-based curve clustering approaches using polynomial regression mixtures (PRM) and spline regression mixtures (SRM) \citep{Gaffney99trajectoryclustering, Gaffneythesis} and recalls the $K$-means like algorithm for curve clustering and optimal segmentation based on polynomial piecewise regression \citep{hebrailEtal:2010}. Section \ref{sec: piecewise polynomial regression mixture} introduces the proposed piecewise regression mixture model (PWRM) and its unsupervised learning by deriving both the estimation approach and the classification approach, and the dedicated EM and CEM algorithms. Finally, Section \ref{sec: Experimental study} deals with the experimental study carried out on simulated curves and  real-world curves to assess the proposed approach by comparing it to the regression mixtures, the $K$-means like algorithm 
and the standard GMM clustering. 


\subsection{Model-based clustering}
\label{ssec: Model-based clustering}

Model-based clustering \citep{banfield_and_raftery_93,mclachlan_basford88,Fraley2002_model-basedclustering}, generally used in a multivariate analysis context, is based on a finite mixture model formulation \citep{mclachlanFiniteMixtureModels, titteringtonBookMixtures}. In the finite mixture approach for cluster analysis, the data probability density function is assumed to be a finite mixture density, each component density being associated with a cluster. The problem of clustering therefore becomes the one of estimating the parameters of the supposed  mixture model. 
In this way, two main approaches are possible, as follows.
\paragraph{The mixture approach}
In the mixture (or estimation) approach, the parameters of the mixture density are estimated by maximizing the observed-data likelihood. This is generally achieved via the expectation-maximization (EM) algorithm \citep{dlr,mclachlanEM}. After performing the model estimation, the posterior cluster probabilities, which represent a fuzzy partition of the data,  are then used to determine the cluster memberships through the maximum a posteriori (MAP) principle. 
  \paragraph{The classification approach}
   The classification approach, also referred to as the maximum likelihood classification approach, consists in optimizing  
   the complete-data likelihood. 
   This maximization can be performed by using the classification version of the EM algorithm, known as the classification EM (CEM) algorithm  \citep{celeuxetgovaert92_CEM}.
The CEM algorithm integrates a classification step between the E and the M steps of the EM algorithm which computes the cluster memberships in a hard way by using the MAP principle. 
  

\subsection{Model-based curve clustering}

Mixture model-based curve clustering approaches have also been introduced as to generalize the standard multivariate mixture model to the case of analysis of curves where the individuals are presented as curves rather than vectors. 
Indeed, when the data are curves which are in general very structured, relying on standard multivariate mixture analysis may lead to unsatisfactory results in terms of classification accuracy or modeling accuracy \cite{chamroukhi_et_al_NN2009, chamroukhi_PhD_2010, chamroukhi_et_al_neurocomputing2010}.
However, addressing the problem from a functional data analysis prospective, that is formulating functional mixture models, allows for overcoming this limitation \cite{chamroukhi_et_al_NN2009, chamroukhi_PhD_2010, chamroukhi_et_al_neurocomputing2010, same_chamroukhi_Adac, Gaffney99trajectoryclustering, Gaffneythesis, gaffneyANDsmythNIPS2004, liuANDyangFunctionalDataClustering}.
In this case of model-based curve clustering,  one can distinguish the  regression mixture approaches \citep{Gaffney99trajectoryclustering, Gaffneythesis}, including polynomial regression and spline regression, or random effects polynomial regression as in \citep{gaffneyANDsmythNIPS2004} or spline regression as in \citep{liuANDyangFunctionalDataClustering}.  Another approach based on splines is concerned with clustering sparsely sampled curves \citep{garetjamesJASA2003}. All these approaches use the mixture (estimation) approach with the EM algorithm to estimate the model parameters. 


\section{Related work}
\label{sec: Related work}

In this section, we first describe the model-based curve clustering based on regression mixtures and the EM algorithm \citep{Gaffneythesis, gaffneyANDsmythNIPS2004} as in \cite{chamroukhi_et_al_neurocomputing2010, chamroukhi_ijcnn_2011}. Then we describe the piecewise regression approach for curve clustering and optimal segmentation of \cite{hebrailEtal:2010} and the associated $K$-means-like algorithm.
%

\subsection{Regression mixtures and the EM algorithm for curve clustering} 
\label{ssec: polynomial and spline regression mixture}

As stated in \cite{chamroukhi_ijcnn_2011}, regression mixtures, namely polynomial regression mixture
models (PRM) and polynomial spline regression mixtures (PSRM) \citep{Gaffneythesis, gaffneyANDsmythNIPS2004}, 
assume that each curve is drawn from one of $K$ clusters of curves with mixing proportions $(\alpha_1,\ldots,\alpha_K)$. Each cluster of curves is modeled by either a  polynomial regression model or a spline regression model. Thus, the mixture density of a curve $\by_i$ $(i=1,\ldots,n)$ can be written as:  
\begin{eqnarray}
p(\by_i|\bx_i;\bsPsi)&=& \sum_{k=1}^K \alpha_k \  \N (\by_{i};\bX_i \bsbeta_k,\sigma_k^2\Identity_m), 
\label{eq: PRM or PSRM model}
\end{eqnarray}
where the $\alpha_k$'s defined by $\alpha_k = p(z_i=k)$ are the non-negative mixing proportions that sum to one, $\bsbeta_{k}$ is the coefficient  vector of the $k$th regression model and $\sigma_{k}^2$ the associated noise variance, and $\bX_i$ the design matrix whose construction depends on the adopted model (i.e., polynomial, or polynomial spline, etc). 
%
 The  regression mixture model is therefore fully described by the parameter vector $\bsPsi = (\alpha_1,\ldots,\alpha_k,\bsPsi_1,\ldots,\bsPsi_K)$ with  $\bsPsi_k=(\bsbeta_{k},\sigma_{k}^2)$. The unknown parameter vector $\bsPsi$ can be estimated by maximizing the observed-data log-likelihood given by:
\begin{eqnarray}
\cL(\bsPsi) &=& \sum_{i=1}^n  \log \sum_{k=1}^K \alpha_k \ \N (\by_{i};\bX_i \bsbeta_k,\sigma_k^2\Identity_m)
\label{eq: log-lik for the PRM and PSRM}
\end{eqnarray}
 via the EM algorithm \citep{Gaffneythesis, dlr}. The EM algorithm for the regression mixture models and the corresponding updating formula can be found in \citep{Gaffney99trajectoryclustering, Gaffneythesis}.
Once the model parameters have been estimated, a partition of the data into $K$ clusters can then be computed by maximizing the posterior cluster probabilities (MAP principle). 

The regression mixture model however  does not address the problem of regime changes within the curves. Indeed, it assumes that each cluster present a stationary behavior described by a single polynomial mean function. This approach is therefore not well adapted to handle the problem of  segmenting curves with regime changes.
An alternative way is to use polynomial spline regression rather than polynomial regression as in  \cite{Gaffneythesis, garetjamesJASA2003, liuANDyangFunctionalDataClustering} where the curves are represented by using a combination of several polynomial bases at different time range locations rather than a single polynomial basis. Splines are indeed based on constrained piecewise polynomial fitting with predefined piecewise locations. Therefore, it should be noticed that in spline regression models, the placement of the knots are generally either fixed by the user or placed uniformly over the range of the input $\bx_i$. 
The optimization of the knots locations, which are assumed to be related to the locations of regime changes (the transition points) in this case of curve segmentation,  requires relaxing the regularity constraints for the splines. 
This leads to the  piecewise polynomial regression \citep{McGee, brailovsky, chamroukhi_PhD_2010} model in which the placement of the knots can be optimized  using dynamic programming \citep{bellman, stone}.

The piecewise regression model can be used to perform simultaneous curve clustering and optimal segmentation. In \cite{HugueneyEtAl:ESANN2009,hebrailEtal:2010}, the authors proposed a $K$-means-like algorithm involving a dynamic programming procedure for  simultaneous curve clustering and optimal segmentation based on the piecewise regression model. The idea we propose in this paper is in the same spirit, however it provides a general probabilistic framework to address the problem. Indeed,  in our proposed approach, the piecewise regression model is included into a mixture framework to generalize the deterministic $K$-means like approach. Both fuzzy clustering and hard clustering techniques are possible.
We notice that another possible way for this task of curve clustering and segmentation is to proceed as in the case of sequential data modeling in which it is assumed that the observed sequence (in this case a curve) is governed by a hidden process which enables for switching from one configuration to another among $K$ configurations. The used process in general is a $K$-state homogeneous Markov chain.  
This leads to the mixture of hidden Markov models \cite{Smyth96} or mixture of hidden Markov model regressions \citep{chamroukhi_ijcnn_2011}. 


\subsection{Curve clustering and optimal segmentation with $K$-means-like algorithm}

\label{ssec: piecewise polynomial regression and K-means like}

In \cite{hebrailEtal:2010}, the authors proposed a $K$-means-like algorithm to simultaneously perform curve clustering and optimal segmentation of each cluster of curves. 
This is achieved by minimizing a Euclidean distance criterion similarly as in the standard $K$-means for multivariate data clustering, while in their functional approach the computations are performed in the space of curves.  
The curves are partitioned into $K$ clusters and each cluster $k$ is modeled by a piecewise constant regression model and  segmented into $R_k$ regimes. The segmentation is performed in an optimal way by using dynamic programming thanks to the  additivity of the distance criterion over the set of segments for each cluster. 
In the following, we recall this  technique in order to later show the difference compared to the proposed approach.

\subsubsection{The optimized distance criterion}

The clustering and segmentation algorithm proposed in \citep{HugueneyEtAl:ESANN2009,hebrailEtal:2010} simultaneously minimizes the following error (distance) criterion:
\begin{equation}
E\left(\bz,\{I_{kr}\}, \{\mu_{kr}\}\right)= \sum_{k=1}^{K}  \sum_{i|z_i=k}  \sum_{r=1}^{R_k} \sum_{j\in I_{kr}} (y_{ij} - \mu_{kr})^2  
\label{eq: distance criterion for clustering with piecewise regression}
\end{equation}with respect to the partition $\bz$ and the piecewise cluster parameters  $\{\mu_{kr}\}$ and $\{I_{kr}\}$, where $I_{kr} = (\xi_{kr},\xi_{k,r+1}]$ represent the element indexes of segment (regime) $r$ ($r=1,\ldots,R_k$) for cluster $k$ and $\mu_{kr}$ its constant mean, $R_k$ being the corresponding number of segments.  
The $m\times1$ piecewise constant mean curve $\bg_k = (g_{k1},\ldots,g_{km})$ where $g_{kj}=\mu_{kr}$ if $j\in I_{kr}$ for all $j=1,\ldots,m$ (i.e., the $j$th observation $y_{ij}$ belongs to segment $r$ of cluster $k$) can be seen as the mean curve or the ``centroid'' of cluster $k$ $(k=1,\ldots,K)$. Thus the criterion (\ref{eq: distance criterion for clustering with piecewise regression}) can be seen as the optimized distortion criterion  by the standard $K$-means for multivariate data clustering, and can then be iteratively minimized by the following $K$-means-like algorithm \citep{hebrailEtal:2010}. 

\subsubsection{The $K$-means-like algorithm}
\label{ssec: k-means like algorithm}

After starting with an initial cluster partition $\bz^{(0)}$ (e.g., initialized randomly), the $K$-means-like algorithm alternates between the two following steps, at each  iteration $q$, until convergence.

\paragraph{Relocation step} This step consists in finding the optimal  piecewise constant prototype for a given cluster $k$ as follows. 
Based on the current partition  $\bz^{(q)}$, $q$ being the current iteration number, find the segmentation of each cluster $k$ into $R_k$ regimes by minimizing the following additive criterion : 
\begin{equation}
E_k(\bz^{(q)}, \{I_{kr}\},\{\mu_{kr}\}) = \sum_{r=1}^{R_k} \sum_{i|z_i^{(q)}=k} \sum_{j\in I_{kr}} (y_{ij} - \mu_{kr})^2
\label{eq: Ek kmeans like algorithm}
\end{equation}w.r.t the segment boundaries  $\{I_{kr}\}$ and the constant means $\{\mu_{kr}\}$ for each segment. 
Since (\ref{eq: Ek kmeans like algorithm}) is additive over the segments $r$, the segmentation can 
 be performed in an optimal way by using  dynamic programming \citep{bellman, stone, hebrailEtal:2010}. 
Then, each cluster representative is   relocated to the piecewise constant prototype  $\bg^{(q)}_k$ representing the
mean of all data points assigned to it. 
\paragraph{Assignment step} 
This step updates the curves partition $\bz$ by assigning each curve $\by_i$ to the  nearest piecewise constant prototype $\bg^{(q)}_k$ in the sense of the Euclidean distance, that is:
$z_i^{(q+1)} = \arg \min_{1\leq k\leq K} \parallel\by_i - \bg^{(q)}_{k}\parallel^2.$

However, 
this approach is not probabilistic. It can be seen as deterministic as it does not define a density model on the data. 
As we will show it later in Section \ref{sec: piecewise polynomial regression mixture}, it represents a particular case of a more general probabilistic model, the one which we propose.
 Having a probabilistic formulation has numerous advantages and relies on a sound statistical background. 
It is indeed more advantageous to formulate a probabilistic generative approach for easy interpretation and to help understanding the process governing the data generation.
 In addition, for this clustering task, formulating a latent data model allows to consider naturally the clustering within the missing data framework. Furthermore, as we will see, the general probabilistic framework will still be more adapted to the structure of the data, rather than the $K$-means-like approach which may fail if some constraints on the structure of the data are not satisfied. 
 Another advantage is that the probabilistic approach allows for performing soft clustering, which is not generally the case in deterministic approaches. In addition, in probabilistic model-based clustering, we have the possibility to naturally incorporate prior knowledge on the model parameters through prior distributions. 

Thus, in the next section we present the proposed piecewise regression mixture model (PWRM) and its unsupervised learning by using to variants of parameter estimation: The first one uses a dedicated EM algorithm and the second one uses a dedicated classification EM (CEM) algorithm. We show how the CEM algorithm used for clustering and optimal segmentation constitutes a probabilistic version of the deterministic approach recalled previously.

\section{The piecewise  regression mixture (PWRM)}
\label{sec: piecewise polynomial regression mixture}

In the proposed approach, the piecewise regression model  is stated into a probabilistic framework for model-based curve clustering and optimal segmentation, rather than into a deterministic approach as described previously.  
First, we present the extension of the standard piecewise regression model for modeling a homogeneous set of independent curves rather than a single curve. 
 Then we derive our piecewise regression mixture model (PWRM). 

\subsection{Piecewise  regression for curve modeling and optimal segmentation}
\label{ssec: piecewise regression} 

As stated in \cite{chamroukhi_et_al_neurocomputing2010}, piecewise polynomial regression \citep{McGee, brailovsky, ferrari1, hebrailEtal:2010, picardetal2007} is a modeling and segmentation method that can be used to partition a curve or curves into $R$ regimes (segments). Each segment is characterized by its constant or polynomial mean  curve and its variance. The model parameters can be estimated in an optimal way by  using a dynamic programming procedure \citep{bellman, stone} thanks to the additivity of the optimized criterion 
over the regimes \citep{brailovsky, picardetal2007, hebrailEtal:2010, HugueneyEtAl:ESANN2009, chamroukhi_PhD_2010}. 
In the following section, we present the piecewise polynomial regression model, which is generally used for a single curve, in a context of modeling a set of curves. We also describe the algorithm used for parameter estimation by maximizing the likelihood.

\subsubsection{Piecewise  regression for  modeling and optimal segmentation of a set of curves} 
\label{ssec: functional piecewise polynomial regression model}

Piecewise polynomial regression, generally used to model a single curve, \citep{McGee, brailovsky, ferrari1,chamroukhiIJCNN2009}, can be easily used to model a set of curves with regime changes \citep{chamroukhi_et_al_neurocomputing2010, chamroukhi_PhD_2010}. 
The piecewise polynomial regression model assumes that the curves $(\by_1,\ldots,\by_n)$ incorporate $R$ polynomial regimes defined on $R$ intervals $I_1,\ldots,I_R$ whose bounds indexes can be denoted by $\bsxi = (\xi_1,\ldots,\xi_{R+1})$ where  $I_r = (\xi_{r},\xi_{r+1}]$ with $\xi_1=0 <\xi_2 < \hdots <\xi_{R+1}=m$. This defines a partition of the set of curves into $R$ segments of length $m_1,\ldots,m_R$ respectively: 
$\{y_{ij}|j\in I_1\},\ldots,\{y_{ij}|j\in I_R\}, i=1,\ldots,n.$
The piecewise polynomial regression model for the set of curves, in the Gaussian case, can therefore be defined as follows. For $r=1,\ldots,R$: 
\begin{equation}
y_{ij} =  \bsbeta^T_r\bsx_{ij} +\sigma_r\epsilon_{j} \quad  \mbox{ if } j\in I_r \quad (i=1,\ldots,n; j=1,\ldots,m) 
\label{eq: functional piecewise regression}
\end{equation}where the $\epsilon_{j}$ are independent zero mean and unit variance Gaussian variables representing additive noise. The model parameters which can be denoted by $(\bstheta,\bsxi)$ where $\bstheta=(\bsbeta_1,\ldots,\bsbeta_R,\sigma_1^2,\ldots,\sigma_R^2)$ are composed of the regression parameters and the noise variance for each regime, and are estimated  by maximizing the observed-data likelihood.  
We assume that, given the regimes, the data of each curve are independent. Thus, according to the piecewise regression model (\ref{eq: functional piecewise regression}),  the conditional density of a curve $\by_i$  is given by: 
\begin{equation}
p(\by_i|\bx_i;\bstheta,\bsxi) = \prod_{r=1}^R \prod_{j \in I_r}\N\left(y_{ij};\bsbeta_r^T \bsx_j,\sigma_r^2\right),
\label{eq: single curve distribution for the piecewise regression model}
\end{equation}and the log-likelihood of the model parameters $(\bstheta,\bsxi)$  
given an independent set of curves $(\by_1,\ldots,\by_n)$ is given by:  
{\begin{eqnarray}
\cL(\bstheta,\bsxi)& =  & \log \prod_{i=1}^n p(\by_i|\bx_{i};\bstheta,\bsxi) 
=  -\frac{1}{2}\sum_{r=1}^R \!  \Big[\frac{1}{\sigma_r^2}\sum_{i=1}^n \!  \sum_{j\in I_r} \!  (y_{ij}-\bsbeta_r^{T}\bsx_{ij})^2 \! + \! n m_r \log \sigma_r^2 \Big] \! \!  + \!  c  
\label{eq: loglik for the functional piecewise regression}
\end{eqnarray}}where $m_r$ is the cardinal number of $I_r$ (the indexes of points belonging to regime $r$) and $c$ is a constant term independent of $(\bstheta,\bsxi)$. Maximizing this log-likelihood is equivalent to minimizing the following criterion 
\begin{eqnarray}
\cJ(\bstheta,\bsxi) &=& \sum_{r=1}^R  \Big[\frac{1}{\sigma_r^2}\sum_{i=1}^n \sum_{j\in I_r}\left(y_{ij}-\bsbeta_r^{T}\bsx_{ij}\right)^2 + n m_r \log \sigma_r^2 \Big].
\label{eq: functional piecewise regression criterion J}
\end{eqnarray}This can be performed by a using dynamic programming procedure thanks to the additivity of the criterion $\cJ$ over the segments $r$ over the segments \citep{bellman,stone}. Thus, thanks to dynamic programming, the segmentation can be performed in an optimal way.
The next section shows how the parameters $\bstheta$ and $\bsxi$ can be estimated by using dynamic programming to minimize the criterion $\cJ$ given by (\ref{eq: functional piecewise regression criterion J}).  

\subsubsection{Parameter estimation of the piecewise regression model by dynamic programming} 
\label{ssec: dynamic programming for the functional pwr}
\label{ssec: dynamic programming for the functional pwr}
 
A dynamic programming procedure can be used to minimize the additive criterion (\ref{eq: functional piecewise regression criterion J}) with respect to $(\bstheta, \bsxi)$ or equivalently to minimize the following criterion (\ref{eq: dynamic programming criterion C for a set of curves}) with respect to $\bsxi$:
\begin{eqnarray}
C(\bsxi) &=& \min \limits_{\substack {\bstheta}} \cJ(\bstheta,\bsxi) 
=\sum_{r=1}^R \min \limits_{\substack {(\bsbeta_r,\sigma_r^2)}} \Big[\frac{1}{\sigma_r^2}\sum_{i=1}^n\sum_{j=\xi_r+1}^{\xi_{r+1}}\left(y_{ij}-\bsbeta_k^{T}\bsx_{ij} \right)^2 + n m_r \log \sigma_r^2 \Big] \nonumber \\
& = & \sum_{r=1}^R\Big[ \frac{1}{\hat{\sigma}_r^2} \sum_{i=1}^n \sum_{j=\xi_r+1}^{\xi_{r+1}}(y_{ij}-\hat{\bsbeta}_r^{T}\bsx_{ij})^2 + n m_r \log \hat{\sigma}_r^2 \Big],
\label{eq: dynamic programming criterion C for a set of curves}
\end{eqnarray}
where $\hat{\bsbeta}_r$ and $\hat{\sigma}_r^2$ are the solutions of a polynomial regression problem for segement $r$ and are respectively given by:
{\begin{eqnarray}
{\hat{\bsbeta}}_r &= & \arg \min_{\bsbeta_r} \sum_{i=1}^n  \sum_{j=\xi_r+1}^{\xi_{r+1}}(y_{ij}-\bsbeta_r^{T}\bsx_{ij})^2 
\nonumber \\
& = & 
 \Big[ \sum_{i=1}^n \sum_{j=\xi_r+1}^{\xi_{r+1}} \bsx_{ij}\bsx_{ij}^T \Big]^{-1}\! \sum_{i=1}^n \sum_{j=\xi_r+1}^{\xi_{r+1}}  \bsx_{ij}y_{ij}
\label{eq: estimation of beta_r in piecewise reg}
\end{eqnarray}}
and
\begin{eqnarray}
\!\!\!\! \hat{\sigma}_r^2 &=& \arg \min_{\sigma^2_r} \frac{1}{\sigma_r^2} \sum_{i=1}^n \sum_{j=\xi_r+1}^{\xi_{r+1}}(y_{ij}-\hat{\bsbeta}_r^{T}\bsx_{ij})^2 + n m_r \log \sigma_r^2 
\nonumber \\
&=& 
  \frac{1}{ n m_r} \sum_{i=1}^n\sum_{j=\xi_r+1}^{\xi_{r+1}} (y_{ij}-{\hat{\bsbeta}}_r^T \bsx_{ij})^2. 
\label{eq: estimation of sigma_r in piecewise reg}
\end{eqnarray}
The matrix form of these solutions can be written as:
{\begin{eqnarray}
 {\hat{\bsbeta}}_r  
& = & \Big[ \sum_{i=1}^n \bX_{ir}^T \bX_{ir}  \Big]^{-1} \sum_{i=1}^n \bX_{ir}\by_{ir}\\
\label{eq: estimation of beta_r in piecewise reg : matrix form} 
\hat{\sigma}_r^2 &=& \frac{1}{ n m_r} \sum_{i=1}^n |\!|(\by_{ir}-\bX_{ir}{\hat{\bsbeta}}_r)|\!|^2
\label{eq: estimation of sigma_r in piecewise reg : matrix form}
\end{eqnarray}where $\by_{ir}$ is the segment (regime) $r$ of the $i$th curve, that is the observations $y_{ij}, j=(\xi_r+1, \ldots, \xi_{r+1})$ and
and $\bX_{ir}$ its associated design matrix with rows $\bsx_{ij},  j=(\xi_r+1, \ldots, \xi_{r+1})$ for $i=1,\ldots,n$.

It can be seen that the criterion $C(\bsxi)$ given by Equation (\ref{eq: dynamic programming criterion C for a set of curves})  is additive over the $R$ segments. Thanks to its additivity, this criterion  can be optimized globally using a dynamic programming procedure \citep{bellman,stone,brailovsky}. 
The piecewise approach provides therefore an optimal segmentation of a homogeneous set of curves into $R$ polynomial segments, each segment being associated with a regime. To handle non-homogeneous sets of curves and at the same time take benefit from the efficient segmentation provided by piecewise regression, the model can therefore be integrated in a mixture framework, each component density will represent a set of curves with a specified number of regimes. This results into the piecewise regression mixture model presented in the next section.

\subsection{Piecewise  regression mixture model (PWRM) for curve clustering and optimal segmentation} 

In this section, we integrate the piecewise polynomial regression model presented previously into a mixture model-based curve clustering framework. Thus, the resulting  model is a piecewise regression mixture model which will be abbreviated as PWRM. According to the PWRM model, each curve $\by_i$ $(i=1,\ldots,n)$ is assumed to be generated by a piecewise regression model among $K$ models defined by (\ref{eq: single curve distribution for the piecewise regression model}), with a prior probability $\alpha_k$.
The distribution of a curve is given by the following piecewise polynomial regression mixture (PWRM) model:
\begin{equation}
p(\by_i|\bx_i;\bsPsi) = \sum_{k=1}^K \alpha_k \prod_{r=1}^{R_k} \prod_{j \in I_{kr}}\N (y_{ij};\bsbeta_{kr}^T\bsx_{ij},\sigma_{kr}^2), 
\label{eq: piecewise regression mixture}
\end{equation}where $I_{kr}$  is the set of elements indexes of polynomial segment (regime) $r$ for the cluster $k$, $\bsbeta_{kr}$ is the $(p+1)$-dimensional vector of its polynomial coefficients and the $\alpha_k$ are the non-negative mixing proportions that sum to one. The  parameters of the PWRM model can therefore be denoted by:
$$\bsPsi = (\alpha_1,\ldots,\alpha_K,\bstheta_1,\ldots,\bstheta_K,\bsxi_1,\ldots,\bsxi_K)$$
 where $\bstheta_k=(\bsbeta_{k1},\ldots,\bsbeta_{k{R_k}},\sigma_{k1}^2,\ldots,\sigma_{k{R_k}}^2)$ and $\bsxi_k=(\xi_{k1},\ldots,\xi_{k,{R_k}+1})$ are respectively
the set of polynomial coefficients and noise variances, and the set of transition points which correspond to the segmentation of cluster $k$. 
 
 The proposed mixture model is therefore suitable for clustering and optimal segmentation of complex shaped curves. More specifically,   by integrating the piecewise polynomial regression into a mixture framework, the resulting model is able to
perform curve clustering. The problem of regime changes within each cluster of curves will be addressed as well thanks to the optimal segmentation provided by dynamic programming for each piecewise regression component model.
These two simultaneous outputs are clearly not provided by the standard generative curve clustering approaches namely the regression mixture and spline regression mixtures. On the other hand, the PWRM is a probabilistic model and as it will be shown in the following, generalizes the deterministic $K$-means-like algorithm for curve clustering and optimal segmentation.

\medskip
Once the model is defined, now we have to estimate its parameters from data and show how it is used for clustering and optimal segmentation. 
We present two approaches to learn the model parameters. The former is an estimation approach and is based on maximizing the observed-data log-likelihood via a dedicated EM algorithm. The latter however is  a classification approach and  maximizes the completed-data log-likelihood through a specific CEM algorithm. In the next section we derive the first approach and then we present the second one.  

\section{Maximum likelihood estimation via a dedicated EM algorithm}
\label{ssec: ML estimation vi EM for the piecewise polynomial regression mixture}

As seen in the introduction, in the estimation (maximum likelihood) approach, the parameter estimation is performed by maximizing the observed-data (incomplete-data) log-likelihood. Assume we have a set of $n$ i.i.d  curves $\bY=(\by_1,\ldots,\by_n)$ regularly sampled at the time points $\bx_i$. According to the model (\ref{eq: piecewise regression mixture}), the log-likelihood of $\bsPsi$ given the observed data can be written as:
\begin{eqnarray}
 \cL(\bsPsi) = \log \prod_{i=1}^n p(\by_i|\bx_i;\bsPsi)    =   \sum_{i=1}^n  \log \sum_{k=1}^K \alpha_k \prod_{r=1}^{R_k} \prod_{j \in I_{kr}}\N\left(y_{ij};\bsbeta_{kr}^T\bsx_{ij},\sigma_{kr}^2\right).
\label{eq: log-lik for the mixture of piecewise regression}
\end{eqnarray}
The maximization of this log-likelihood can not be performed in a closed form. The  EM algorithm \citep{dlr,mclachlanEM} is generally used to iteratively maximize it similarly as in standard mixtures. In this framework, the complete-data log-likelihood, for a particular partition $\bz=(z_1,\ldots,z_n)$, where $z_i$ is the cluster label of the $i$th curve, is given by: 
\begin{equation}
\cL_c(\bsPsi,\bz) = \sum_{i=1}^{n}\sum_{k=1}^{K} z_{ik} \log \alpha_k + \sum_{i=1}^{n} \sum_{k=1}^{K} \sum_{r=1}^{R_k} \sum_{j\in I_{kr}}  z_{ik} \log  \N  (y_{ij};\bsbeta^T_{kr} \bsx_{ij},\sigma^2_{kr}) 
 \label{eq: complete log-lik for the piecewise regression mixture}
\end{equation}  
where  $z_{ik}$ is an indicator binary-valued variable such that $z_{ik}=1$ iff $z_i=k$ (i.e., if the curve $\by_i$ is generated by the cluster $k$). The next paragraph shows how the observed-data log-likelihood (\ref{eq: log-lik for the mixture of piecewise regression}) of the proposed model is maximized by the EM algorithm to perform curve clustering and optimal segmentation. 

\subsection{The EM algorithm for  piecewise  regression mixture (EM-PWRM)}
\label{sssec: EM for  piecewise regression mixture}
The EM algorithm for the polynomial piecewise regression mixture model (EM-PWRM) starts with an initial solution $\bsPsi^{(0)}$ (e.g., computed from a random partition and uniform segmentation) and alternates between the two following steps until convergence (e.g., when there is no longer change in the relative variation of the log-likelihood):
\paragraph{E-step}
\label{ssec: E-step of the EM algorithm for the piecewise regression mixture}
The E-step computes the expected complete-data log-likelihood given the observed curves $\cD=((\bx_1,\by_1),\ldots,(\bx_n,\by_n))$ and the current value of the  model parameters denoted by  $\bsPsi^{(q)}$, $q$ being the current iteration number: 
\begin{eqnarray}
Q(\bsPsi,\bsPsi^{(q)})& = & \E\big[\cL_c(\bsPsi;\cD,\bz)|\cD;\bsPsi^{(q)}\big] \nonumber\\ 
 &=&  \sum_{i=1}^{n}\sum_{k=1}^{K} \E \big[z_{ik}|\cD;\bsPsi^{(q)}\big] \log \alpha_k  
+ \sum_{i=1}^{n} \sum_{k=1}^{K} \sum_{r=1}^{R_k} \sum_{j\in I_{kr}}  \E \big[z_{ik}|\cD;\bsPsi^{(q)}\big] \log  \N  (y_{ij};\bsbeta^T_{kr} \bsx_{ij},\sigma^2_{kr})  \nonumber \\
& =& \sum_{i=1}^{n}\sum_{k=1}^{K} \tau_{ik}^{(q)} \log \alpha_k \! + \! \sum_{i=1}^{n} \sum_{k=1}^{K} \sum_{r=1}^{R_k} \!\!  \sum_{j\in I_{kr}} \!\!  \tau_{ik}^{(q)} \! \log  \N  (y_{ij};\bsbeta^T_{kr} \bsx_{ij},\sigma^2_{kr})  
\label{eq: Q-function for the mixture of piecewise regression}
\end{eqnarray}
where 
\begin{eqnarray}
%
\tau_{ik}^{(q)} = p(z_i=k|\by_{i},\bx_i;\bsPsi^{(q)}) =  \frac{\alpha_k^{(q)}\prod_{r=1}^{R_k}  \prod_{j\in I^{(q)}_{kr}} \N\big(y_{ij};\bsbeta^{T(q)}_{kr}\bsx_{ij},\sigma^{2(q)}_{kr}\big)}{\sum_{k'=1}^K \alpha_{k'}^{(q)}\prod_{r=1}^{R_{k'}}  \prod_{j\in I^{(q)}_{k'r'}} \N(y_{ij};\bsbeta^{T(q)}_{k'r'}\bsx_{ij},\sigma^{2(q)}_{k'r'})} 
\label{eq: post prob tau_ik of cluster k for the mixture of piecewise regression}
\end{eqnarray}
is the posterior probability that the curve $\by_i$ belongs to  the cluster $k$. This step therefore only requires  the computation of the posterior cluster probabilities $\tau^{(q)}_{ik}$ $(i=1,\ldots,n)$ for each of the $K$ clusters.

\paragraph{M-step}
\label{par: M-step of the EM algorithm for the PWRM}  The M-step computes the parameter update $\bsPsi^{(q+1)}$ by maximizing the $Q$-function (\ref{eq: Q-function for the mixture of piecewise regression})  with respect to $\bsPsi$, that is: 
\begin{equation}
\bsPsi^{(q+1)} = \arg \max_{\bsPsi} Q(\bsPsi,\bsPsi^{(q)})\cdot
\label{eq: parameter update by EM for the mixture of rhlp}
\end{equation}
To perform this maximization, it can be seen that the $Q$-function can be decomposed as 
\begin{equation}
Q(\bsPsi,\bsPsi^{(q)})= Q_{\alpha}(\alpha_1,\ldots,\alpha_K,\bsPsi^{(q)}) + \sum_{k=1}^{K}Q_{\Psi_{k}}\big(\{I_{kr},\bsbeta_{kr},\sigma_{kr}^2\}_{r=1}^{R_k},\bsPsi^{(q)}\big),
\label{eq: decomposition of the Q-function for the PWRM}
\end{equation}
where
\begin{equation}
Q_{\alpha}(\alpha_1,\ldots,\alpha_K,\bsPsi^{(q)})= \sum_{i=1}^{n}\sum_{k=1}^{K} \tau_{ik}^{(q)} \log \alpha_r,
\label{eq: Q(alpha cluster prior prob) for the PWRM}
\end{equation}
and
\begin{eqnarray}
Q_{\bsPsi_{k}}(\{I_{kr},\bsbeta_{kr},\sigma_{kr}^2\}_{r=1}^{R_k},\bsPsi^{(q)}) &=& \sum_{r=1}^{R_k} \sum_{i=1}^{n} \sum_{j\in I_{kr}} \tau_{ik}^{(q)} \log \mathcal{N} \left(y_{ij};{\bsbeta}^{T}_{rk}\bsx_{ij},\sigma^2_{rk} \right).
\label{eq: Q(Ik,beta_kr,sigma2_kr) for the PWRM}
\end{eqnarray} The maximization of $Q(\bsPsi,\bsPsi^{(q)})$ can therefore be performed by separate  maximizations of $Q_{\alpha}$ (\ref{eq: Q(alpha cluster prior prob) for the PWRM}) with respect to the mixing proportions $\alpha_k$'s and $Q_{\bsPsi_{k}}$ (\ref{eq: Q(Ik,beta_kr,sigma2_kr) for the PWRM}) with respect to the parameters of each piecewise polynomial regression model $\bsPsi_{k}=\{I_{kr},\bsbeta_{kr},\sigma_{kr}^2\}_{r=1}^{R_k}$  for $k=1,\ldots,K$, as follows.  
The function  $Q_{\alpha}(\alpha_1,\ldots,\alpha_K,\bsPsi^{(q)})$ is maximized with respect to $(\alpha_1,\ldots,\alpha_K) \in[0,1]^R$ subject to the constraint $\sum_{k=1}^K \alpha_k = 1$ using Lagrange multipliers and the updates are given by:
\begin{eqnarray}
\alpha_k^{(q+1)} 
&=& \frac{\sum_{i=1}^n \tau_{ik}^{(q)}}{n},\quad (k=1,\ldots,K).
\label{eq: EM estimate of cluster prior prob alpha_k for the mixture of piecewise regression}
\end{eqnarray}

The maximization of (\ref{eq: Q(Ik,beta_kr,sigma2_kr) for the PWRM}) 
 corresponds to finding the new update of $\bsPsi_k$, that is the piecewise segmentation $\{I_{kr}\}$ of cluster $k$ and the corresponding piecewise regression representation through $\{\bsbeta_{kr},\sigma_{kr}^2\}$, $(r=1,\ldots,R_k)$, to the fuzzy cluster $k$ which is composed of the $n$ curves weighted by their posterior probabilities relative to cluster $k$. 
Thus, one can observe that each of the maximizations of (\ref{eq: Q(Ik,beta_kr,sigma2_kr) for the PWRM}) 
corresponds to a weighted version of the piecewise regression problem for a set of curves given by Equation (\ref{eq: loglik for the functional piecewise regression}), the weights being the posterior cluster probabilities $\tau_{ik}^{(q)}$.
Optimizing $Q_{\bsPsi_{k}}$ therefore simply consists in solving a weighted piecewise regression problem where the curves are weighted by  
 the posterior cluster probabilities $\tau_{ik}^{(q)}$.
The optimal segmentation of each cluster $k$, represented by the parameters $\{\bsxi_{kr}\}$ is performed by running a dynamic programming procedure similarly as in Section \ref{ssec: dynamic programming for the functional pwr}
Equation (\ref{eq: dynamic programming criterion C for a set of curves}) by weighting the optimization problem. The updating rules for the regression parameters for each cluster of curves correspond to weighted versions of (\ref{eq: estimation of beta_r in piecewise reg}) and (\ref{eq: estimation of sigma_r in piecewise reg}), and are given by:
{\begin{eqnarray}
 \bsbeta^{(q+1)}_{kr} & = & \Big[ \sum_{i=1}^n  \tau_{ik}^{(q)} \bX_{ir}^T \bX_{ir}  \Big]^{-1} \sum_{i=1}^n \bX_{ir}\by_{ir} \\
\label{eq: EM estimate of reg coeff beta_kr for the PWRM}
\sigma_{kr}^{2(q+1)} &=& \frac{1}{ \sum_{i=1}^n\sum_{j\in I_{kr}^{(q)}} \tau_{ik}^{(q)}}  \sum_{i=1}^n \tau_{ik}^{(q)} |\!|(\by_{ir}-\bX_{ir} \bsbeta^{(q+1)}_{kr})|\!|^2
\label{eq: EM estimeate of variance sigma^2_kr for the PWRM}
\end{eqnarray}where $\by_{ir}$ is the segment (regime) $r$ of the $i$th curve, that is the observations $\{y_{ij}|j\in I_{kr}\}$ and
 $\bX_{ir}$ its associated design matrix with rows $\{\bsx_{ij}|j\in I_{kr}\}$. 
Thus, the proposed EM algorithm for the PWRM model provides a fuzzy partition of the curves into $K$ clusters through the posterior cluster probabilities $\tau_{ik}$, each fuzzy cluster is optimally segmented into regimes with indexes $\{I_{kr}\}$. 
At convergence of the EM algorithm, a hard partition of the curves can then be deduced by assigning each curve to the cluster which maximizes the posterior probability (\ref{eq: post prob tau_ik of cluster k for the mixture of piecewise regression}), that is:
\begin{equation}
\hat{z}_i = \arg \max_{1\leq k \leq K} \tau_{ik}(\hat{\bsPsi}), \quad (i=1,\ldots,n).
\end{equation}
where $\hat{z}_i$ denotes the estimated class label for the $i$th curve.

\medskip
To summarize, the proposed EM algorithm computes the maximum likelihood (ML) estimate of the PWRM model. It simultaneously updates a fuzzy partition of the curves into $K$ clusters and an optimal segmentation of each cluster into regimes.  At convergence, we obtain the model parameters that include the segments boundaries and the fuzzy clusters. A hard partition of the curves into $K$ clusters is then deduced according to the MAP principle by maximizing the posterior cluster probabilities.
%

We notice that a similar algorithm for segmentation clustering is proposed in \citep{picardetal2007}. This approach uses a dynamic programming procedure with the EM algorithm to segment the temporal gene expression data and the clustering is performed on the segments to assign each set of  homogeneous segments to a cluster relative to the spatial behavior of such data. The PWRM model proposed here is quite different from its mixture formulation in the sense that here the curves are supposed to be mixed at random rather than the segments, so that each cluster is composed of a set of homogeneous temporal curves segmented into heterogeneous segments.

\medskip

As noticed in the introduction, we propose another scheme to achieve both the model estimation (including the segmentation) and the clustering by using a dedicated Classification EM (CEM) algorithm. In the next section we present the classification approach with the corresponding classification likelihood criterion, and derive  the CEM algorithm to maximize it.  

\section{Maximum classification likelihood estimation via a dedicated Classification EM algorithm}
\label{ssec: ML estimation vi EM for the piecewise polynomial regression mixture}

The maximum classification likelihood approach simultaneously performs  the clustering and  the parameter estimation, which includes the curves segmentation, by maximizing the completed-data log-likelihood given by Equation (\ref{eq: complete log-lik for the piecewise regression mixture}) for the proposed PWRM model. The maximization is performed through a dedicated Classification EM (CEM) algorithm. 

\subsection{The CEM algorithm for piecewise regression mixture (CEM-PWRM)}
\label{sssec: CEM for  piecewise regression mixture}

The CEM algorithm \citep{celeuxetgovaert92_CEM} was initially proposed for model-based clustering of multivariate data. We adopt it here in order to perform model-based curve clustering with the proposed PWRM model.
The resulting CEM simultaneously estimates both the PWRM parameters and the classes' labels by maximizing the complete-data log-likelihood given by Equation (\ref{eq: complete log-lik for the piecewise regression mixture}) w.r.t the model parameters $\bsPsi$ and the partition represented by the vector of cluster labels $\bz$, in an iterative manner as follows. 
After starting with an initial mixture model parameters $\bsPsi^{(0)}$ (e.g., computed from a randomly chosen partition and a uniform segmentation), the CEM-PWRM algorithm alternates between the two following steps at each iteration $q$ until convergence (e.g., when there is no longer change  in the partition or in the relative variation of the complete-data log-likelihood):

\paragraph{Step 1}
\label{par: Step 1 of CEM for the piecewise regression mixture}

The first step updates the cluster labels for the current  model defined by  $\bsPsi^{(q)}$ by maximizing the complete-data log-likelihood   (\ref{eq: complete log-lik for the piecewise regression mixture}) w.r.t to the cluster labels $\bz$, that is:
\begin{equation}
\bz^{(q+1)} = \arg \max_\bz \cL_c(\bz,\bsPsi^{(q)}).
\end{equation}

\paragraph{Step 2}
\label{par: Step 2 of CEM for the piecewise regression mixture}
Given the estimated partition defined by $\bz^{(q+1)}$, the second step updates the model parameters by maximizing the complete-data log-likelihood w.r.t to the PWRM parameters $\bsPsi$:
\begin{equation}
\bsPsi^{(q+1)} = \arg \max_{\bsPsi} \cL_c(\bz^{(q+1)},\bsPsi).
\end{equation}
Equivalently, the CEM algorithm therefore consists in integrating a classification step (C-step) between the E- and the M- steps of the EM algorithm presented previously.
In this case of the proposed PWRM model, the dedicated CEM-PWRM algorithm runs as follows. It consists in starting with an initial model parameters $\bsPsi^{(0)}$ and alternating between the three following steps at each iteration $q$ until convergence. 
\paragraph{E-step}
\label{par: E-step mixture of piecewise regression CEM}
The E-step computes the posterior probabilities $\tau^{(q)}_{ik}$ $(i=1,\ldots,n)$ given by Equation (\ref{eq: post prob tau_ik of cluster k for the mixture of piecewise regression}), that the $i$th curve belongs to cluster $k$ for $i=1,\ldots,n$ and for each of the $K$ clusters.

\paragraph{C-step}
\label{par: C-step mixture of piecewise regression CEM}

The C-step computes a hard partition of the $n$ curves into $K$ clusters by estimating the cluster labels through the MAP rule: 
\begin{equation}
	z_i^{(q+1)}  = \arg \max_{1\leq k \leq K} \tau^{(q)}_{ik} \  (i=1,\ldots,n). 
\label{eq: C-Step cluster labels update for the mixture of piecewise regression}
\end{equation}

\paragraph{M-step}
\label{par: M-step mixture of rhlp and CEM} 
Finally,  given the estimated cluster labels $\bz^{(q+1)}$, the M-step updates the model parameters by computing the parameter vector $\bsPsi^{(q+1)}$ which maximizes the complete-data log-likelihood  (\ref{eq: complete log-lik for the piecewise regression mixture})  with respect to $\bsPsi$.  
By rewriting the complete-data log-likelihood given the current estimated partition as
\begin{eqnarray} 
\cL_c(\bsPsi,\bz^{(q+1)}) = \sum_{k=1}^{K}\sum_{i|z^{(q)}_i=k}  \log \alpha_k 
+  \sum_{k=1}^{K} \sum_{r=1}^{R_k} \sum_{i|z^{(q)}_i=k} \sum_{j\in I_{kr}}  \log  \N  (y_{ij};\bsbeta^T_{kr} \bsx_{ij},\sigma^2_{kr})   
\label{eq: complete log-lik for thePWRM v-Mstep}
\end{eqnarray}
we can see that this function can be optimized by separately optimizing the two terms of the r.h.s of (\ref{eq: complete log-lik for thePWRM v-Mstep}). 
More specifically, the mixing proportions $\alpha_k$'s are updated by maximizing the function  $\sum_{i=1}^{n}\sum_{k=1}^{K} z_{ik}^{(q+1)} \log \alpha_k$ w.r.t $(\alpha_1,\ldots,\alpha_K) \in[0,1]^K$ subject to the constraint $\sum_{k=1}^K \alpha_k = 1$. This is performed by using Lagrange multipliers and gives the following updates:
\begin{eqnarray}
\alpha_k^{(q+1)} &=& \frac{1}{n}\sum_{i=1}^n z_{ik}^{(q)} \quad (k=1,\ldots,K).
\label{eq: CEM estimate of cluster prior prob alpha_k for the mixture of piecewise regression}
\end{eqnarray} 
The regression parameters and the segmentation which are denoted by $\{\bsPsi_k\}=\{(\bstheta_k,\bsxi_k)\}$ for each of the $K$ clusters are updated by maximizing the second term of the r.h.s of (\ref{eq: complete log-lik for thePWRM v-Mstep}) similarly as in the case of  the EM-PWRM algorithm presented in the previous section. 
%
The only difference is that the posterior probabilities $\tau_{ik}$ in the case of the EM-PWRM algorithm are replaced by the cluster label indicators $z_{ik}$ when using the CEM-PWRM; The curves being assigned in a hard way rather than in a soft way.
This step consists therefore in  estimating a piecewise polynomial regression model for the set of curves of each of the $K$ clusters separately. Each polynomial regression model  estimation for each cluster of curves is  performed using a dynamic programming procedure as in seen in Section \ref{ssec: functional piecewise polynomial regression model}. 


\subsection{The CEM-PWRM algorithm as to generalize the $K$-means-like algorithm}
In this section we show how the proposed PWRM estimated by the CEM algorithm provides a general framework for the  $K$-means-like algorithm of \citep{hebrailEtal:2010} seen in Section \ref{ssec: piecewise polynomial regression and K-means like}.
\begin{prop}
\label{prop. CEM view of the Kmeans like}
The complete-data log-likelihood (\ref{eq: complete log-lik for the piecewise regression mixture}) optimized by the proposed CEM algorithm for the piecewise regression mixture model, is equivalent to the distance criterion (\ref{eq: distance criterion for clustering with piecewise regression}) optimized by the $K$-means-like algorithm of \citep{hebrailEtal:2010} if the following constraints are imposed:
\begin{itemize}
\item  $\alpha_k = \frac{1}{K}$ $\forall K$ (identical mixing proportions)
\item $\sigma^2_{kr} = \sigma^2$ $\forall r=1,\ldots,{R_k}$ and $\forall k=1,\ldots,K$ 
(isotropic and homoskedastic model)
\item piecewise constant approximation of each segment of curves rather than a polynomial fitting.
\end{itemize}
 Therefore, the proposed CEM algorithm for piecewise polynomial regression mixture is the  probabilistic version for hard curve clustering and optimal segmentation of the $K$-means-like algorithm (c.f., Section \ref{ssec: k-means like algorithm}). 
\end{prop} 
\noindent \textit{Proof.} The complete data log-likelihood (\ref{eq: complete log-lik for the piecewise regression mixture}) can be rewritten as:
{\begin{eqnarray} 
\cL_c(\bsPsi,\bz) = \sum_{k=1}^{K} \sum_{i|z_i=k} \log \alpha_k 
 -\frac{1}{2}\sum_{k=1}^{K} \sum_{i|z_i=k} \sum_{r=1}^{R_k} \sum_{j\in I_{kr}}\big[\big(\frac{y_{ij}-\bsbeta^T_{kr} \bsx_{ij}}{\sigma^2_{kr}}\big)^2  + \log(2 \pi\sigma_{kr}^2) \big].
\label{eq: complete log-lik for the PWRM v2}
\end{eqnarray}}
Then, if we consider the constraints in Proposition \ref{prop. CEM view of the Kmeans like}
for the proposed PWRM model, the maximized complete-data log-likelihood takes the following form:
\begin{eqnarray}
\cL_c(\bsPsi,\bz) = \sum_{k=1}^{K} \sum_{i|z_i=k} \log \frac{1}{K} -\frac{1}{2}\sum_{k=1}^{K} \sum_{i|z_i=k} \sum_{r=1}^{R_k} \sum_{j\in I_{kr}}\big[\big(\frac{y_{ij}-\mu_{kr}}{\sigma^2}\big)^2  + \log(2 \pi\sigma^2) \big]. 
\label{eq: complete log-lik for the PWRM v proof}
\end{eqnarray}
Maximizing this function is therefore equivalent to minimizing the following criterion w.r.t the cluster labels $\bz$ and the segments indices $I_{kr}$ and the segments constant means $\mu_{kr}$:   
\begin{equation}
 \cJ \big(\bz, \{ \mu_{kr},I_{kr}\}\big)  = \sum_{k=1}^K \sum_{r=1}^{R_k} \sum_{i|z_i=k} \sum_{j\in I_{kr}} \big(y_{ij}-\mu_{kr}\big)^2 
 \label{eq: criterion J for the PWR mixture and CEM}
\end{equation} which is exactly the distortion  criterion optimized by the $K$-means-like algorithm of \cite{hebrailEtal:2010} (c.f., Equation (\ref{eq: distance criterion for clustering with piecewise regression})).
 

\subsection{Model selection}
\label{ssec. model selection}
The problem of model selection here is equivalent to the one of choosing the optimal number of clusters $K$, the number of regimes $R$ and  the polynomial degree $p$. The optimal value  of the triplet  $(K,R,p)$ can be computed by using some model
selection criteria such as the Bayesian Information
Criterion (BIC) \citep{BIC} similarly as in \cite{liuANDyangFunctionalDataClustering} or the Integrated Classification Likelihood criterion (ICL)  \citep{ICL}, etc. 
Let us recall that BIC is a penalized log-likelihood criterion which can be defined as a function to be maximized that is given by:
$\mbox{BIC}(K,R,p) = \cL(\hat{\bsPsi}) - \frac{\nu_{\bsPsi} \log(n)}{2},$ while ICL consists in a penalized complete-data log-likelihood and can be expressed as follows:
$\ICL(K,R,p) = \cL_c(\hat{\bsPsi}) - \frac{\nu_{\bsPsi} \log(n)}{2}$, where $\cL(\hat{\bsPsi})$ and  $\cL_c(\hat{\bsPsi})$ are respectively the incomplete (observed) data log-likelihood and the complete data log-likelihood, obtained at convergence of the (C)EM algorithm,  $\nu_{\bsPsi} =  \sum_{k=1}^K R_k(p+3) -1$ is the number of free parameters of the model and $n$ is the sample size. 
The number of free model parameters includes 
$K-1$ mixing proportions, $\sum_{k=1}^K R_k(p+1)$ polynomial coefficients, $\sum_{k=1}^K R_k$ noise variances and $\sum_{k=1}^K (R_k - 1)$ transition points.  

\section{Experimental study} 
\label{sec: Experimental study}

In this section, we assess the proposed PWRM with both the EM and CEM algorithms in terms for curve clustering and segmentation. 
We study the performance of the developed PWRM model by comparing it to the polynomial regression mixture models (PRM) \citep{Gaffneythesis}, the standard polynomial spline regression  mixture model (PSRM) \citep{Gaffneythesis, Gui_FMDA, liuANDyangFunctionalDataClustering} and the piecewise regression model used with the $K$-means-like algorithm \citep{hebrailEtal:2010}. We also include comparisons with standard 
model-based clustering of multivariate data using Gaussian mixture models (GMM). For all the compared generative approaches we consider both the EM and the CEM algorithms. Thus, the ten compared approaches can be summarized as follows: 
EM-GMM, EM-PRM, EM-PRM, EM-PSRM, $K$-means-like, EM-PWRM and CEM-PWRM. 
All algorithms have been implemented in Matlab. 
The aim of including the standard multivariate data clustering with Gaussian mixtures models and the EM algorithm is to show that  it is necessary to adapt them to curve clustering approaches as they do not account for the functional structure of the data. 
The algorithms are evaluated using  experiments conducted on both synthetic and real curves.

\subsection{Evaluation criteria}
\label{sssec: evaluation criteria}

The algorithms are evaluated in terms of curves classification and  approximation accuracy.
The used evaluation criteria  are the classification error rate between the true simulated partition and the estimated partition, and the intra-cluster inertia $\sum_{k=1}^K\sum_{i|\hat{z}_i=k} |\!|\by_i-\hat{\by}_k|\!|^2,$ where $\hat{z}_i$ indicates the estimated class label of the curve $\by_i$ and $\hat{\by}_k= (\hat{y}_{kj})_{j=1,\ldots,m}$  is the estimated mean curve of cluster $k$. Each point of the mean curve of cluster $k$ is given by:
\begin{itemize}
\item $\hat{y}_{kj} =  \hat{\bsbeta}^T_{kr} \bsx_{ij}$ if $j \in \hat{I}_{kr}$ for the proposed approach (EM-PWRM, CEM-PWRM) and the $K$-means-like approach 
of \cite{hebrailEtal:2010}, 
\item $\hat{y}_{kj}=  \hat{\bsbeta}^T_k \bsx_{ij}$ for both the polynomial regression mixture (PRM) and the spline regression mixtures (PSRM),
\item $\hat{y}_{kj} = \frac{\sum_{i=1}^n \hat{z}_{ik} y_{ij}}{\sum_{i=1}^n \hat{z}_{ik}} $ for the standard model-based clustering with GMM.
\end{itemize}

\subsection{Experiments with simulated curves} 
\label{ssec: Experiments on simulated curves} 

\subsubsection{Simulation protocol and algorithms setting}

The simulated data consisted of curves issued from a mixture of two classes, each class is simulated as piecewise linear function corrupted by a Gaussian noise.  More specifically,  the simulated curves consisted of $n=100$ curves of $m=160$ regularly sampled observations at the discrete time points $\bt=(1,\ldots,m)$.
The curves are mixed in proportion randomly with mixing proportions $\alpha_k,\ (k=1,2)$. We first considered uniform mixing proportions and then varied the proportions between the two classes as to have a non-uniformly mixed classes. In the simulated curves, we consider variation in mean, variance, and regime shape (constant, linear).
%
 %
 %
%
Table \ref{table. simulation parameters} shows the  used  simulation parameters to generate each curve $\by_i = (y_{ij})_{j=1}^m$
and Figure \ref{fig: example of simulated curves} shows an example of simulated curves for this situation. 
\begin{table}[H]
\centering
{\small
\begin{tabular}{l | l l | l l}
\hline 
regime &  \multicolumn{2}{c}{cluster $k=1$} & \multicolumn{2}{c}{cluster $k=2$} \\
\hline
r=1 & $[5 + \sigma_{11} e_{ij}]  \indic_{[1,20]}$ & $\sigma_{11} = 0.8$ & 
$[5 + \sigma_{11}\ e_{ij}]\indic_{[1,20]}$ & $\sigma_{21} = 0.8$ \\
r=2& $[0.125 j + 2.5 + \sigma_{12} e_{ij}]\indic_{]20,60]}$ & $\sigma_{12} = 0.8$ 
& $[0.1  j + 3 + \sigma_{22} e_{ij}]\indic_{]20,70]}$ & $\sigma_{22} = 0.8$\\
r=3& $[10 + \sigma_{13} e_{ij}]\indic_{]60,115]}$ & $\sigma_{22} = 0.6$
& $[10 + \sigma_{23} e_{ij}]\indic_{]70,90]}$ & $\sigma_{22} = 0.8$\\
r=4& $[10 + \sigma_{14} e_{ij}]\indic_{]115,140]}$ & $\sigma_{22} = 0.8$
& $[10 + \sigma_{24} e_{ij}]\indic_{]90,140]}$ & $\sigma_{22} = 0.6$ \\
r=5& $[6 + \sigma_{15} e_{ij}] \indic_{]140,160]}$ & $\sigma_{22} = 0.8$ 
& $[5.5 + \sigma_{25} e_{ij}]\indic_{]140,160]}$ & $\sigma_{22} = 0.8$\\
& $\bsxi_{1} =  [1, 20, 60, 115, 140, 160]$ & &  $\bsxi_{2} = [1, 20, 70, 90, 140, 160]$ & \\
\hline
\end{tabular}}
\caption{\label{table. simulation parameters}Simulation parameters: $\sigma_{kr}$ represents the noise standard deviation for regime $r$ of cluster $k$, $\bsxi_k$ the transition points within cluster $k$, and  $e_{j}\sim \N(0,1)$ are zero-mean unit-variance Gaussian variables representing an additive noise.}
\end{table}

\subsubsection{Algorithms setting}
\label{ssec: Algorithms setting}

The algorithms are initialized from a random partition 
for the clustering. For the segmentation, the models performing segmentation are initialized from random contiguous segmentations, including a uniform segmentation. The algorithms are stopped when the relative variation of the optimized criterion 
between two iterations is less than a predefined threshold ($10^{-6}$). 
For the same model parameters, the results are computed for 20 different data sets, and for each data, we performed $10$ runs of each algorithm EM and the solution providing the best value of the optimized criterion is chosen.
%
%

\subsubsection{Obtained results}
 We applied the different models on the simulated curves where for the piecewise regression model we trained it with linear polynomial regimes ($p=1$). For the polynomial regression mixture (PRM), it was trained with a polynomial degree $p=10$. For the polynomial spline regression mixture (PSRM), we used cubic splines (of degree $p=3$) with 20 uniformly placed internal knots. 
In terms of numerical results, Table \ref{table. calssif and inertia results for simulations unifPWRM} gives the obtained 
 intra-cluster inertias. 
For this situation, which is extremely difficult, all the algorithms retrieve the actual partition (misclassification error of 0\% for all the algorithms). However, in terms of curves approximation, we can clearly see that, on the one hand, the standard model-based clustering using the GMM is not adapted as it does not take into account the functional structure of the curves and therefore does not account for the smoothness, they rather compute an overall mean curve. On the other hand, the proposed probabilistic approach (EM-PWRM, CEM-PWRM) and the one of \cite{hebrailEtal:2010} which we denoted here by $K$-means-like, as expected, provide the same results in terms of clustering and segmentation. This is attributed to the fact that the  $K$-means PWRM approach is a particular case of our probabilistic approach.
%
\begin{table}[H]
\centering
\small
\begin{tabular}{lcccccc}
\hline
 EM-GMM & EM-PRM 	& EM-PSRM 	& $K$-means-like &  EM-PWRM		& CEM-PWRM		\\
\hline
\hline
19639 &  25317  &  21539 & 17428 &  17428 & 17428\\ 
\hline
\end{tabular}
\caption{\label{table. calssif and inertia results for simulations unifPWRM}Intra-class inertia for the simulated curves}
\end{table}
Figure \ref{fig. clustering results situation 1} shows the different clustering and segmentation results for the simulated curves given in Figure \ref{fig: example of simulated curves}. 
\begin{figure}[H] 	
\centering
\includegraphics[width=4.5cm]{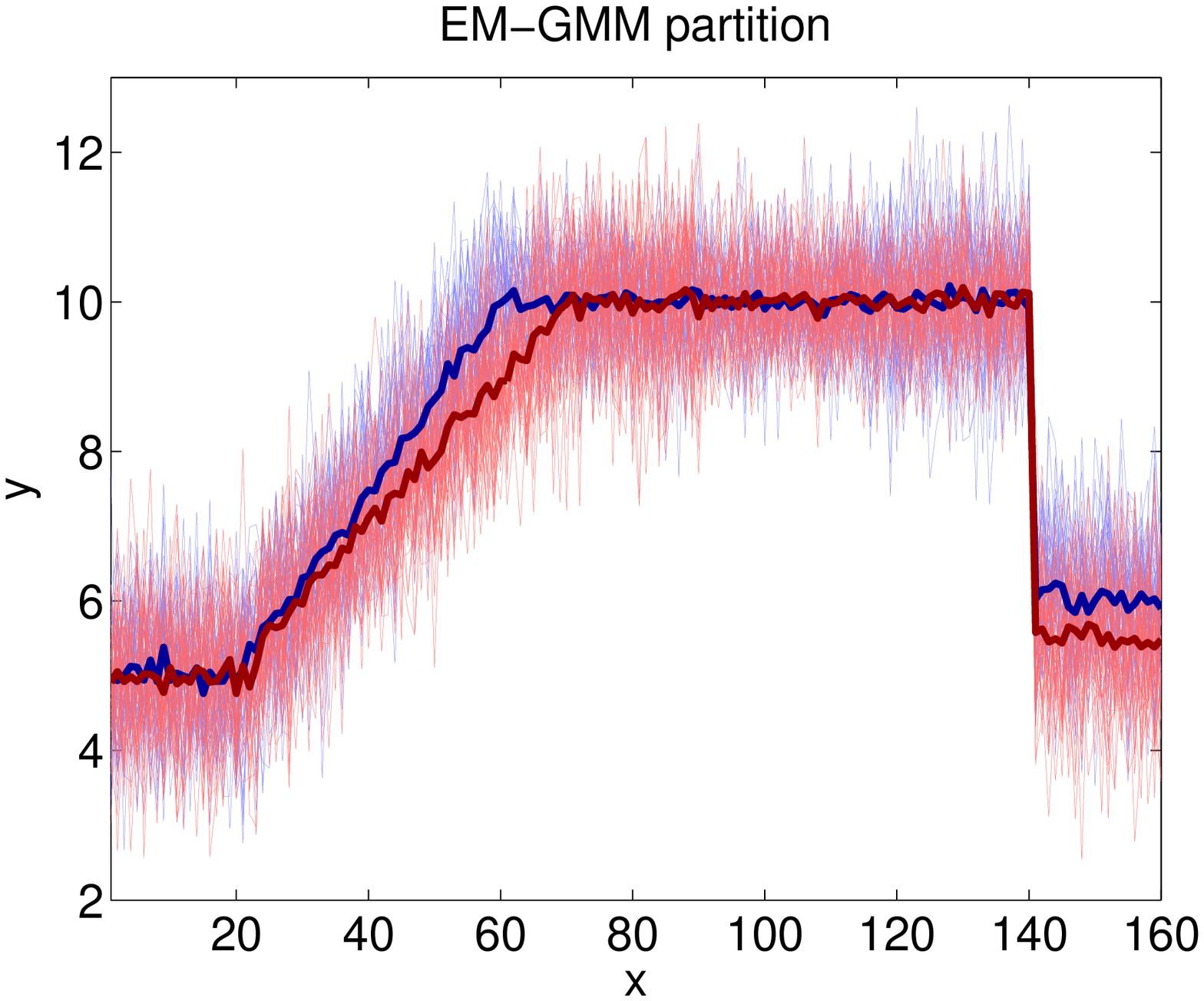}
\includegraphics[width=4.5cm]{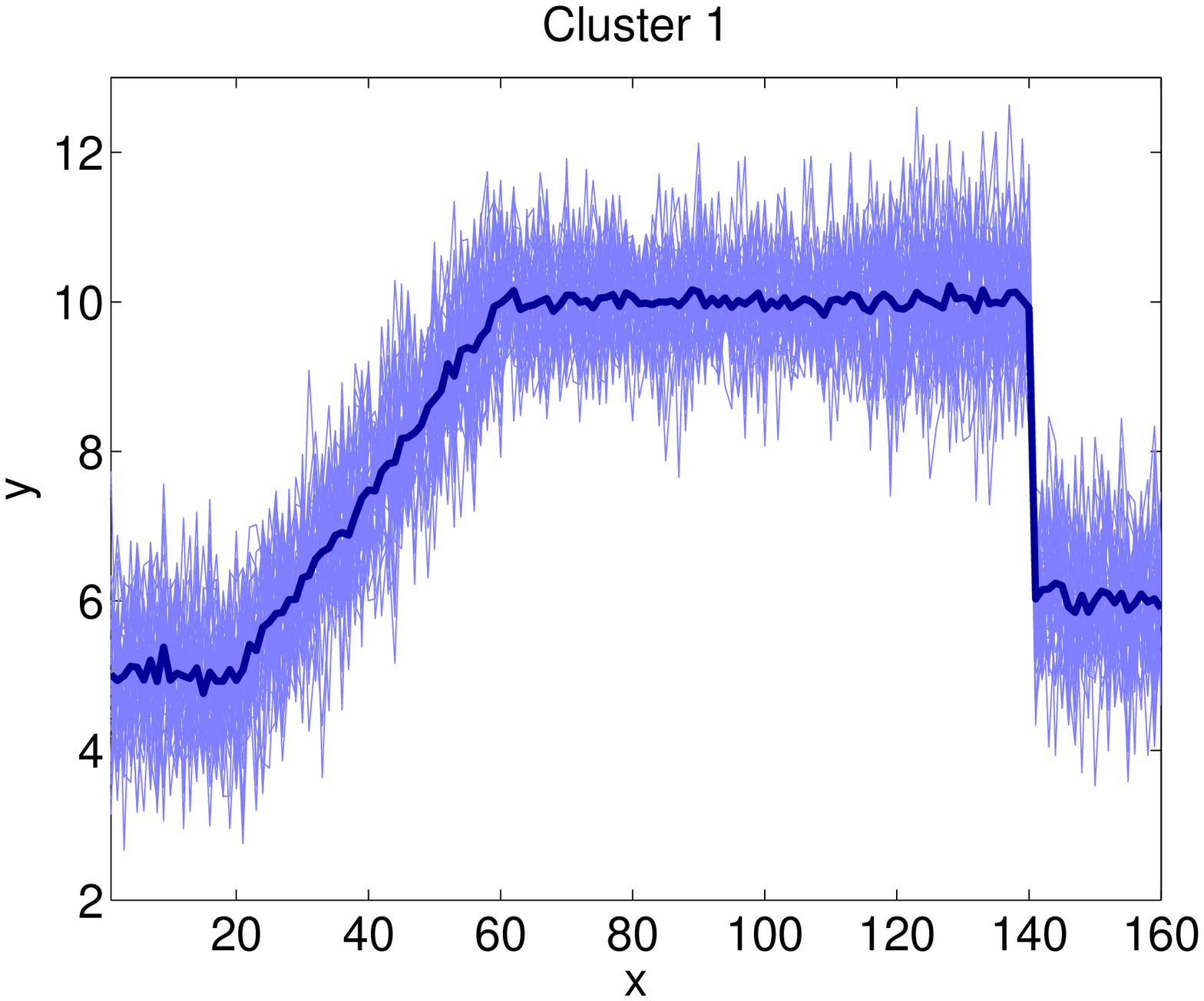}
\includegraphics[width=4.5cm]{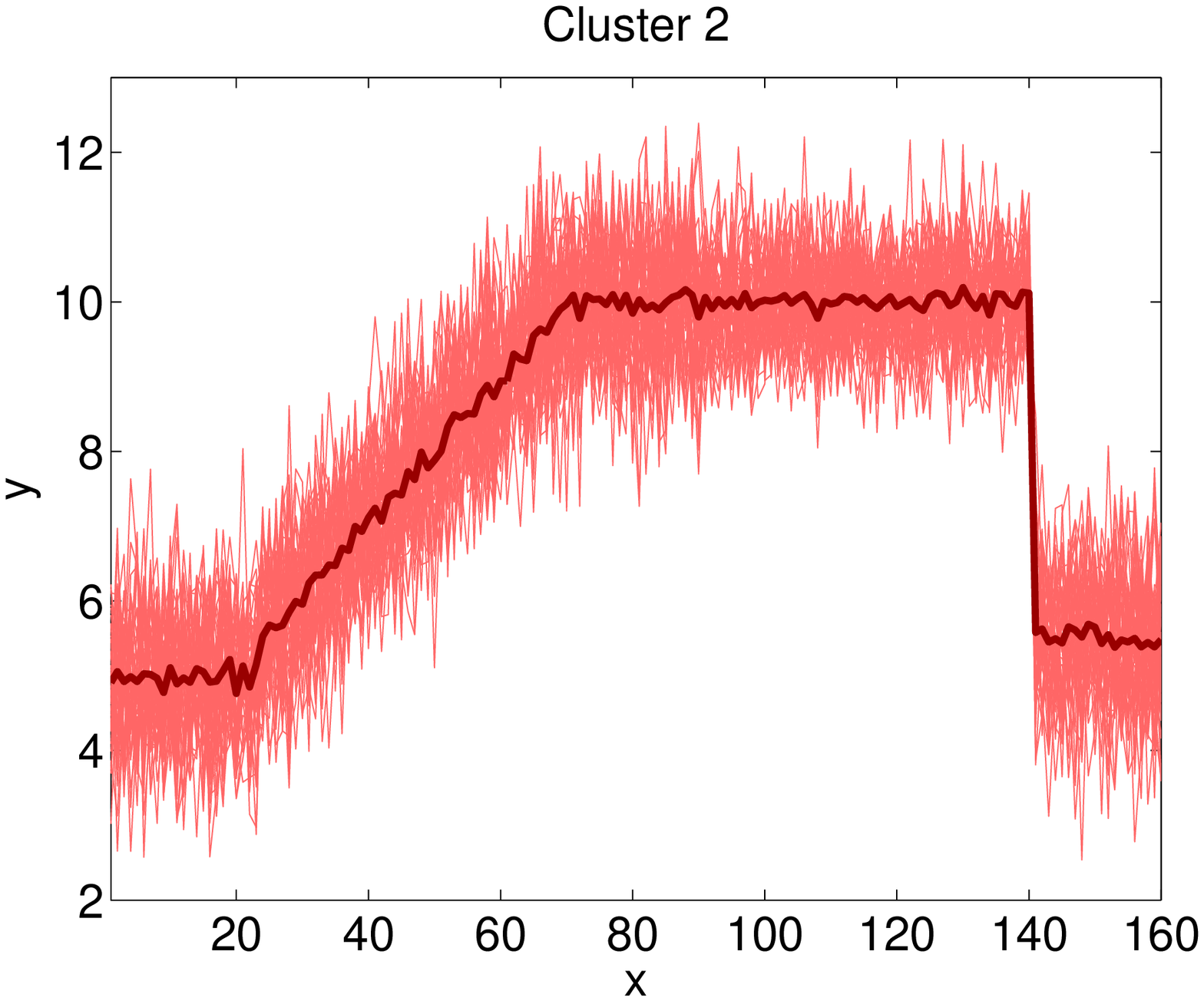}\\
\includegraphics[width=4.5cm]{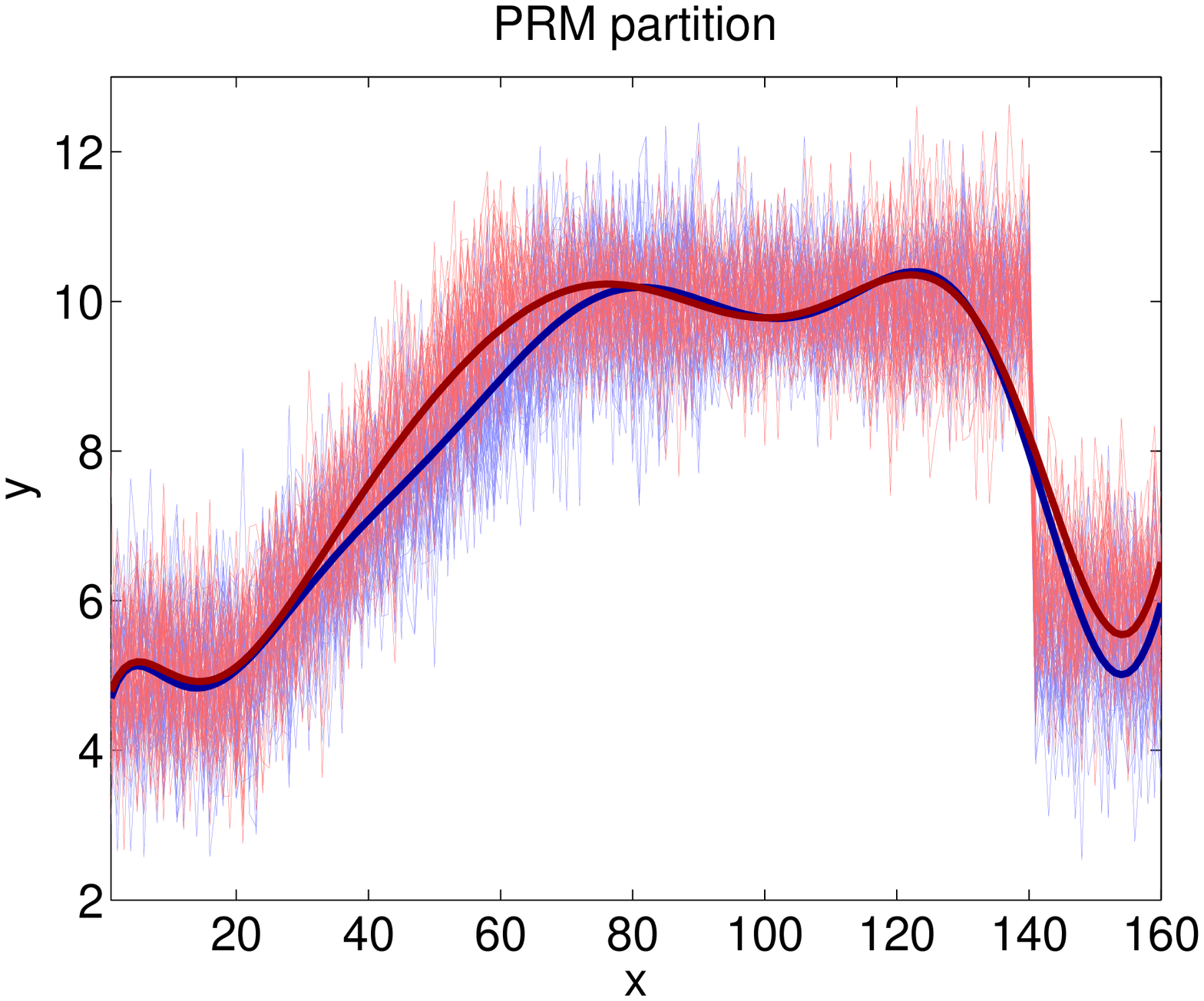}
\includegraphics[width=4.5cm]{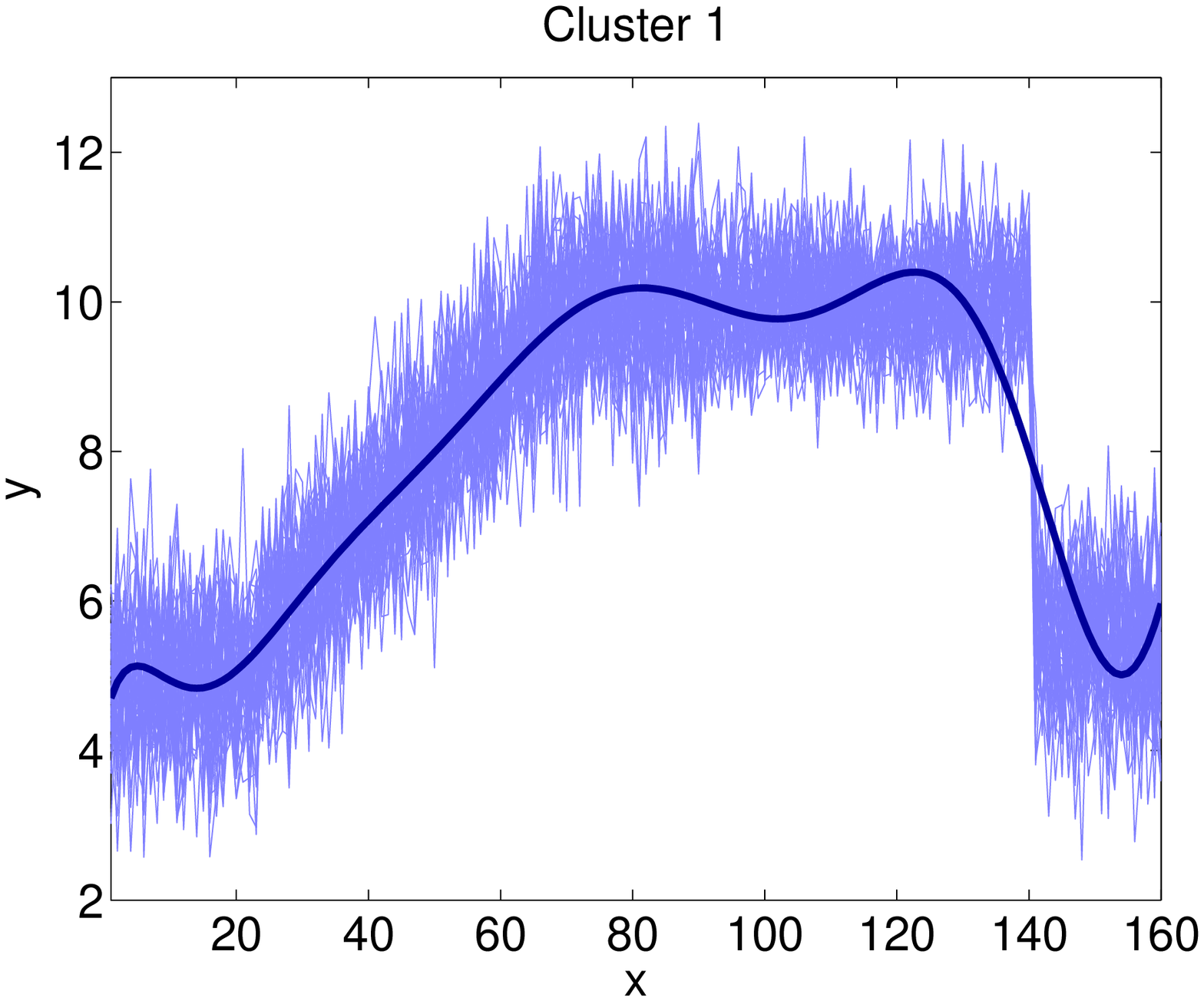}
\includegraphics[width=4.5cm]{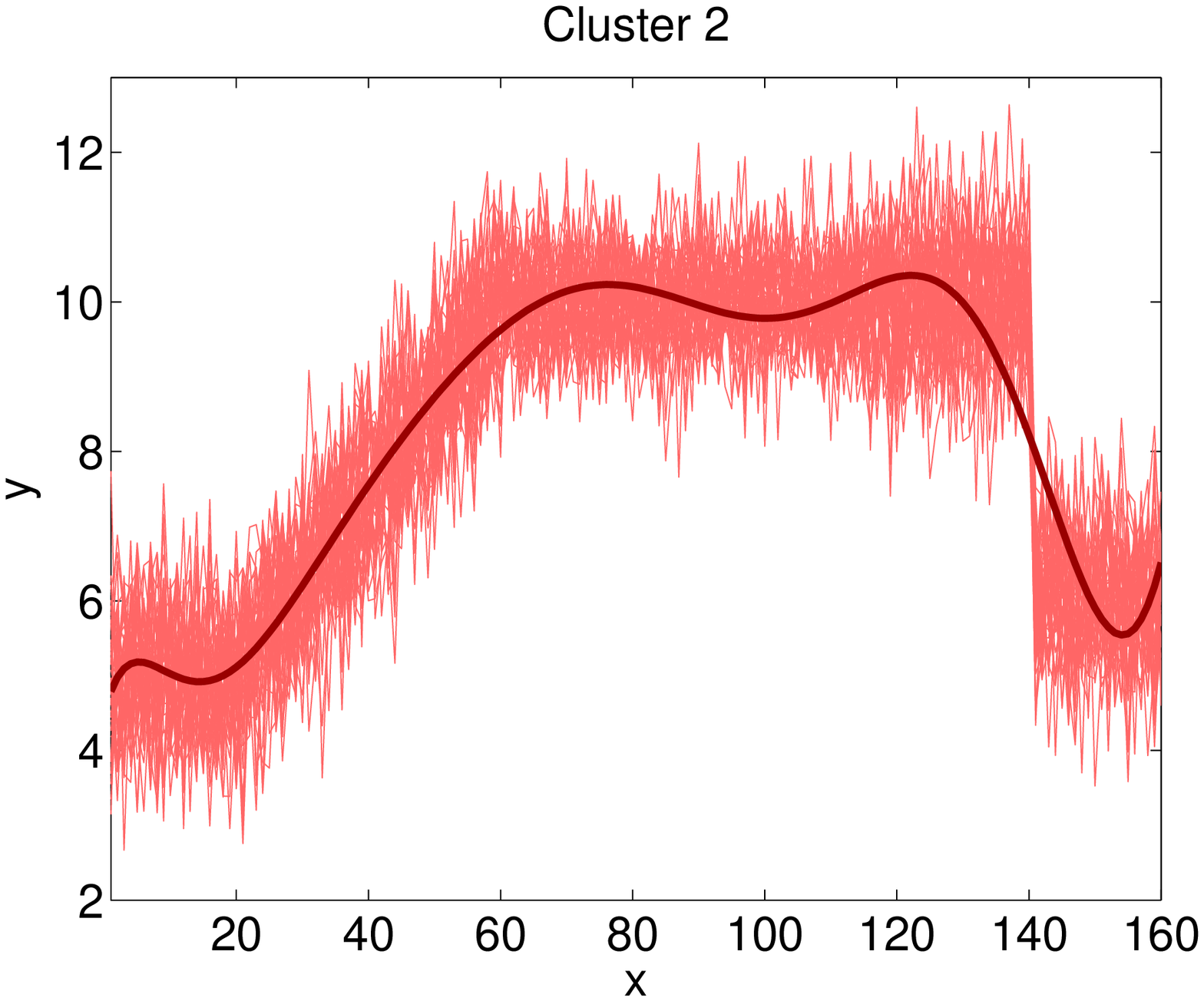}\\
\includegraphics[width=4.5cm]{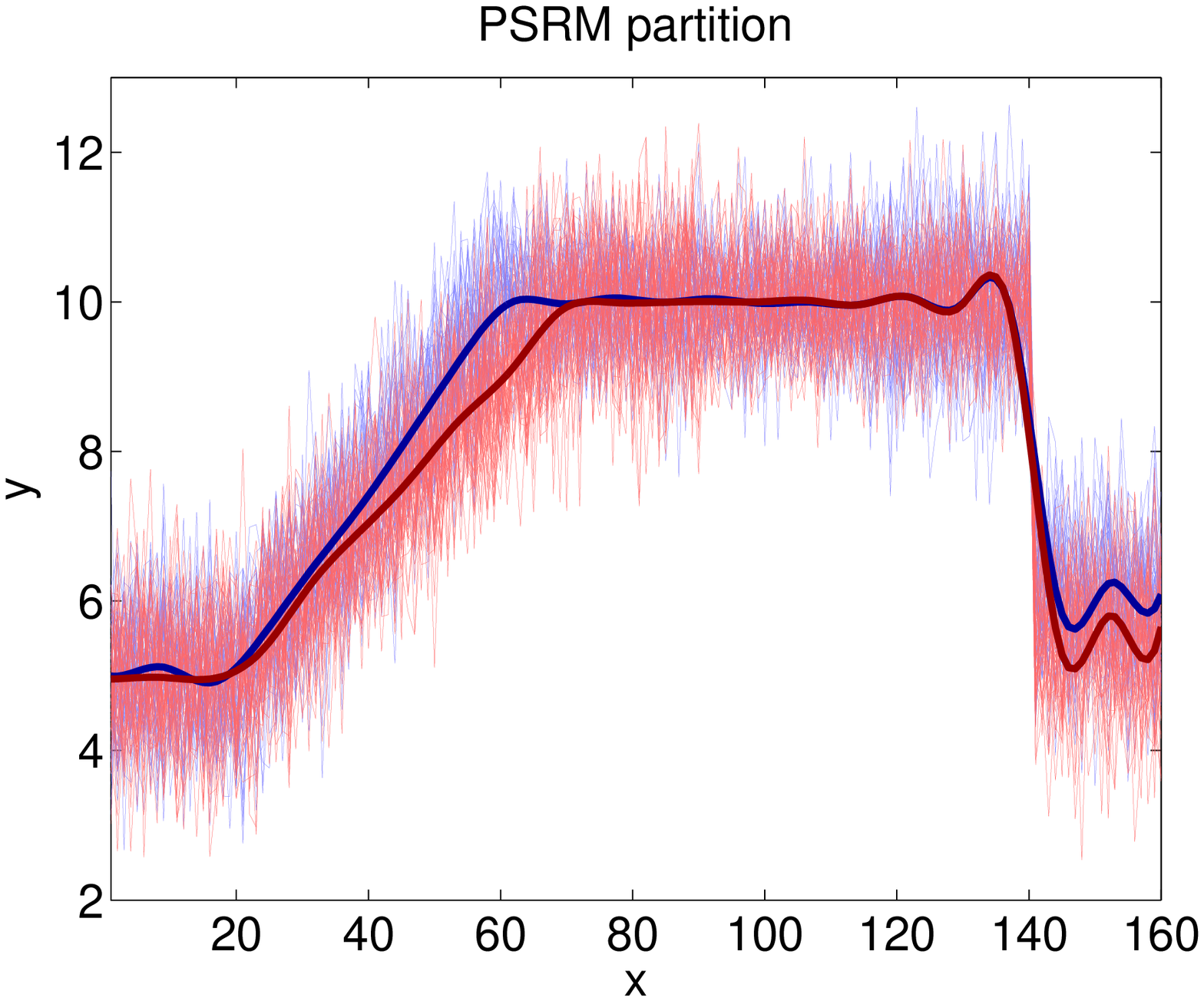}
\includegraphics[width=4.5cm]{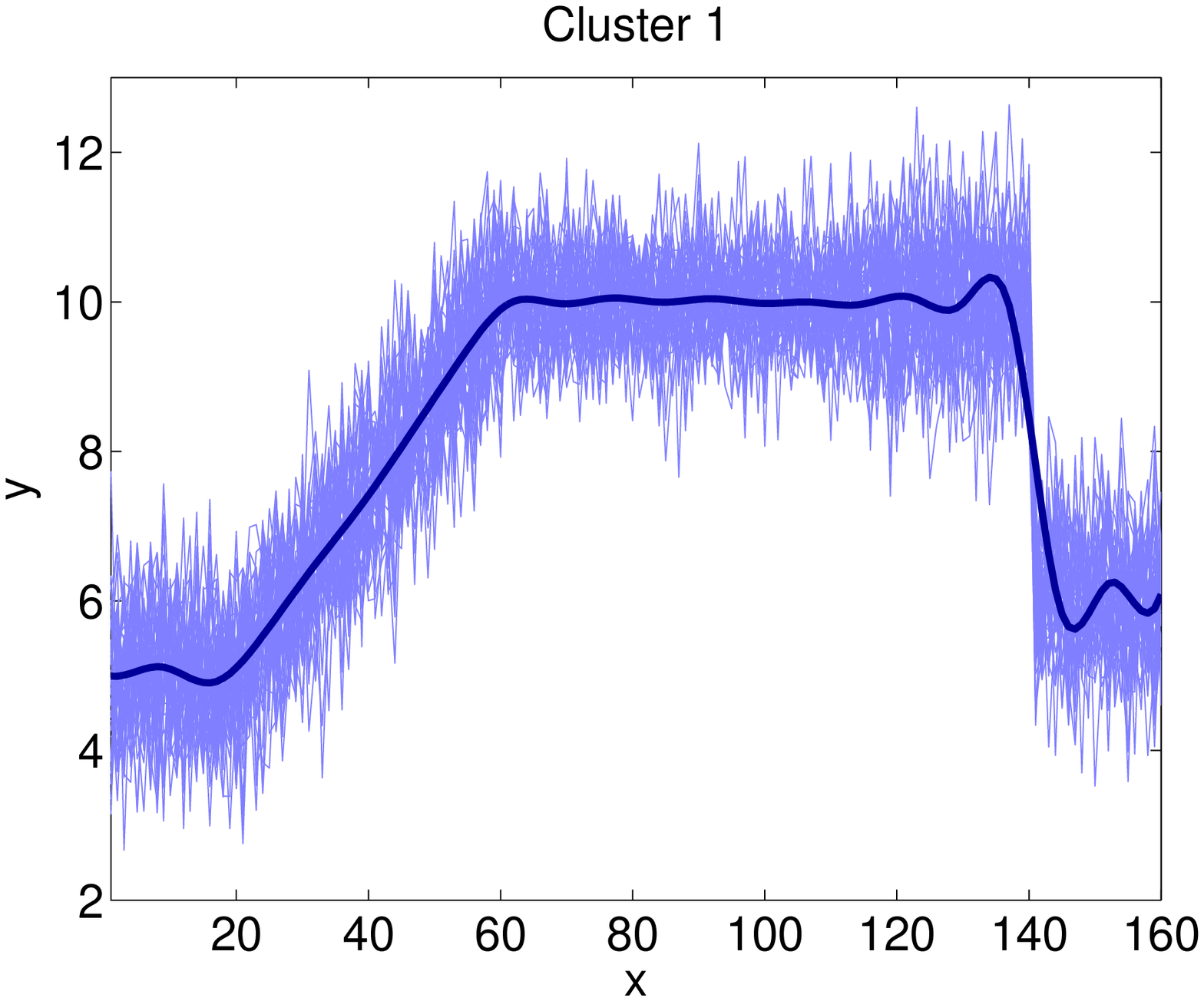}
\includegraphics[width=4.5cm]{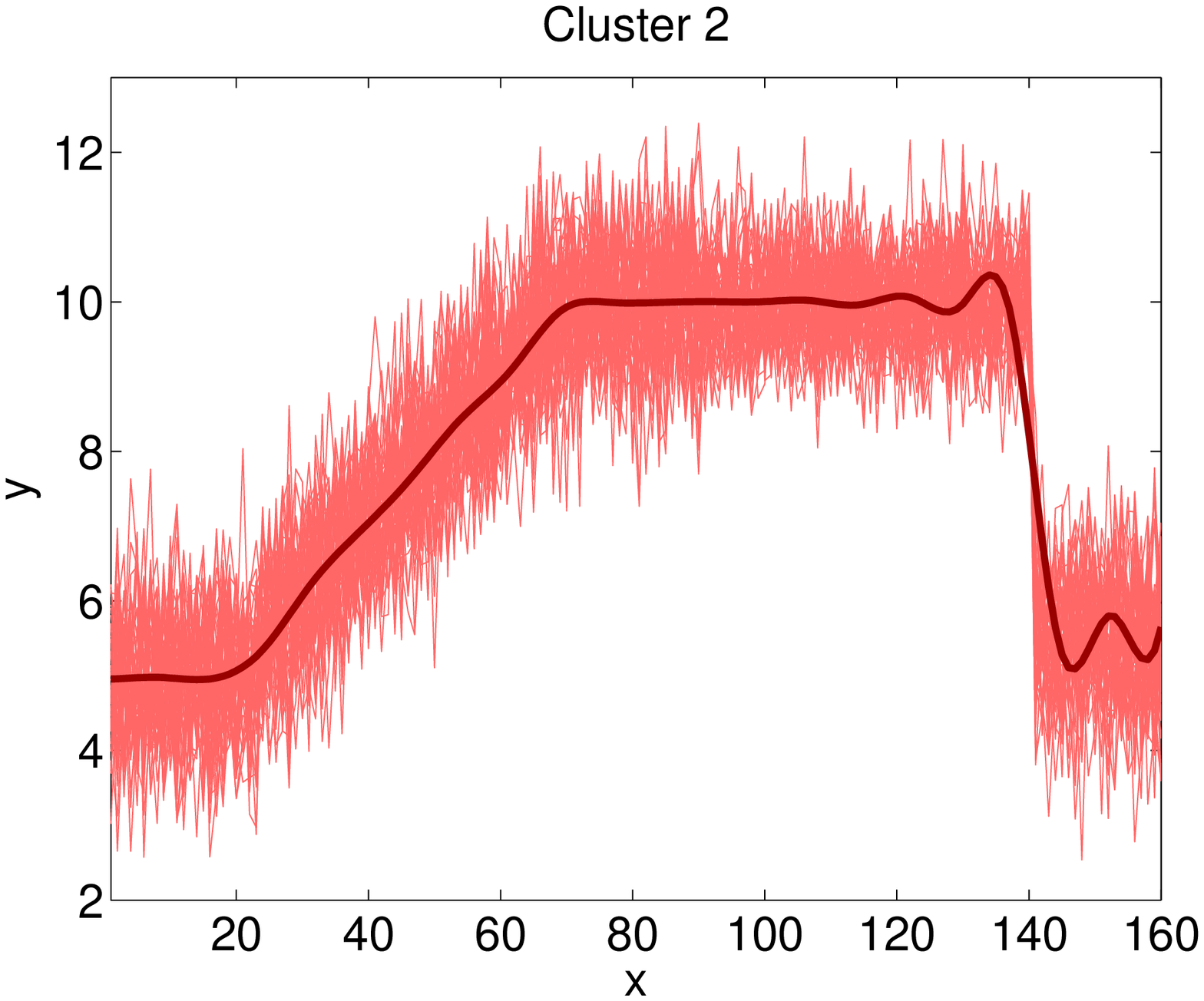}\\
\includegraphics[width=4.5cm]{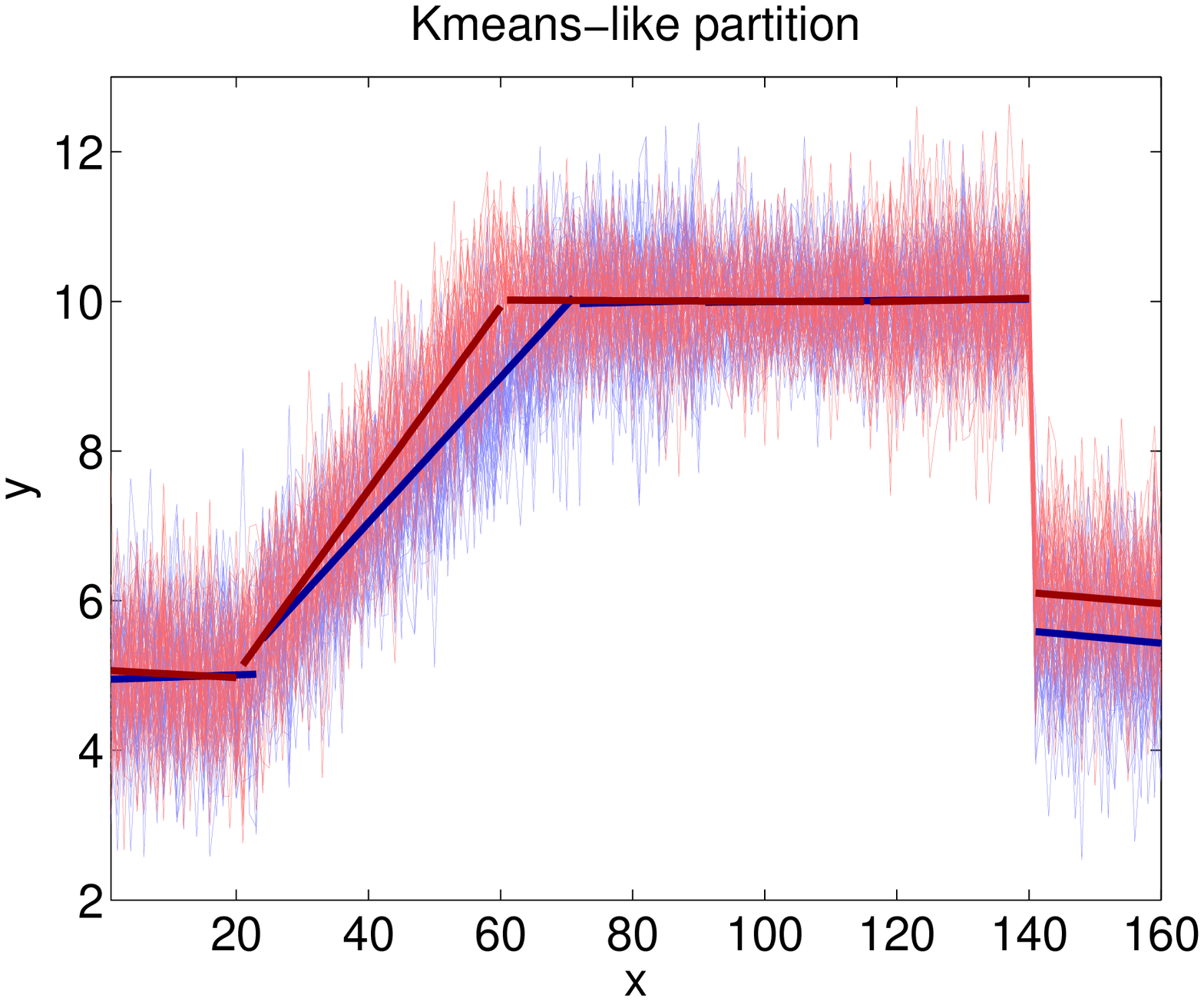}
\includegraphics[width=4.5cm]{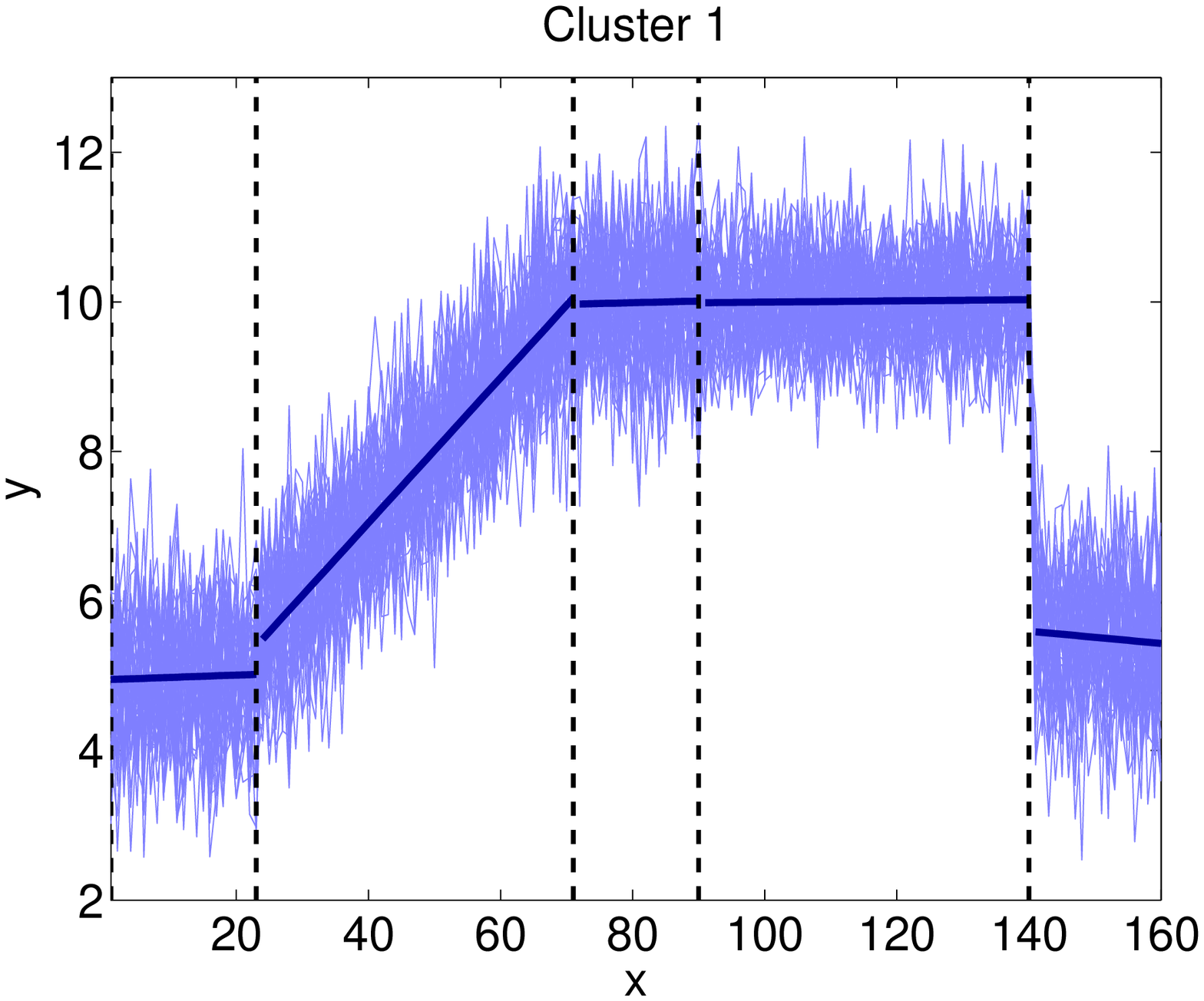}
\includegraphics[width=4.5cm]{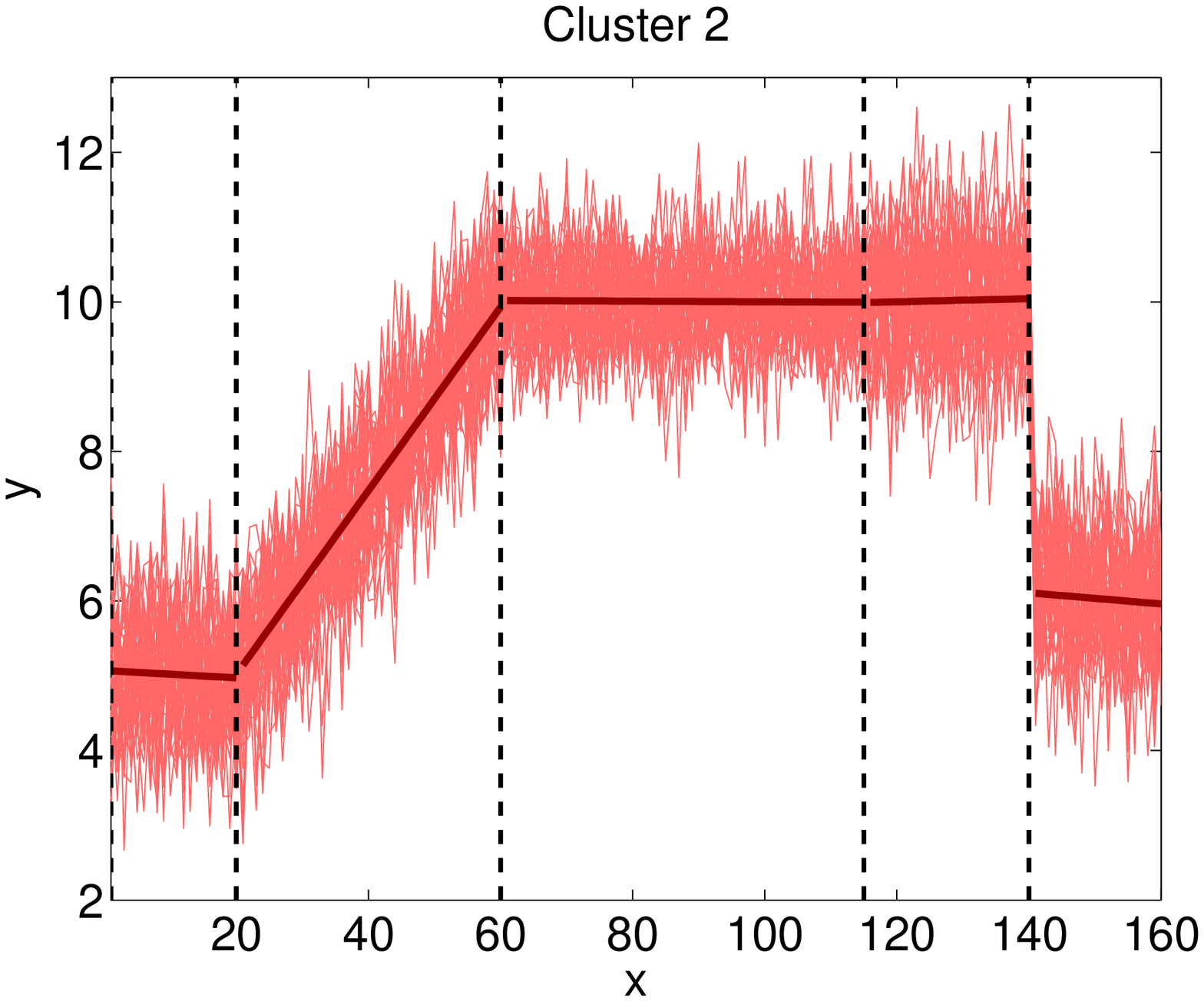}\\
%
\includegraphics[width=4.5cm]{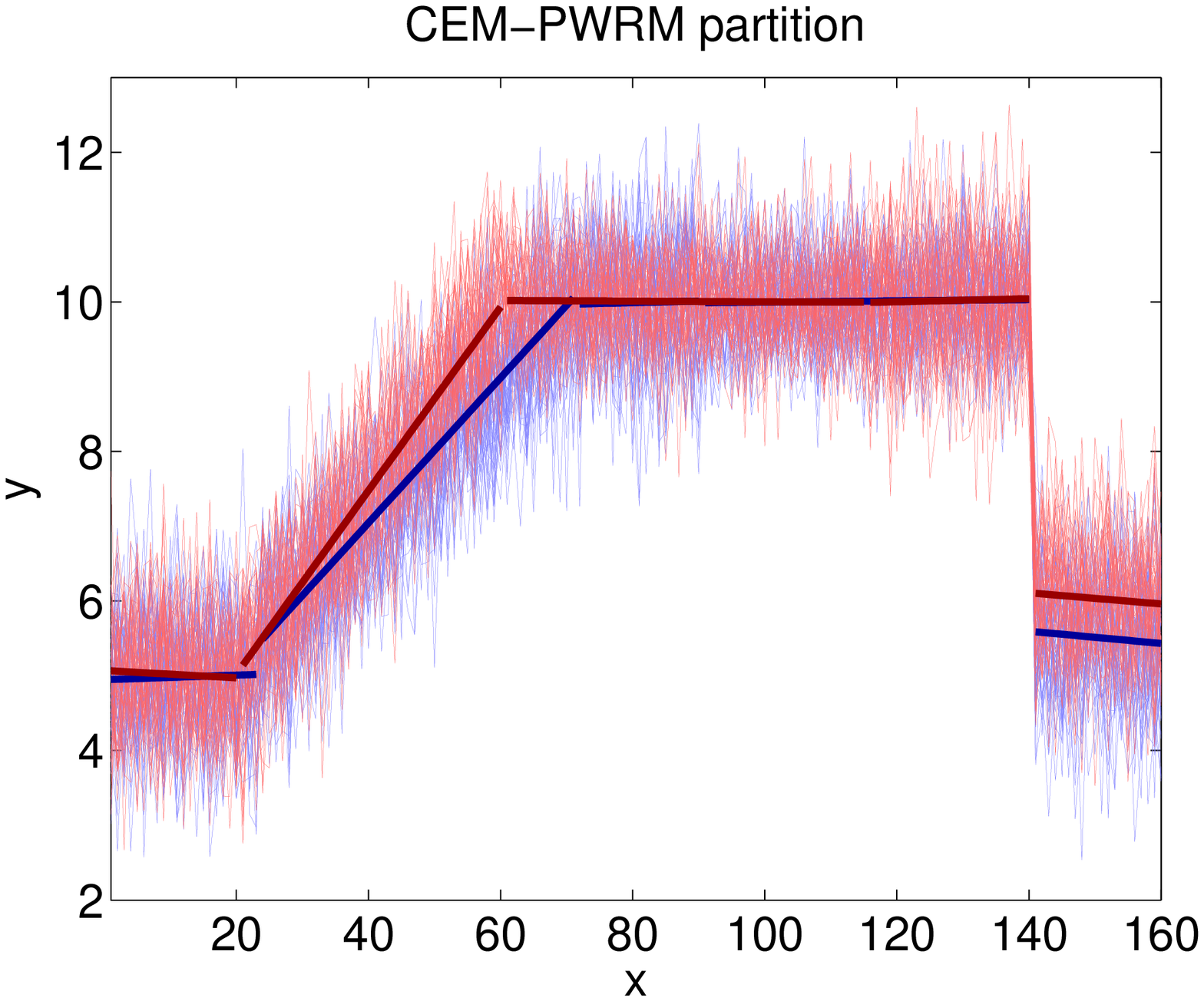}
\includegraphics[width=4.5cm]{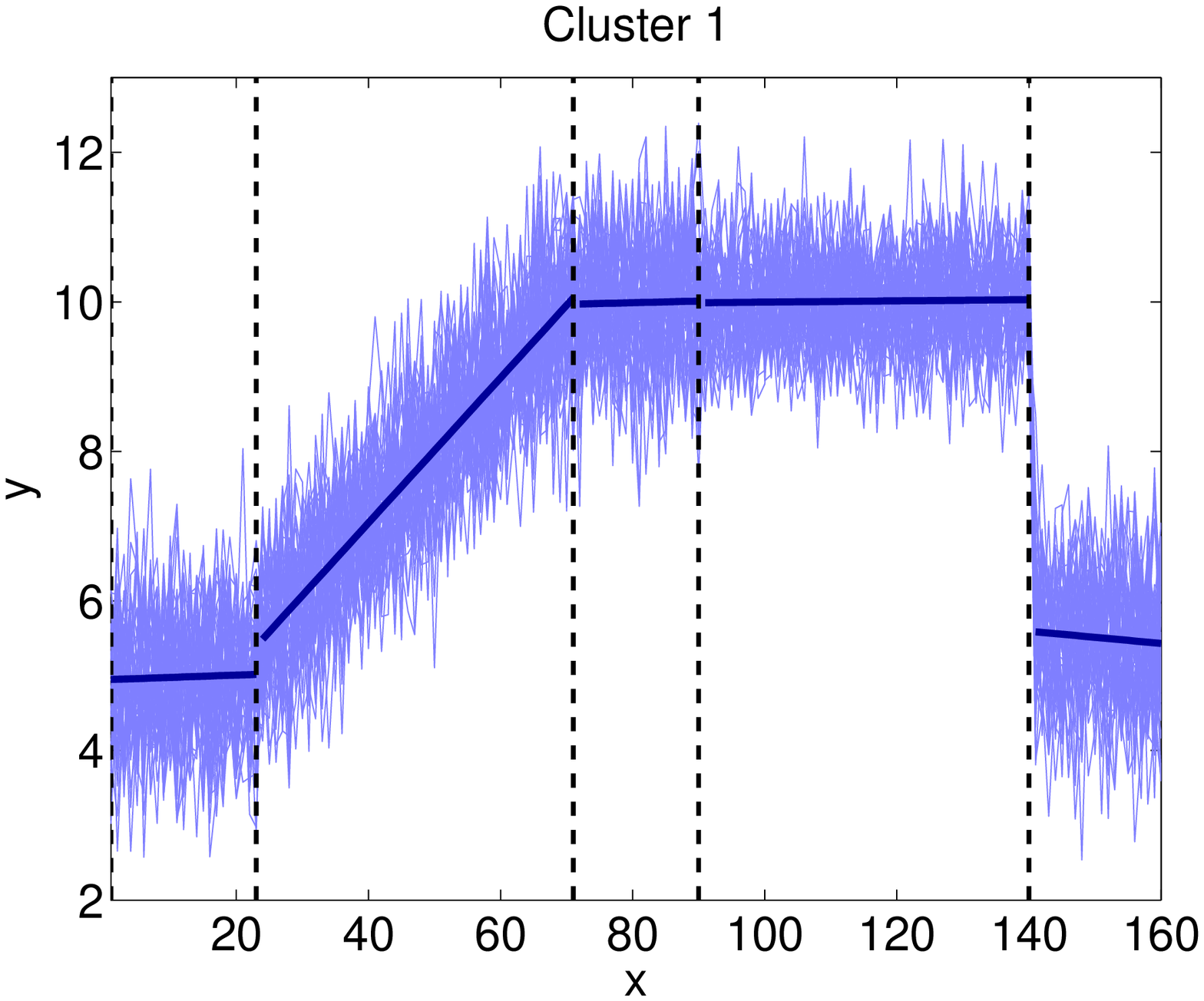}
\includegraphics[width=4.5cm]{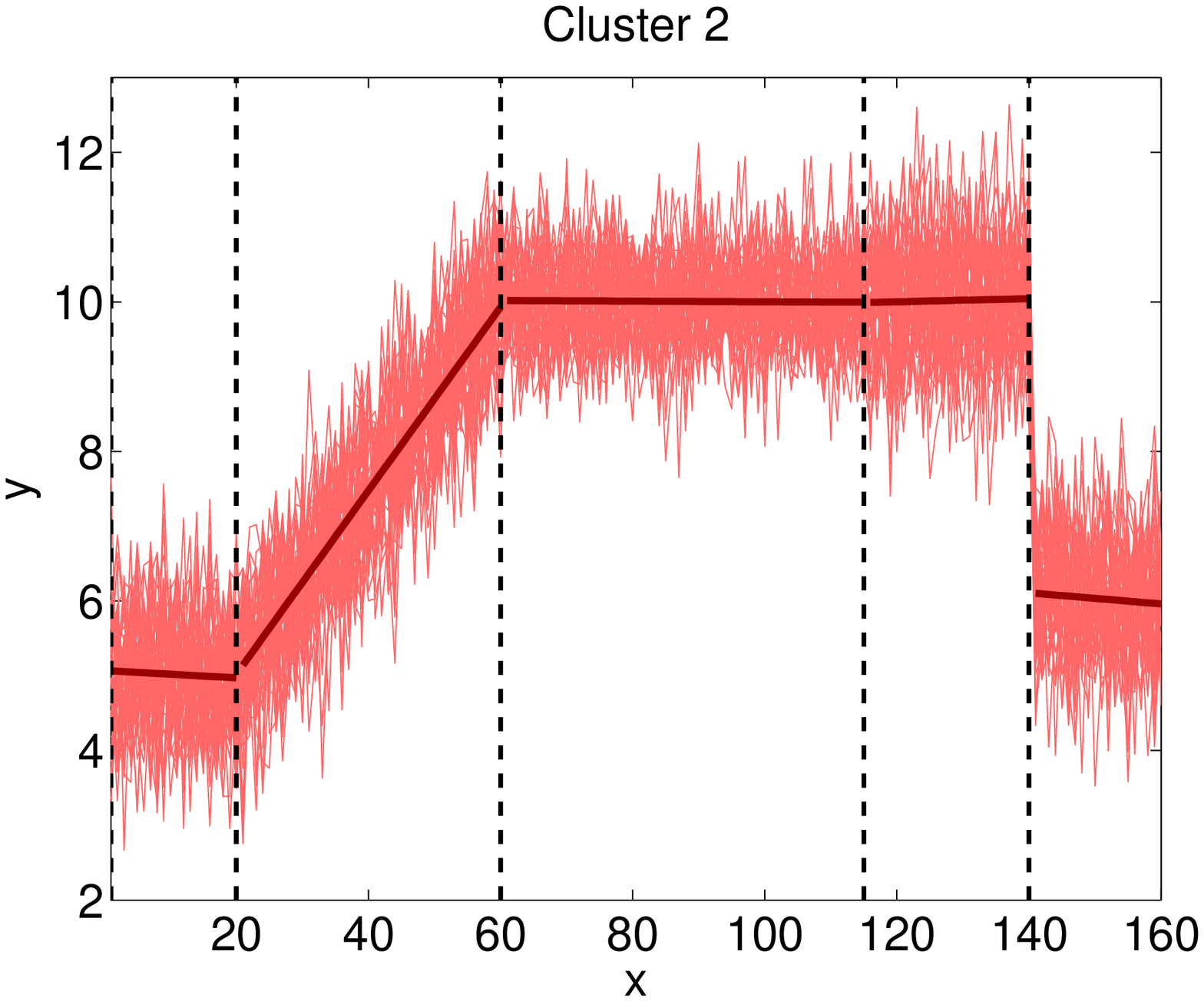}
\caption{\label{fig. clustering results situation 1}Clustering results and the corresponding cluster prototypes obtained with EM-GMM, EM-PRM, EM-PSRM, and the corresponding cluster segmentations obtained with Kmeans-like and CEM-PWRM.}
\end{figure}It can be seen that the best curves approximation are provided by the PWRM models. The GMM mean curves are simply over all means, and the PRM and the PSRM models, as they are based on continuous curve prototypes,  do not account for the segmentation, in contrast to the PWRM models which are well adapted to perform simultaneous curve clustering and segmentation. We note that in all the experiments we included both the EM and the CEM algorithm and the results are not significantly different, we chose to give the results for only one algorithm among the two versions.
 

In the previous situation, the algorithms were mainly evaluated regarding the curves approximation while keeping the clustering task not very difficult. Now, we vary the noise level in order to assess the models in terms of curve clustering. This is performed by computing the misclassification error rate for different noise level values. The curves are still be simulated according to the same parameters of Table \ref{table. simulation parameters} while varying the noise level for all the regimes by adding a noise level variation $s$ to the standard deviation $\sigma_{kr}$.

Figure \ref{fig. MCER results for simulations} shows the obtained misclassification error rate for the different noise levels. 
\begin{figure}[H] 
\centering
\includegraphics[width=7cm]{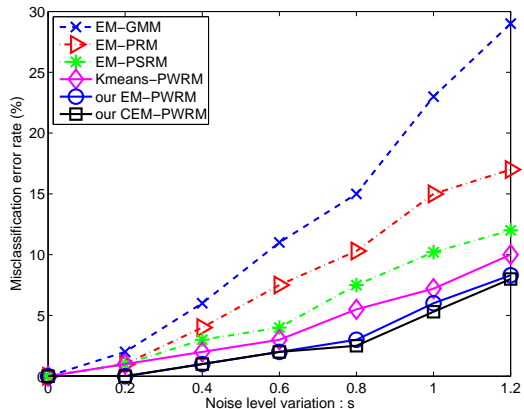} 
\caption{\label{fig. MCER results for simulations}The misclassification error rate versus the noise level variation.}
\end{figure}For a small noise level variation, the results are very similar and comparable to those presented previously. However, as the noise level variation increases, the misclassification error rate increases faster for the other models compared to the proposed PWRM model. The EM and the CEM algorithm for the proposed approach provide very similar results with a slight advantage for the CEM version. 
%

For the previous situations, the data was simulated according to a mixture with equal mixing proportions.  
Now we vary the parameters in order to make the mixture with non-uniform mixing proportions ($\alpha_1 = 0.2 \ \alpha_2 = 0.8$) and the variance change less pronounced than before (namely we set $\sigma_{13} = 0.7$ and $\sigma_{14}=0.6$. Simulated curves according to this situation are shown in Figure \ref{fig: example of simulated curves sit 2}.
\begin{figure}[H]
\centering 
\includegraphics[width=4.8cm]{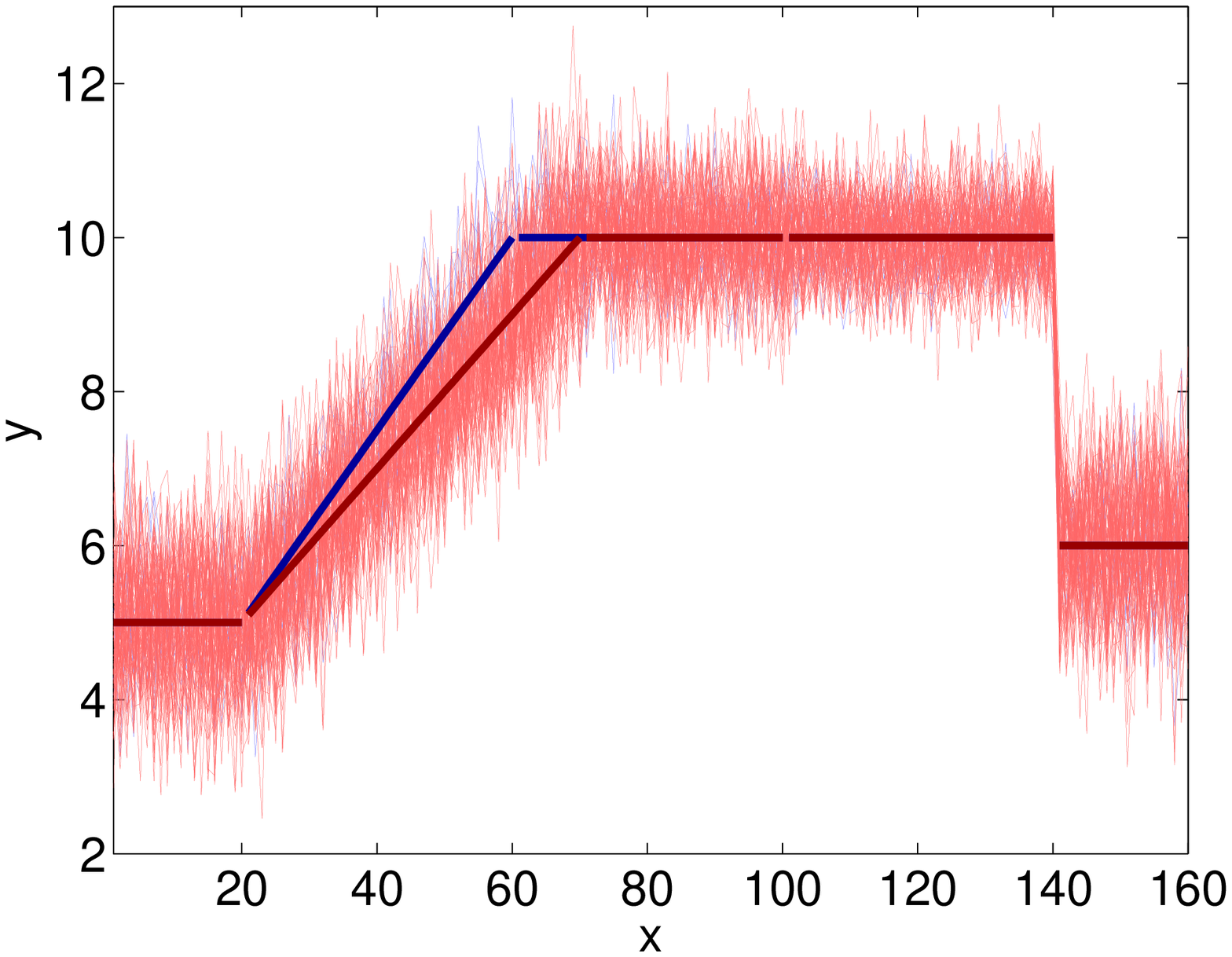} 
\includegraphics[width=4.8cm]{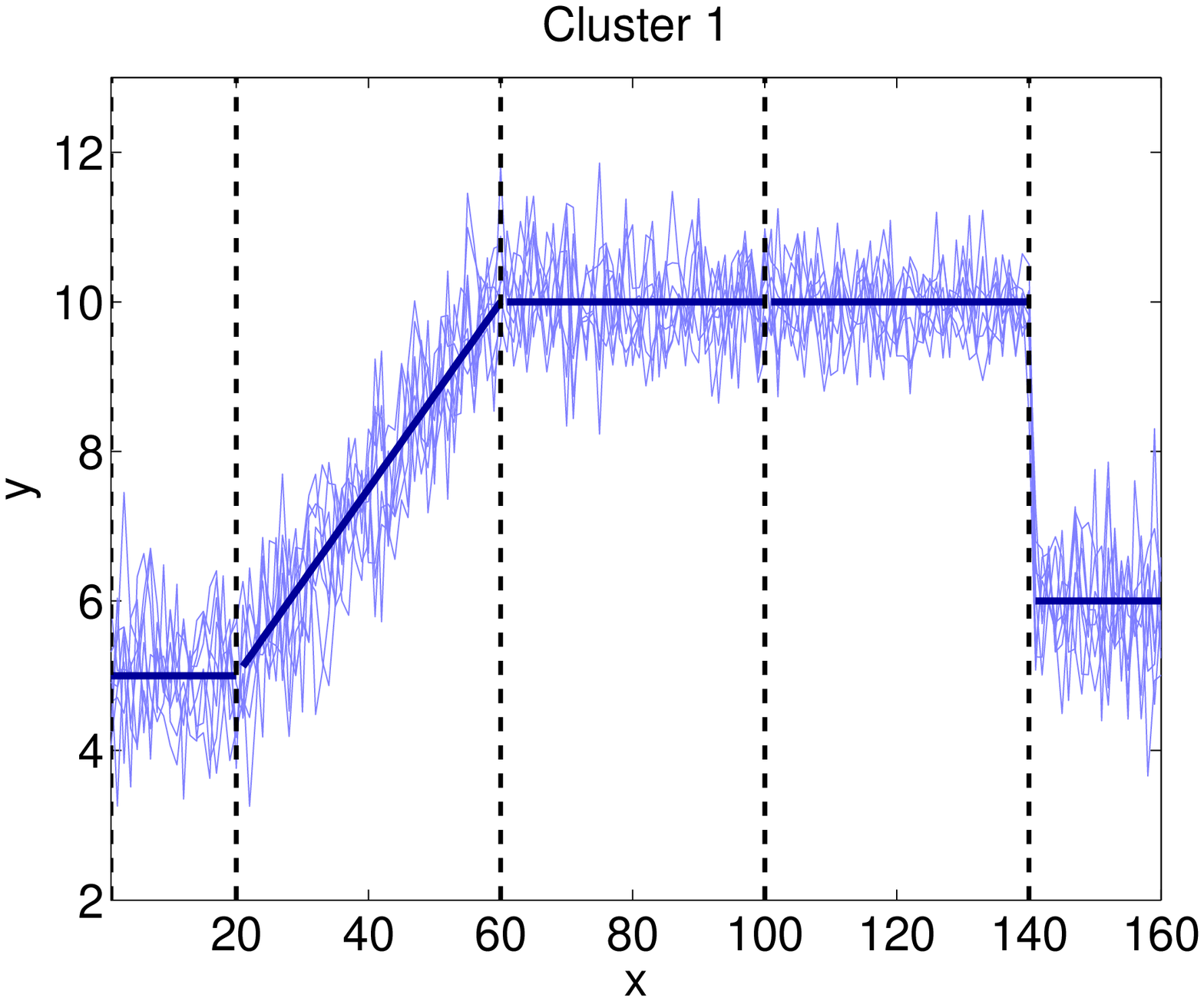}
\includegraphics[width=4.8cm]{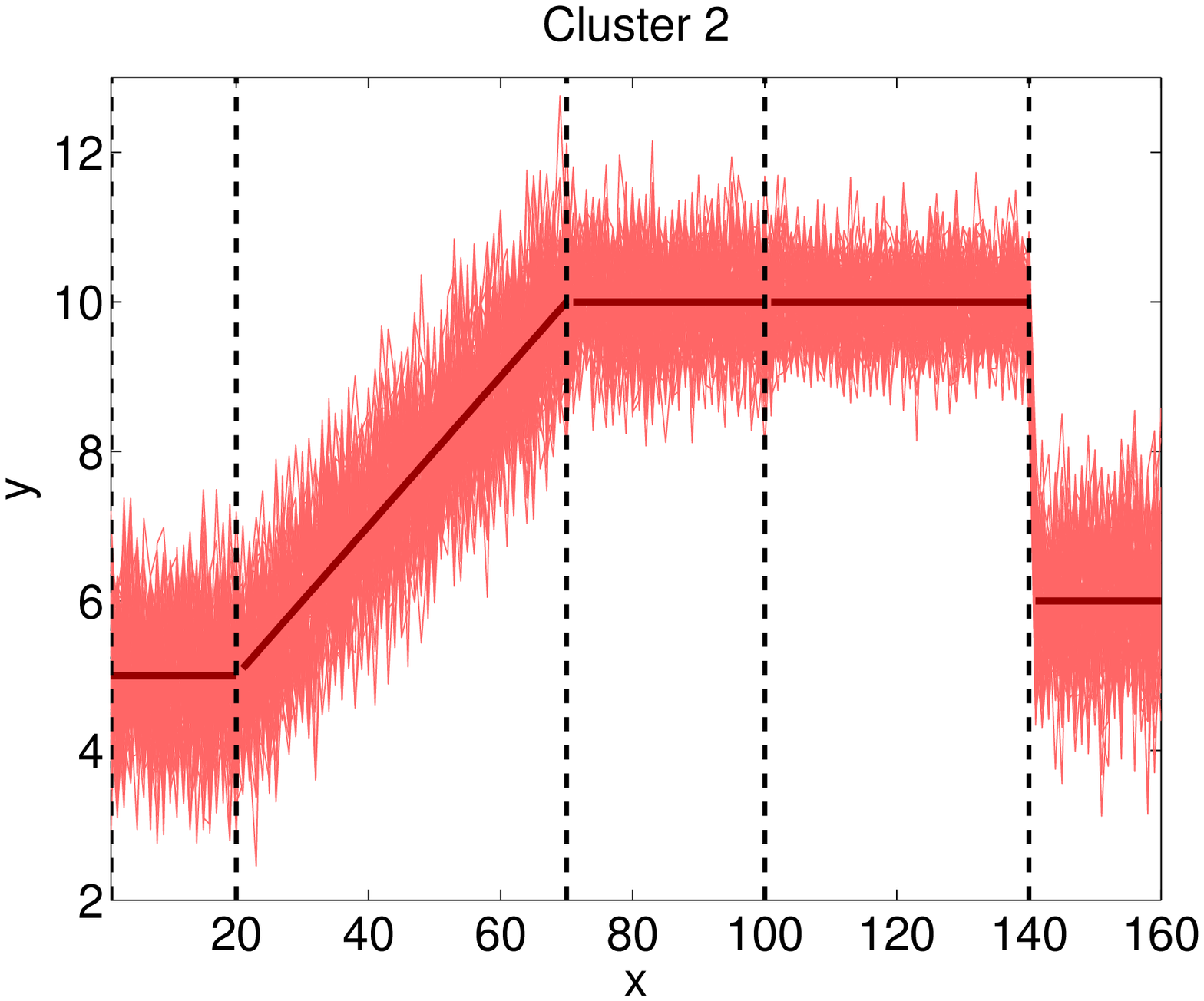}
\caption{\label{fig: example of simulated curves sit 2} A two-class data set of  simulated curves from a PWRM with non-uniform mixing proportions ($\alpha_1 =0.2 \ , \alpha_2 = 0.8$): the clusters colored according to the true partition, and the prototypes (left) and the true segmentation for cluster 1 (middle) and cluster 2 (right).}
 \end{figure}
The clustering results for this example are shown in Figure \ref{fig. clustering results situation 2}.
The misclassification error for this situation is of 7\% for the $K$-means like approach and of 3\% for the proposed PWRM approach. For the other approaches, the misclassification error is around 10\% for both the PRM and the PSRM, while the one of the GMM is of 20\%. 
Another interesting point to see here is that the $K$-means based approach can fail in terms of segmentation.
As it can be seen in Figure \ref{fig. clustering results situation 2} (top, right), the third and the fourth regime do not correspond to the actual ones (see Figure \ref{fig: example of simulated curves sit 2}, middle).
 This is attributed to the fact that the $K$-means-like approach for PWRM is constrained as it assumes the same proportion for each cluster, and does not sufficiently take  into account the heteroskedasticity within each cluster compared to the proposed general probabilistic PWRM model.
\begin{figure}[H] 
\centering
\includegraphics[width=4.8cm]{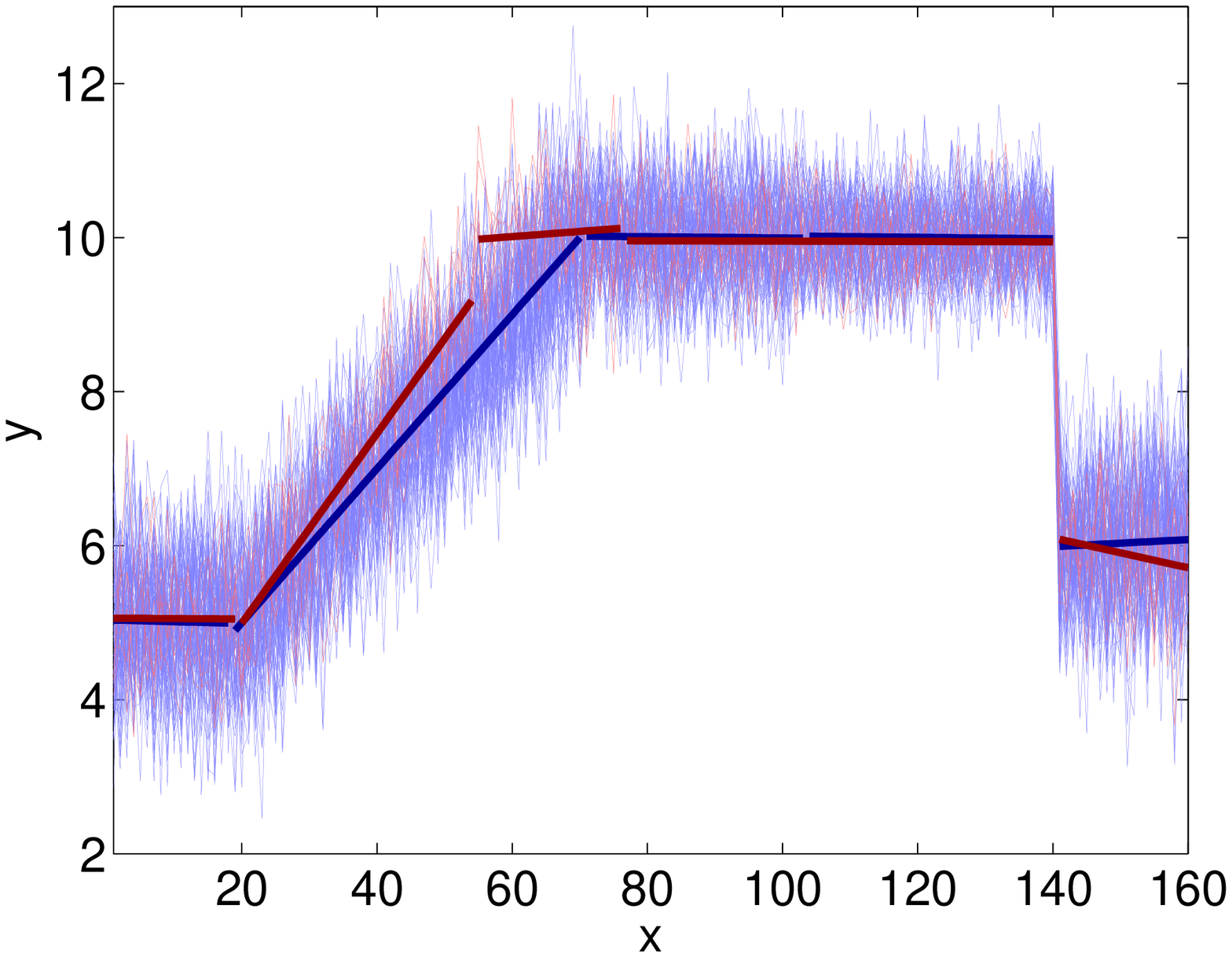}
\includegraphics[width=4.8cm]{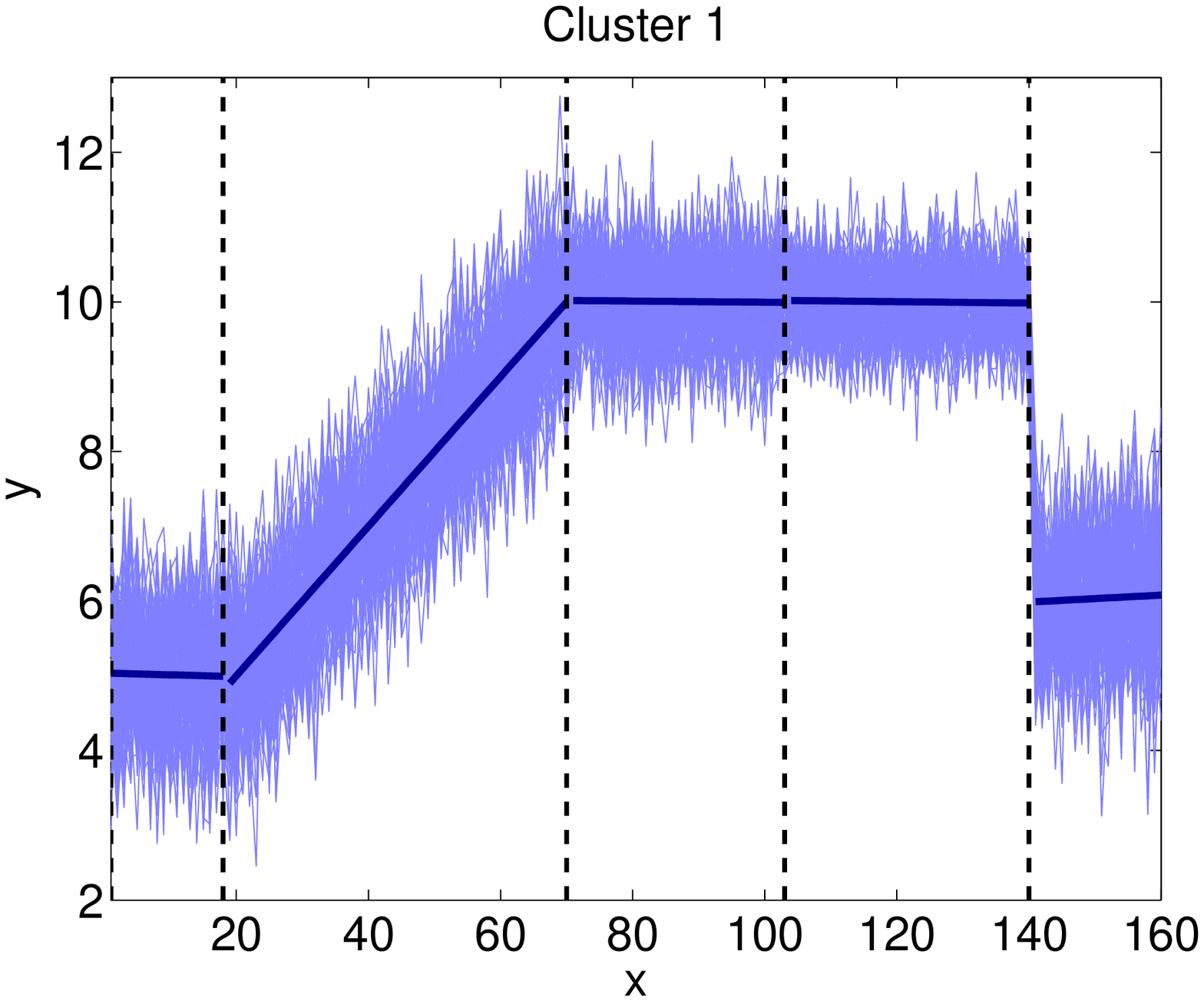}
\includegraphics[width=4.8cm]{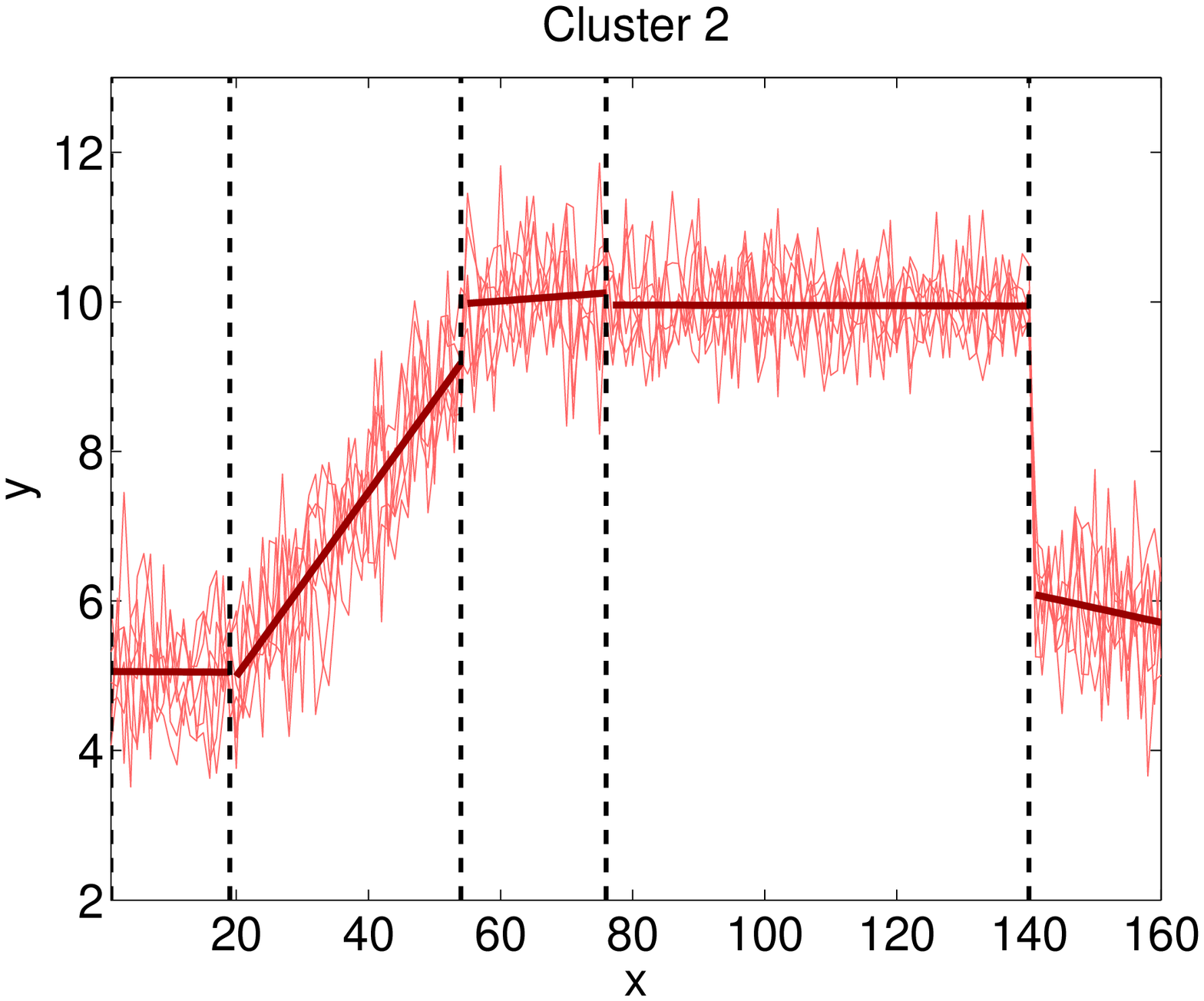}\\
%
\includegraphics[width=4.8cm]{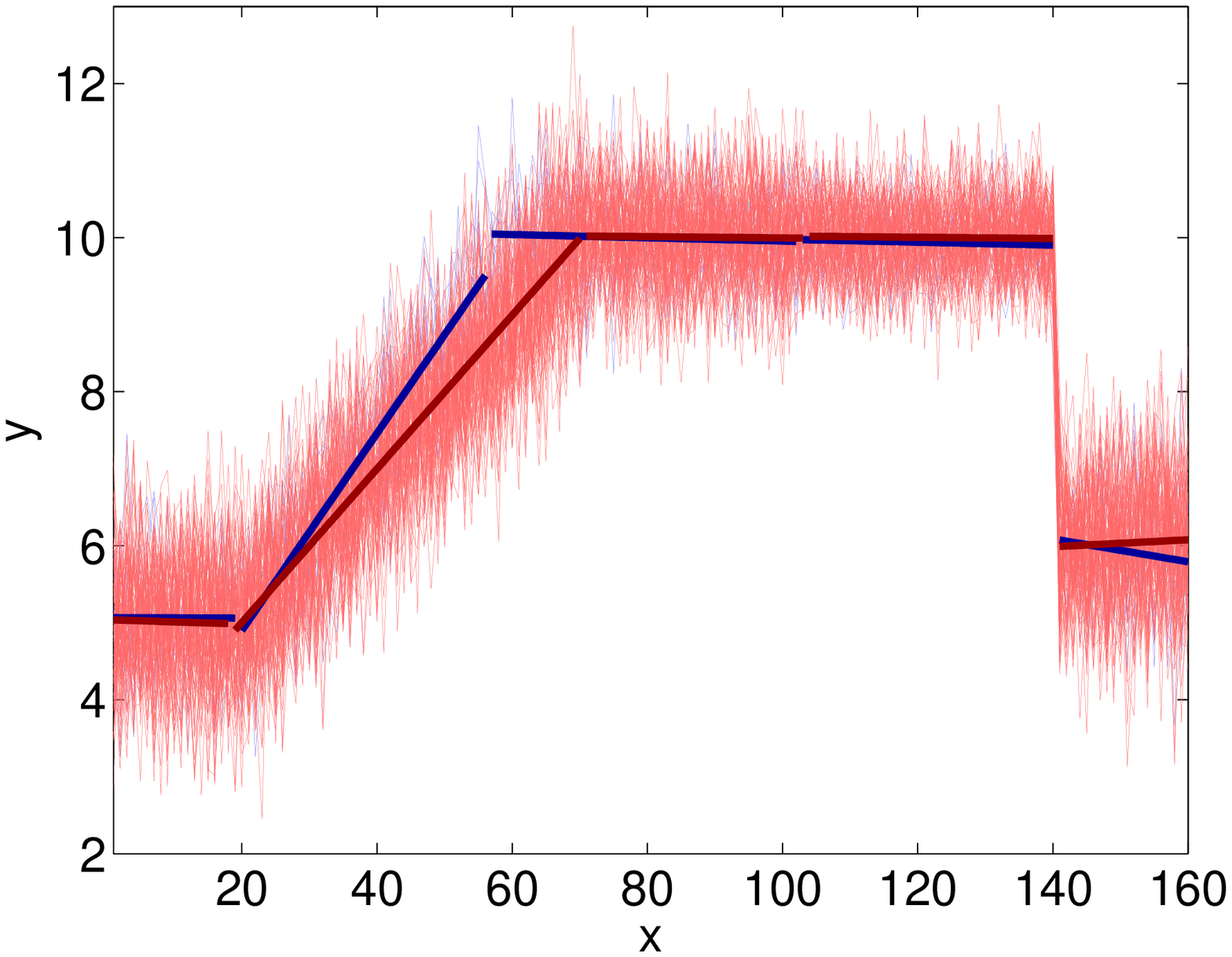}
\includegraphics[width=4.8cm]{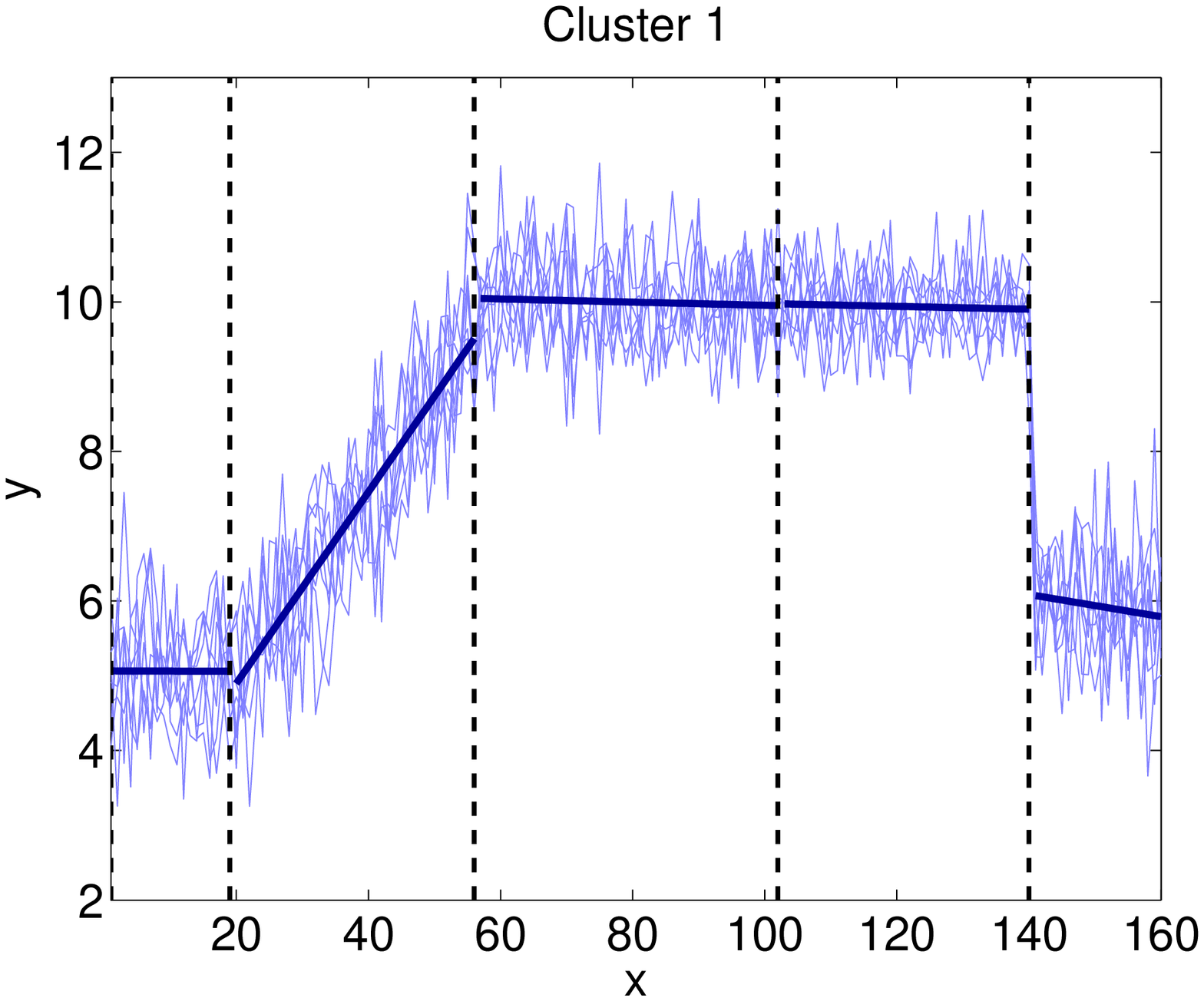}
\includegraphics[width=4.8cm]{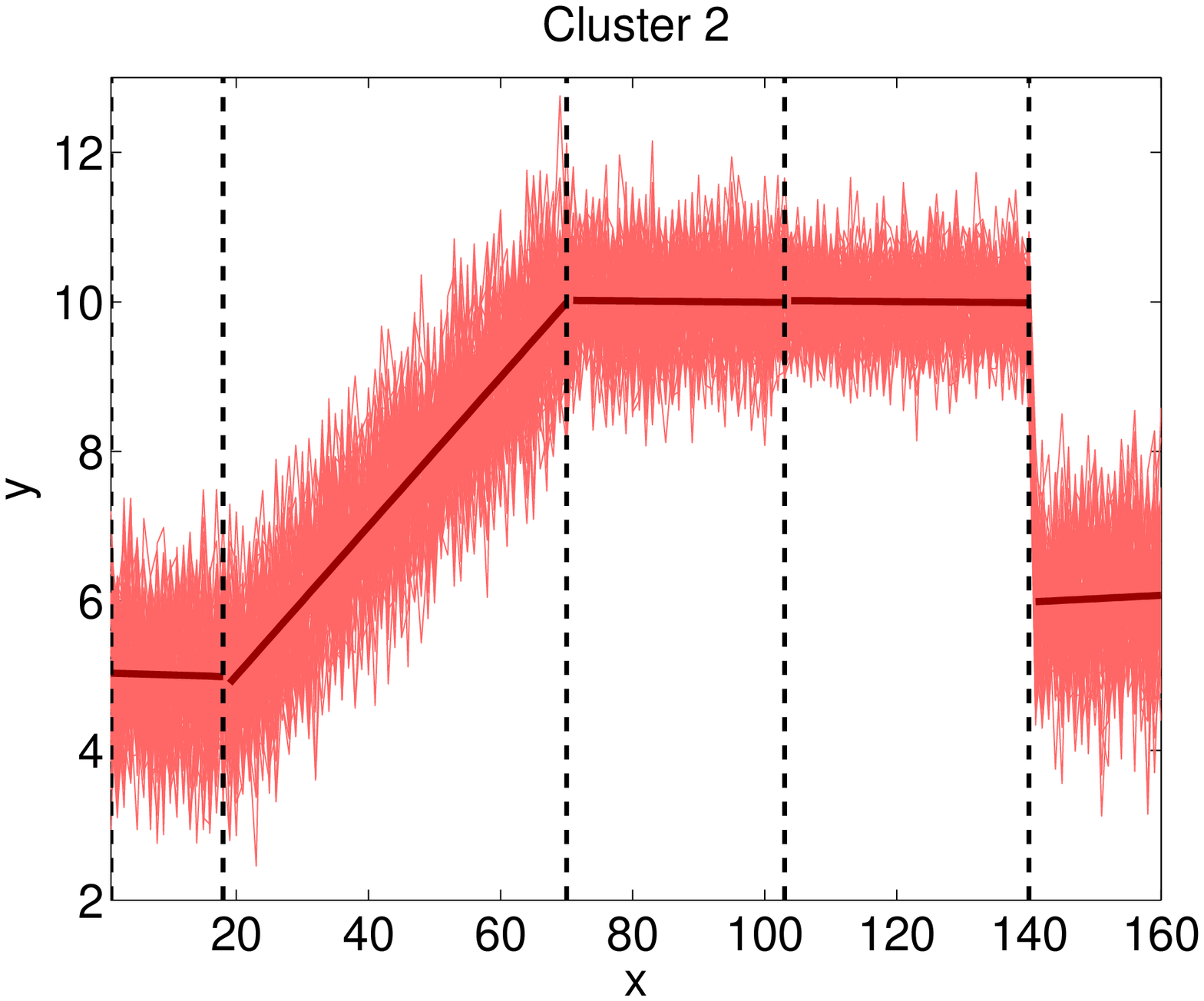}
\caption{\label{fig. clustering results situation 2}Results for the curves shown in Figure \ref{fig: example of simulated curves sit 2} : Clustering results and the corresponding cluster prototypes and cluster segmentations obtained with Kmeans-like (top) and the proposed CEM-PWRM (down).}
\end{figure}

%

\subsubsection{Model selection}

In this section we give the results concerning the selection of the best values of the triplet $(K,R,p)$ by using the ICL criterion as presented in Section \ref{ssec. model selection}. The values of $(K_{max},R_{max},p_{max})$ (respectively $(K_{min},R_{min},p_{min})$) were ($4,6,3$) (respectively ($1,1,0$)).  
We note that for the $K$-means-like algorithm, the 
complete-data log-likelihood is $\cL_c = -\frac{1}{2}E$ up to a constant term (see Equation (\ref{eq: complete log-lik for the PWRM v proof})), where $E$ is the criterion minimized by this approach which is given by Equation (\ref{eq: distance criterion for clustering with piecewise regression}). The ICL criterion for this approach is therefore computed as
$\mbox{ICL}(K,R,p) = -\frac{E}{2} - \frac{\nu_{\bsPsi} \log(n)}{2},$ where $\nu_{\bsPsi} =  \sum_{k=1}^K R_k(p+2) - K$ is the number of free parameters of the model and $n$ is the sample size. The number of free model parameters in this case includes  $\sum_{k=1}^K R_k(p+1)$ polynomial coefficients and $\sum_{k=1}^K (R_k - 1)$ transition points, the model being a constrained PWRM model (isotropic with identical mixing proportions). 

For this experiment, 
we observed that the model with the highest percentage of selection corresponds to $(K,R,p) = (2,5,1)$ for the proposed EM-PWRM and CEM-PWRM approaches with respectively 81\% and 85\% of selection. While for the $K$-means-like approach, the same model $(K,R,p) = (2,5,1)$ has a percentage of selection of only 72\%.
The number of regimes is underestimated with only around 10\% for the proposed approaches, while the number of clusters is correctly estimated. However, the $K$-means-like approach overestimates the number of clusters ($K=3$) in 12\% of cases.
These results illustrate an advantage of the fully probabilistic approach compared to the  one based on the $K$-means-like approach.
We also note that the models with $K=1, 4$ and those with $R=1,2$ were not selected (percentage of 0\%) for all the models.

\subsection{Application on real curves} 

In this section we apply the proposed approach on real curves issued from three different data sets, and compare it to 
 alternatives.  The studied curves are the railway switch curves, the Tecator curves and the Topex/consist satellite data as studied in \cite{hebrailEtal:2010}. The curves of each dataset are respectively shown in Figure  \ref{fig. switch data}, Figure \ref{fig. tecator data} and Figure \ref{fig. satellite data} .
 %

\subsubsection{Railway switch curves}
The first studied curves are the railway switch curves issued from a railway diagnosis application of the railway switch. 
Roughly, the railway switch is the component that enables (high speed) trains to be guided from one track to another at a railway junction, and  is controlled by an electrical motor. The considered  curves are the signals of the consumed power during the switch operations. These curves present several changes in regime due to successive mechanical motions involved in each switch operation (see Figure \ref{fig.  switch data}).
\begin{figure}[H] 
\centering
\includegraphics[width=7cm]{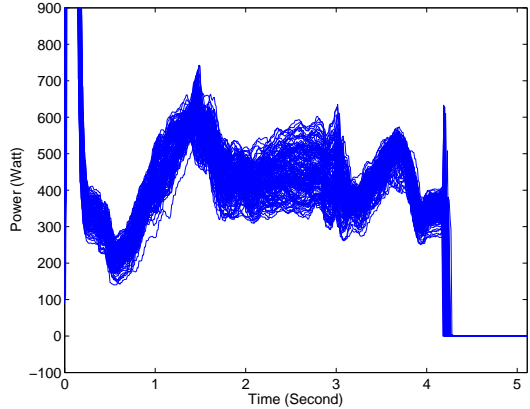}
\caption{\label{fig.  switch data}Railway switch curves.}
\end{figure} 
The diagnosis task can be achieved through the analysis of these curves 
to identify possible faults. However, the large amount of data  makes the manual labeling task onerous for the  experts. Therefore, the main concern of this task is to propose a data preprocessing approach that allows for automatically identifying homogeneous groups (without defect or with possible defect). 
 %
The used database is composed of $n=146$ real curves of $m=511$ observations.  
%
We assume that in the database we have two clusters ($K=2$). The first contains curves corresponding to an operating state without defect and the second contains curves corresponding to an operating state with a possible defect. The number of regression components 
was set to $R = 6$ in accordance with the number of electromechanical phases of a switch operation and the degree of the polynomial regression $p$ was set to 3 which is  appropriate for the different regimes in the curves. 
However, we note that no ground truth for this data set is available, neither regarding the classifications nor regarding the segmentation. This study could provide a preliminary result to help experts labelling the data.


Figure \ref{fig. clustering results for real data} shows the graphical clustering results and the corresponding cluster prototypes for the real switch operation curves. 
\begin{figure}[htbp] 	
\centering
\includegraphics[width = 4.6cm]{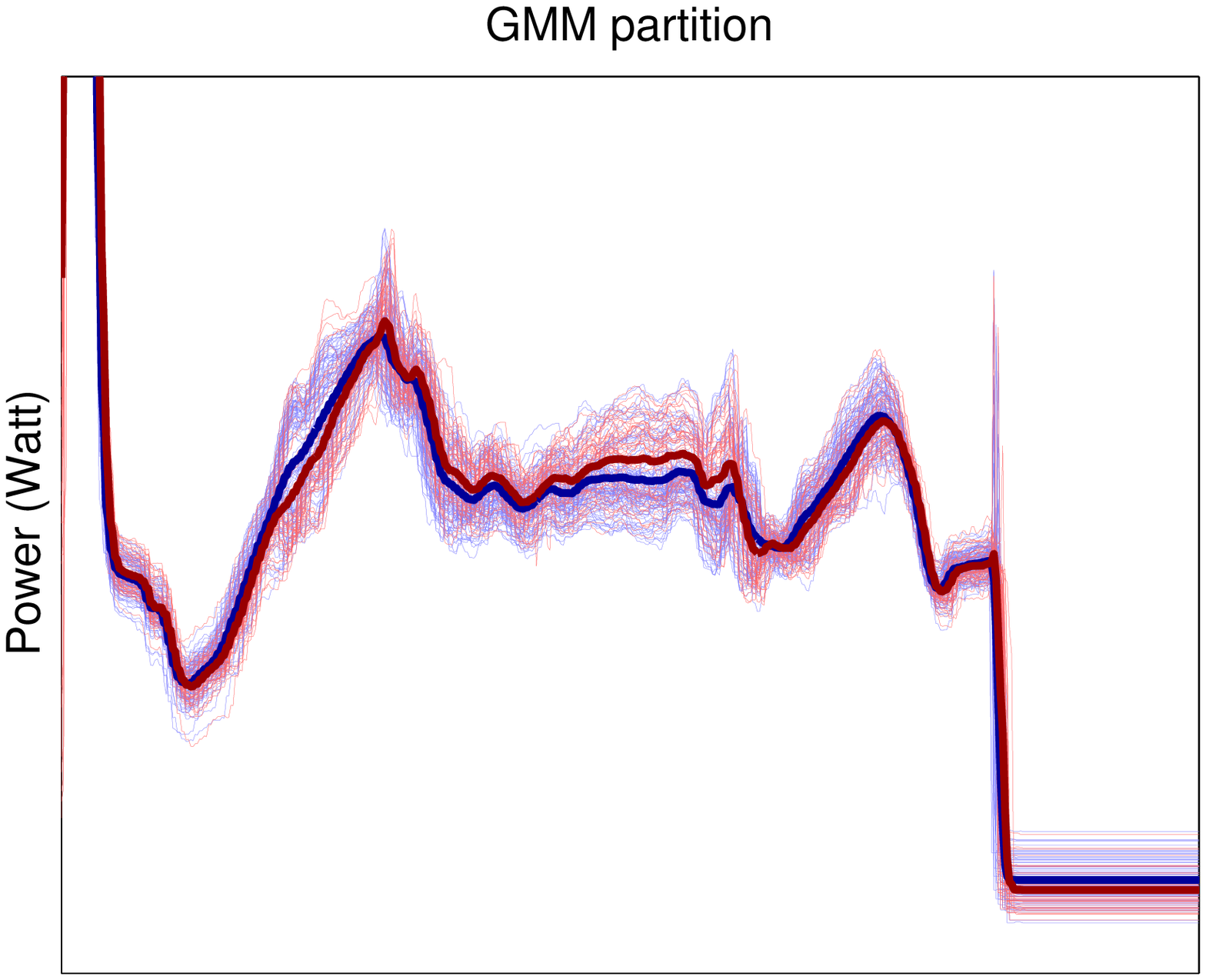}
\includegraphics[width = 4.8cm]{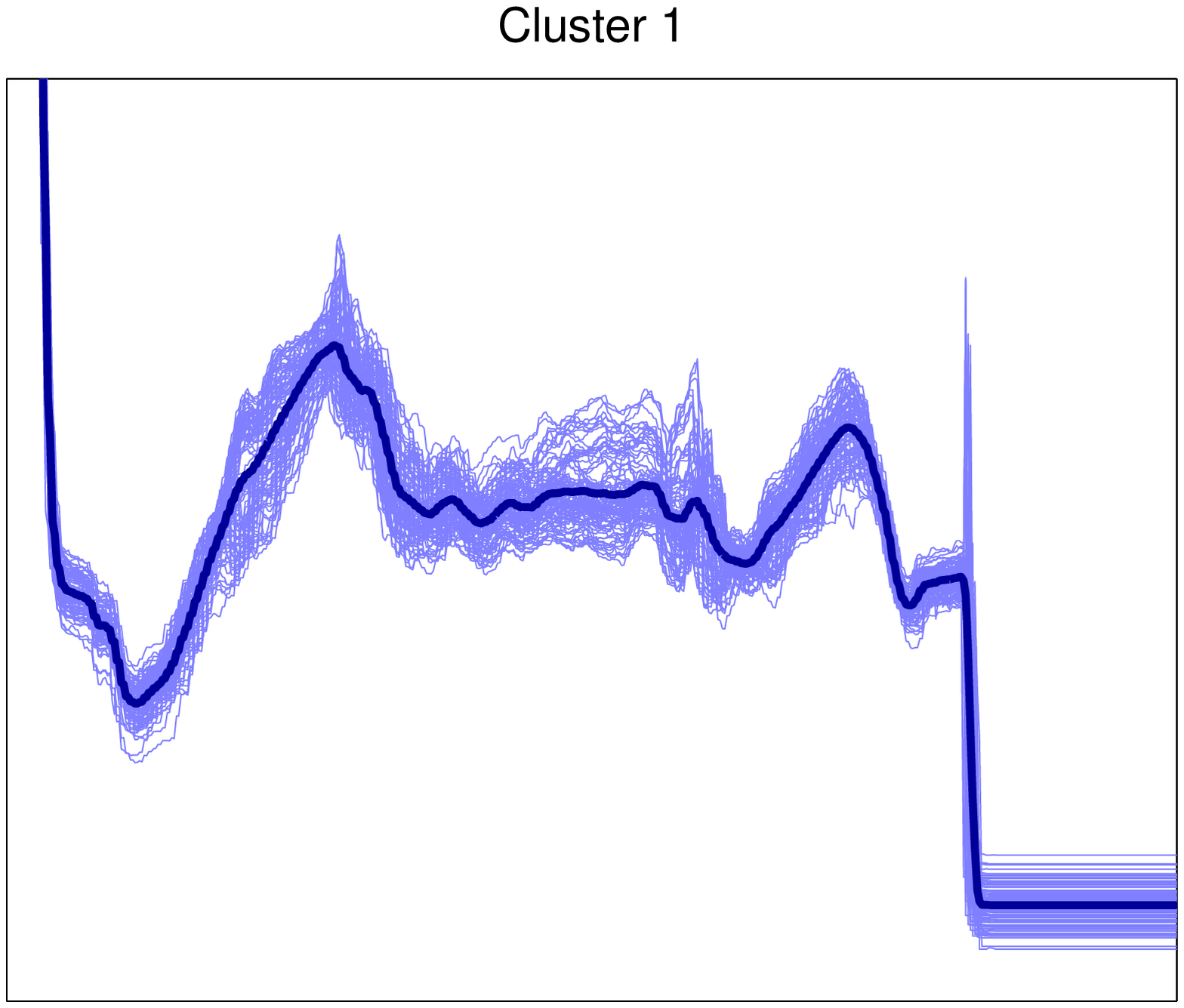}
\includegraphics[width = 4.8cm]{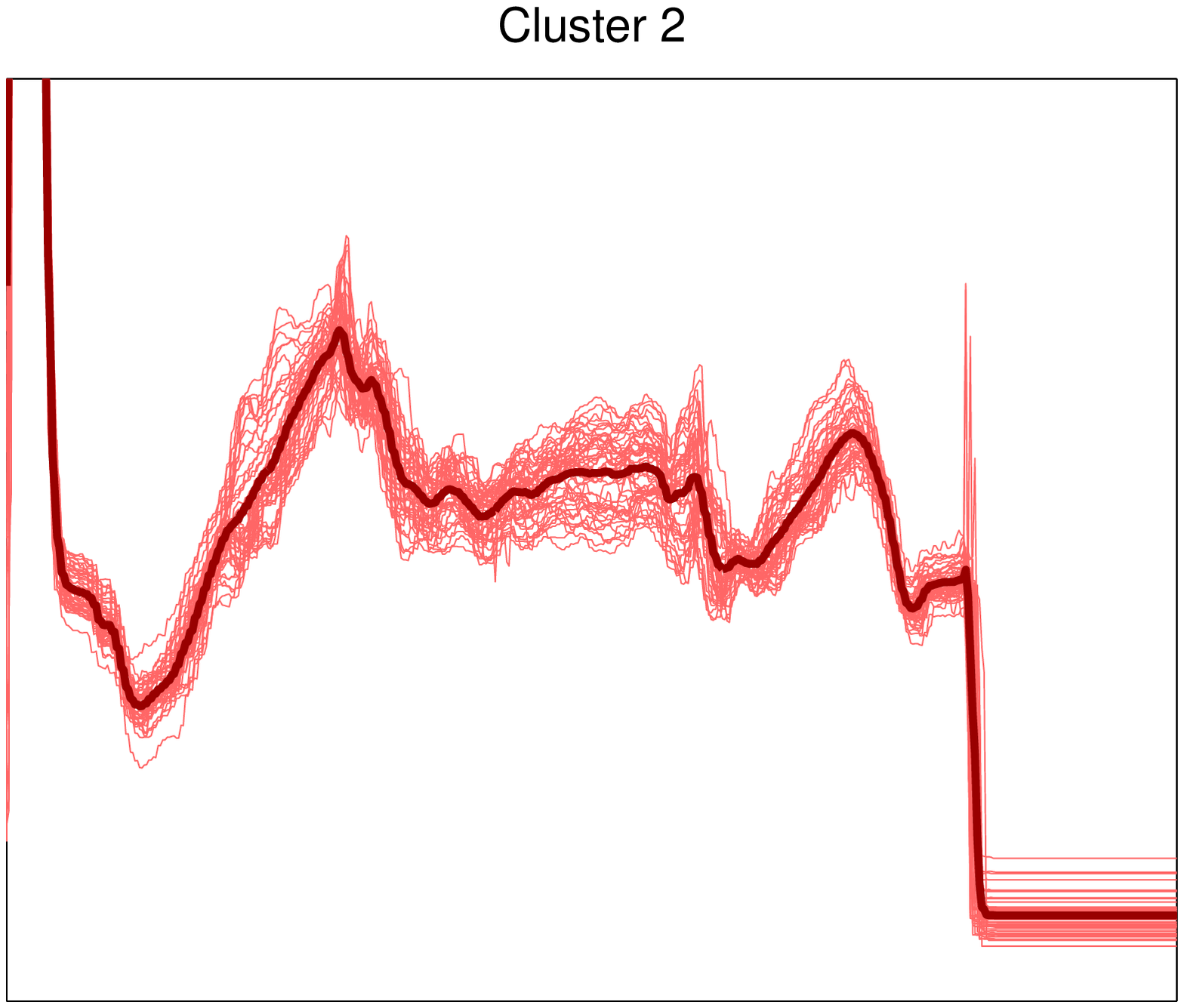}\\
\includegraphics[width = 4.6cm]{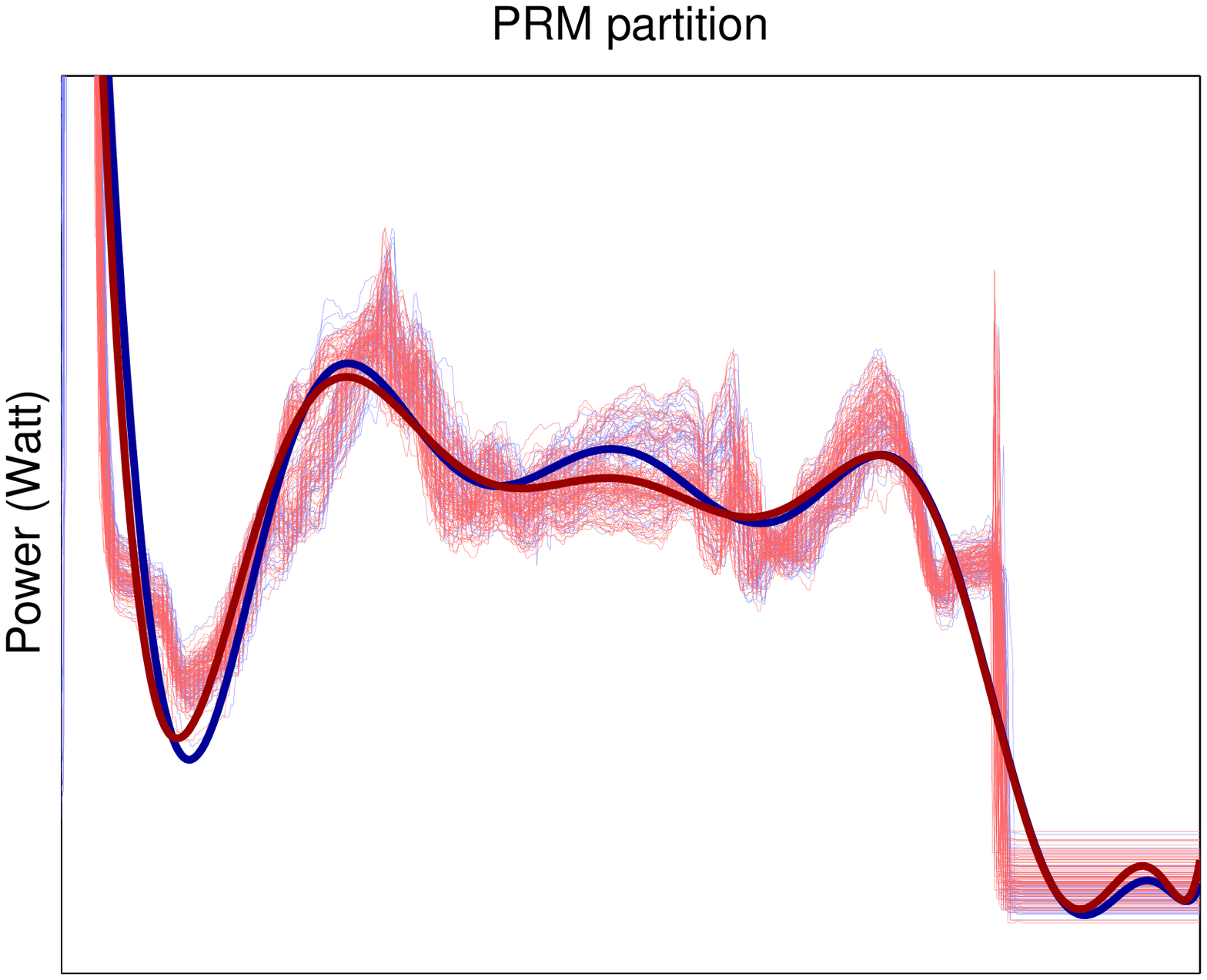}
\includegraphics[width = 4.8cm]{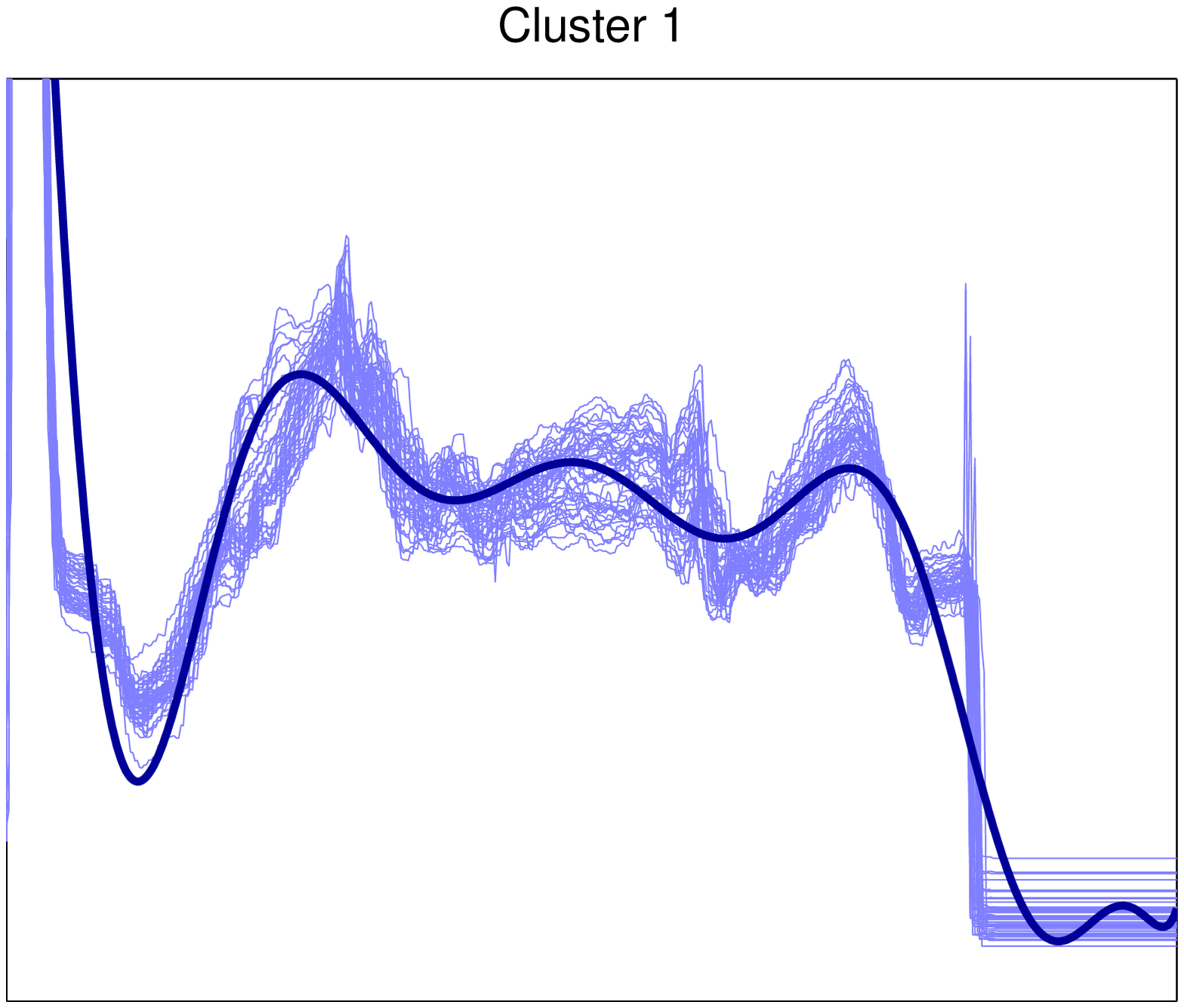}
\includegraphics[width = 4.8cm]{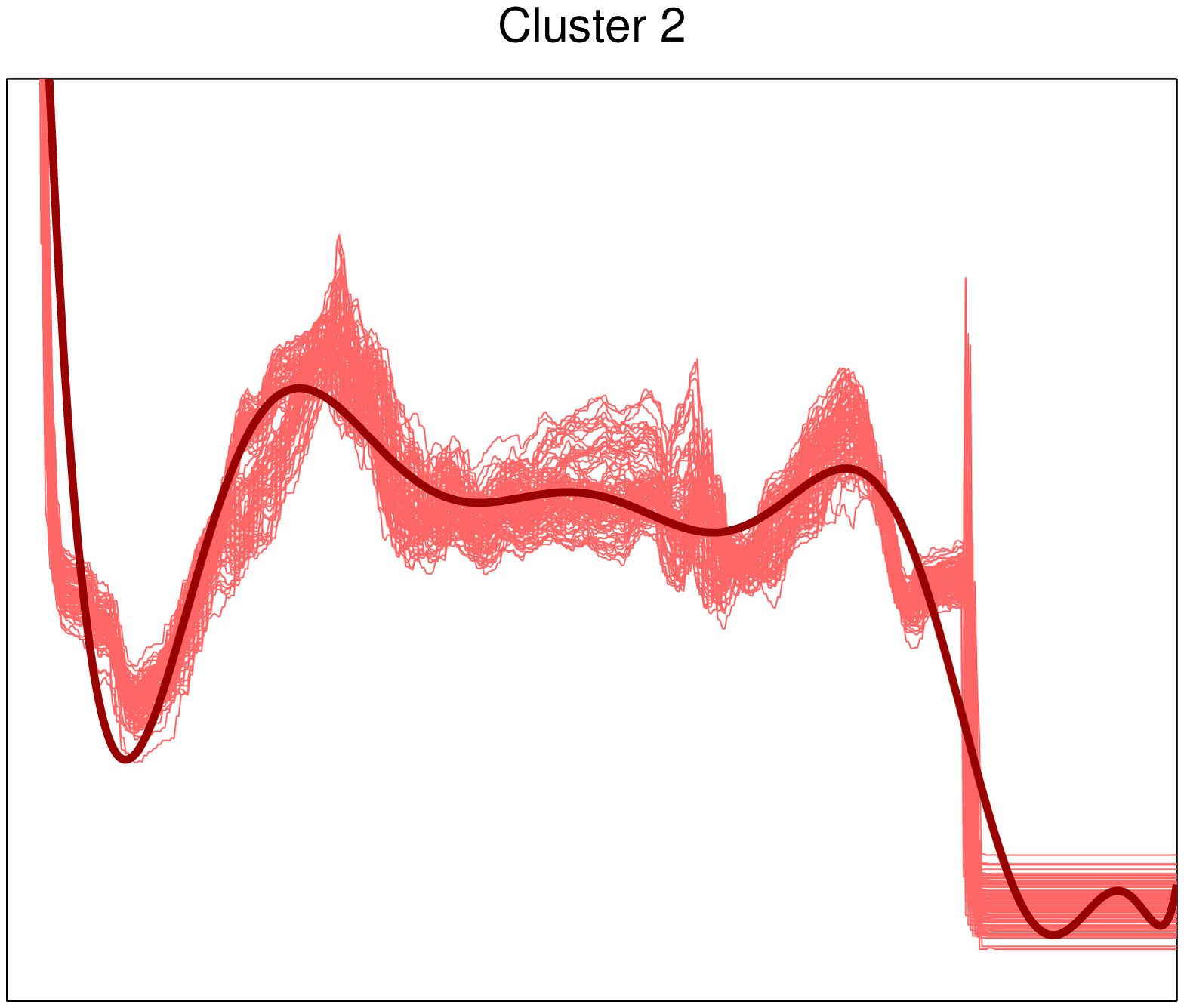}\\
\includegraphics[width = 4.6cm]{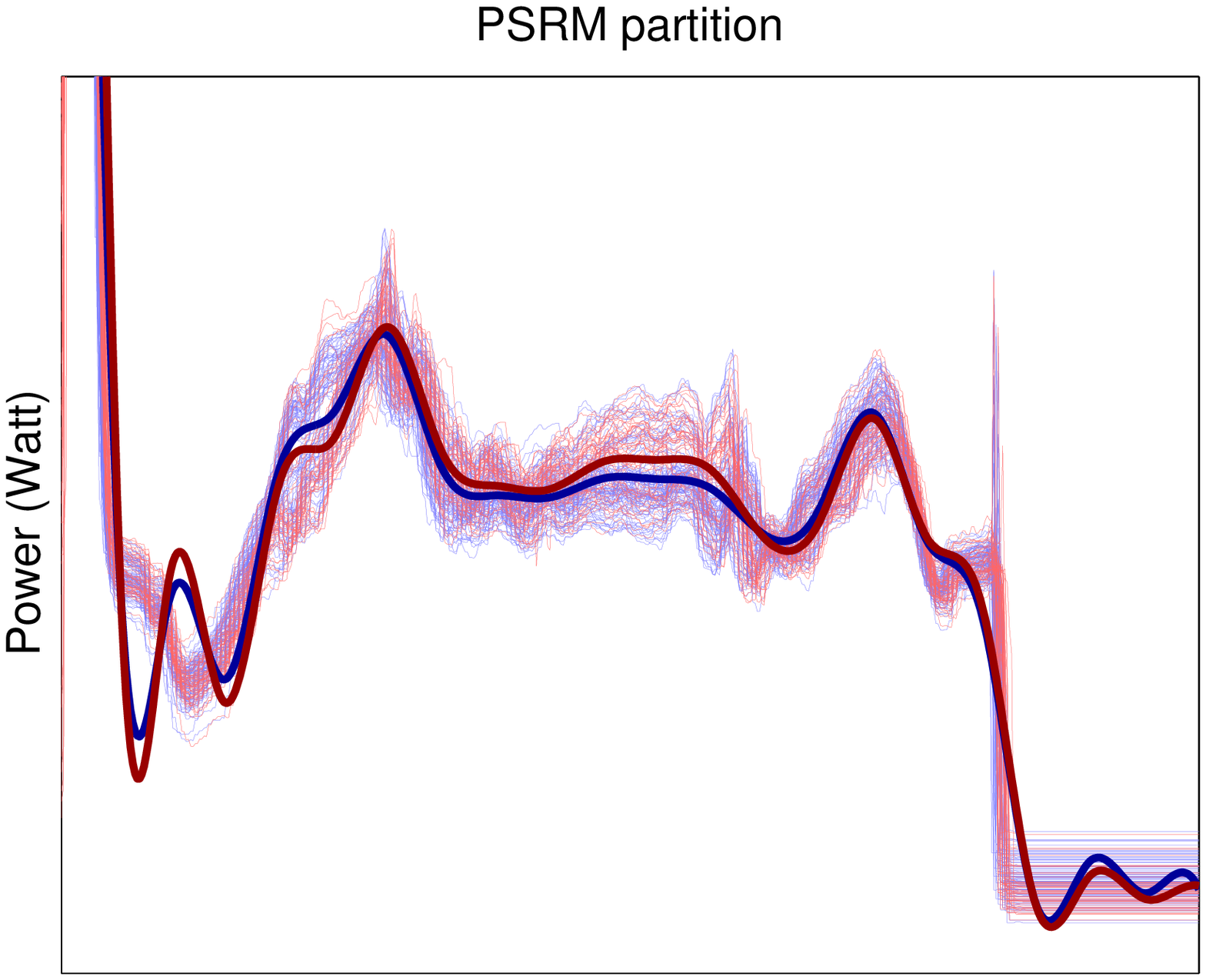}
\includegraphics[width = 4.8cm]{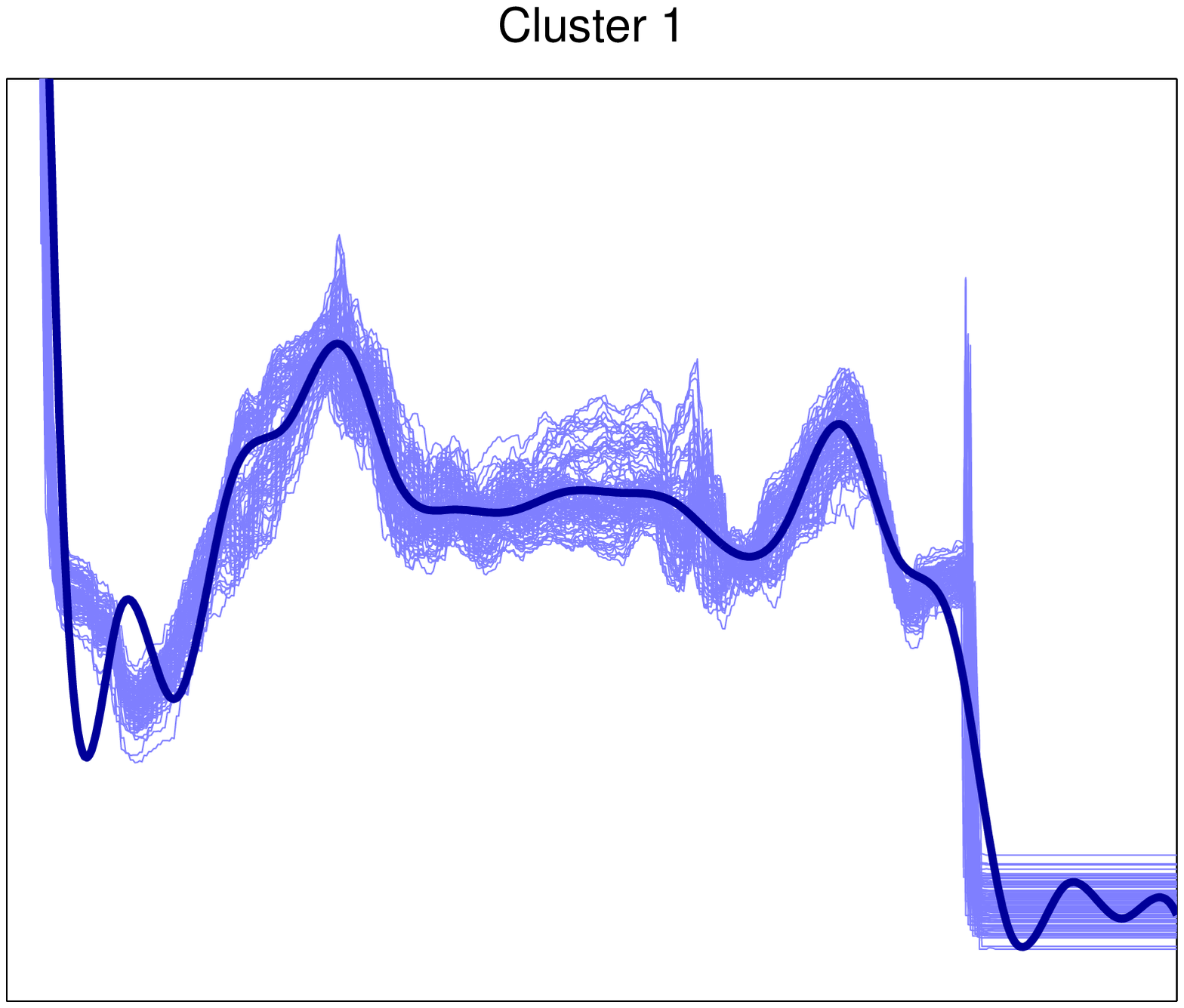}
\includegraphics[width = 4.8cm]{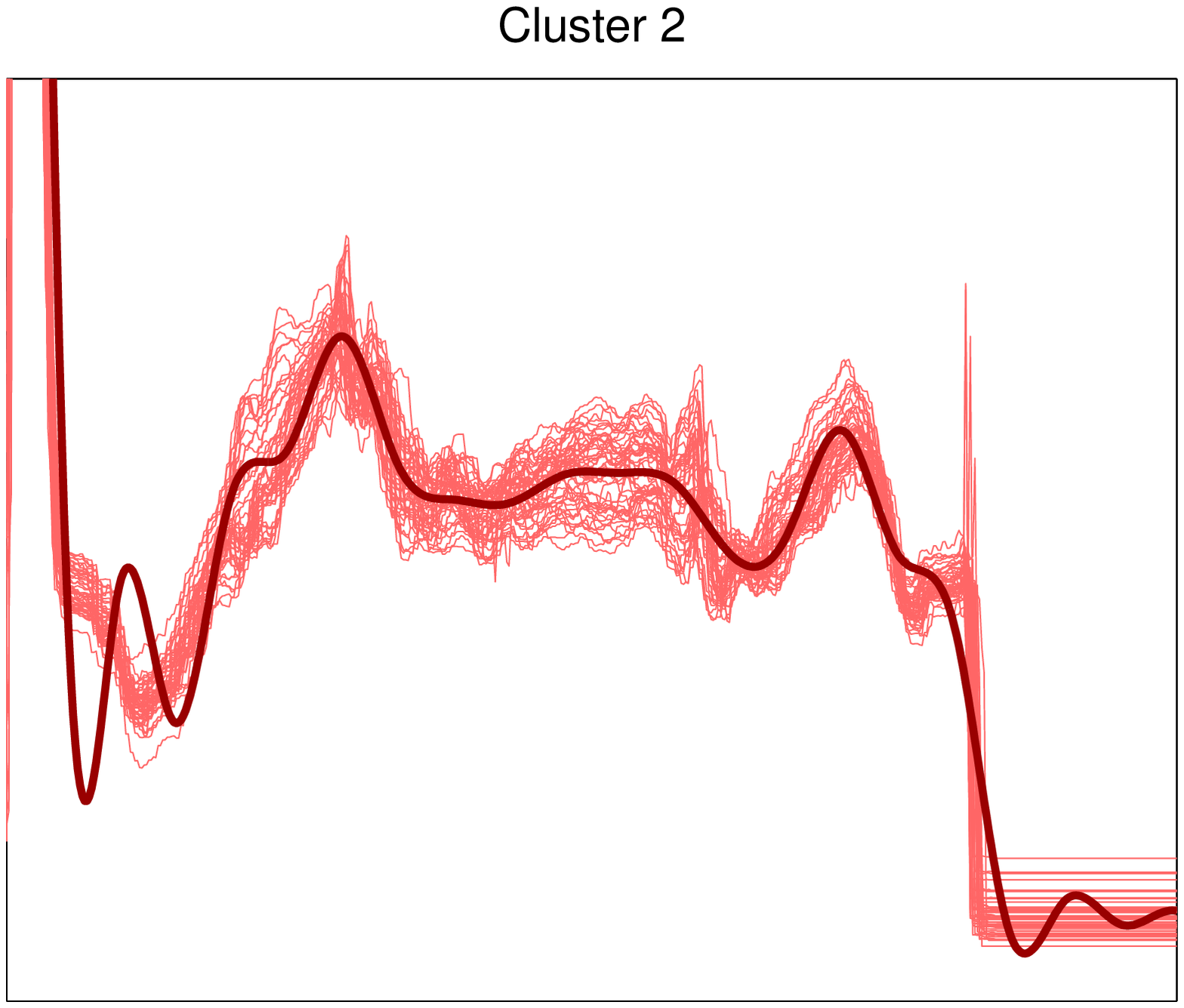}\\
\includegraphics[width = 4.6cm]{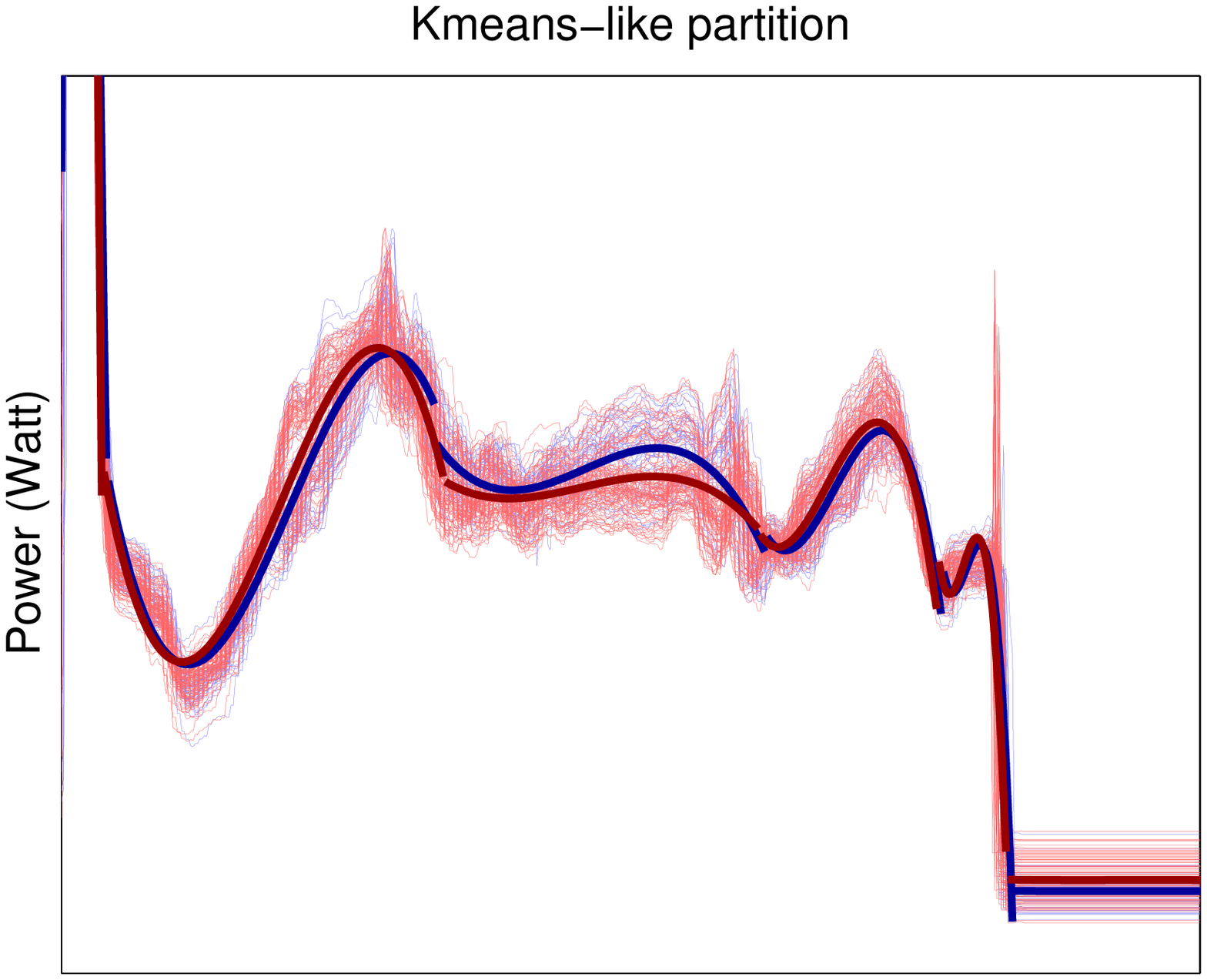}
\includegraphics[width = 4.8cm]{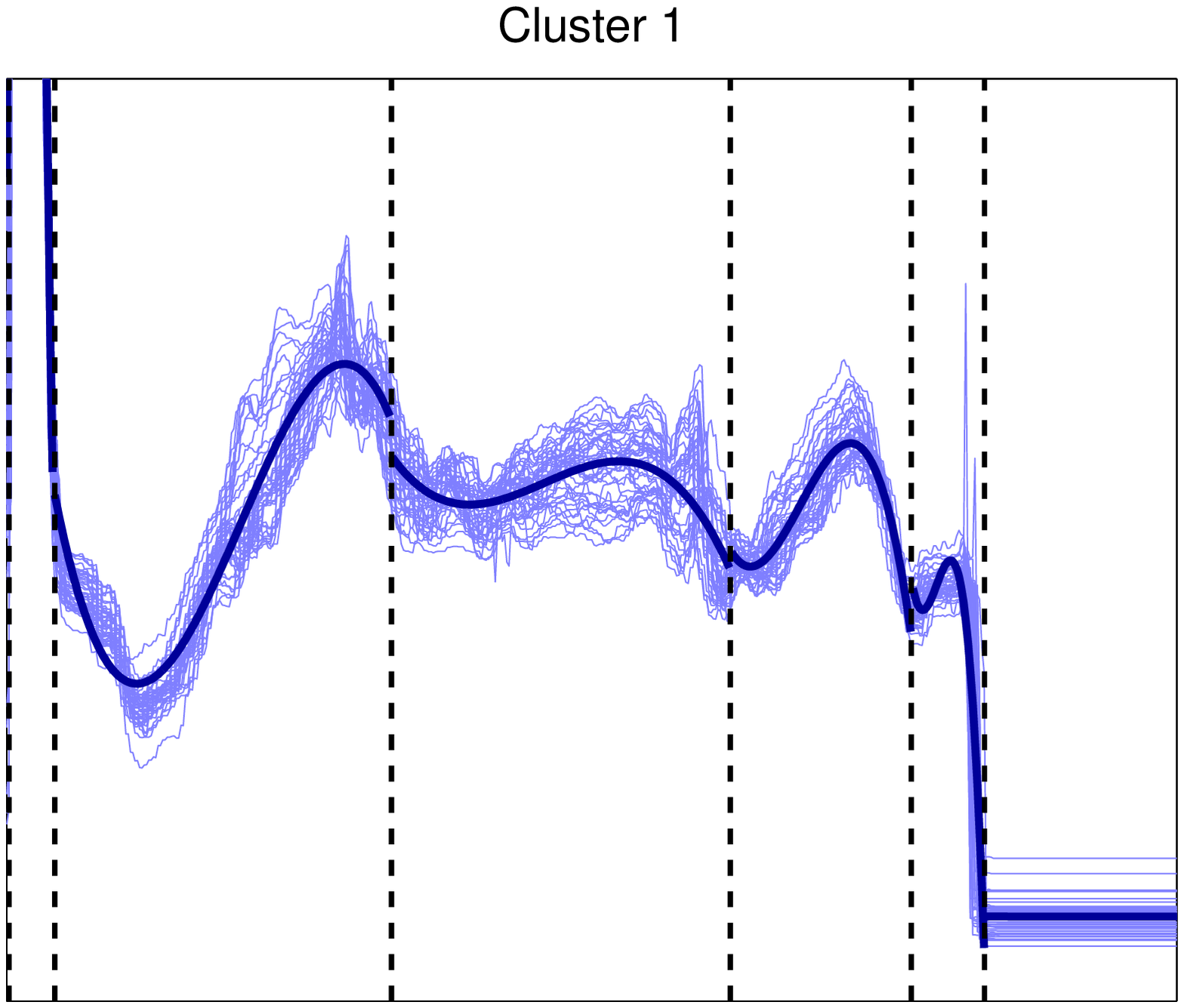}
\includegraphics[width = 4.8cm]{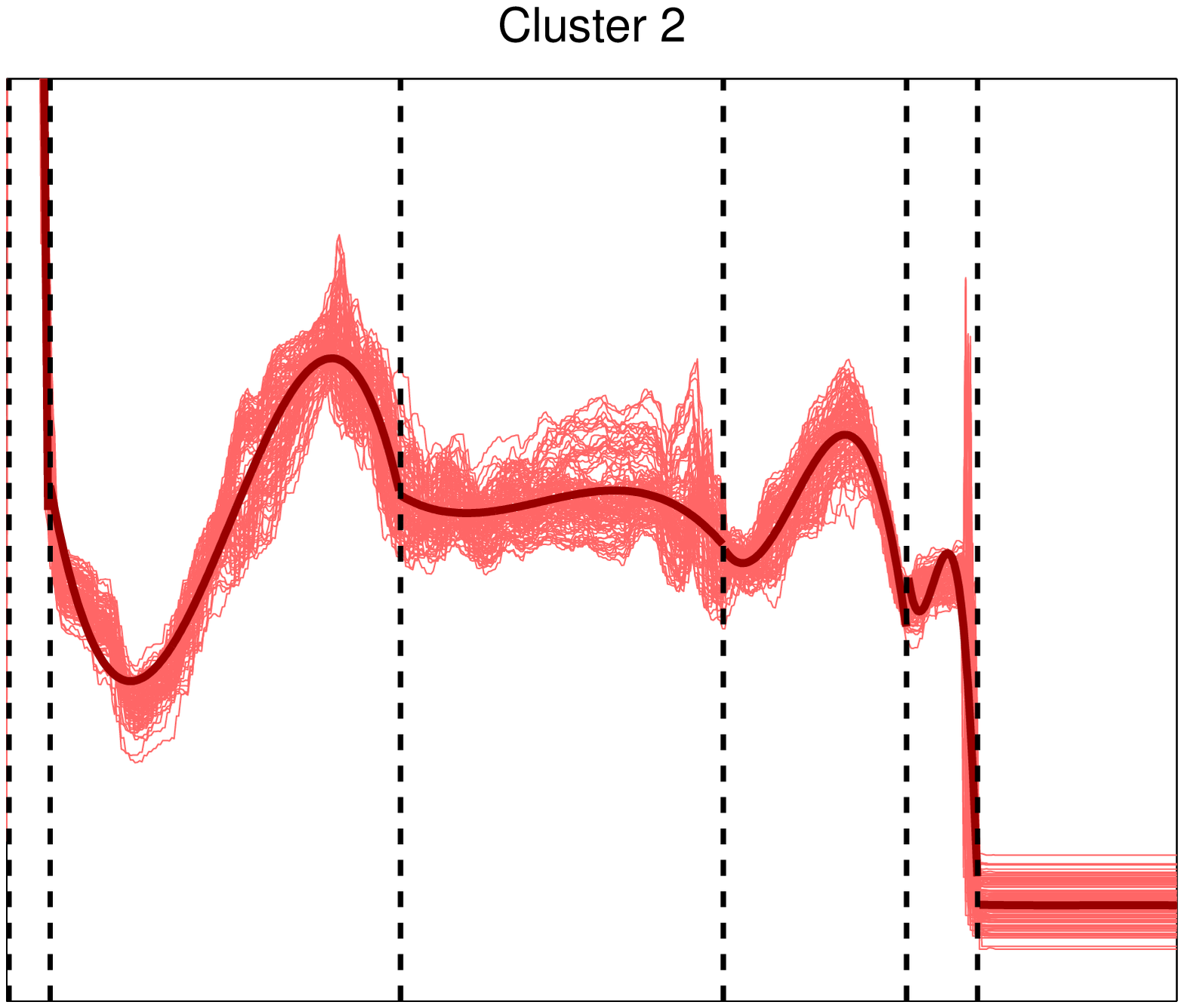}\\
\hspace{.05cm}
\includegraphics[width = 4.6cm]{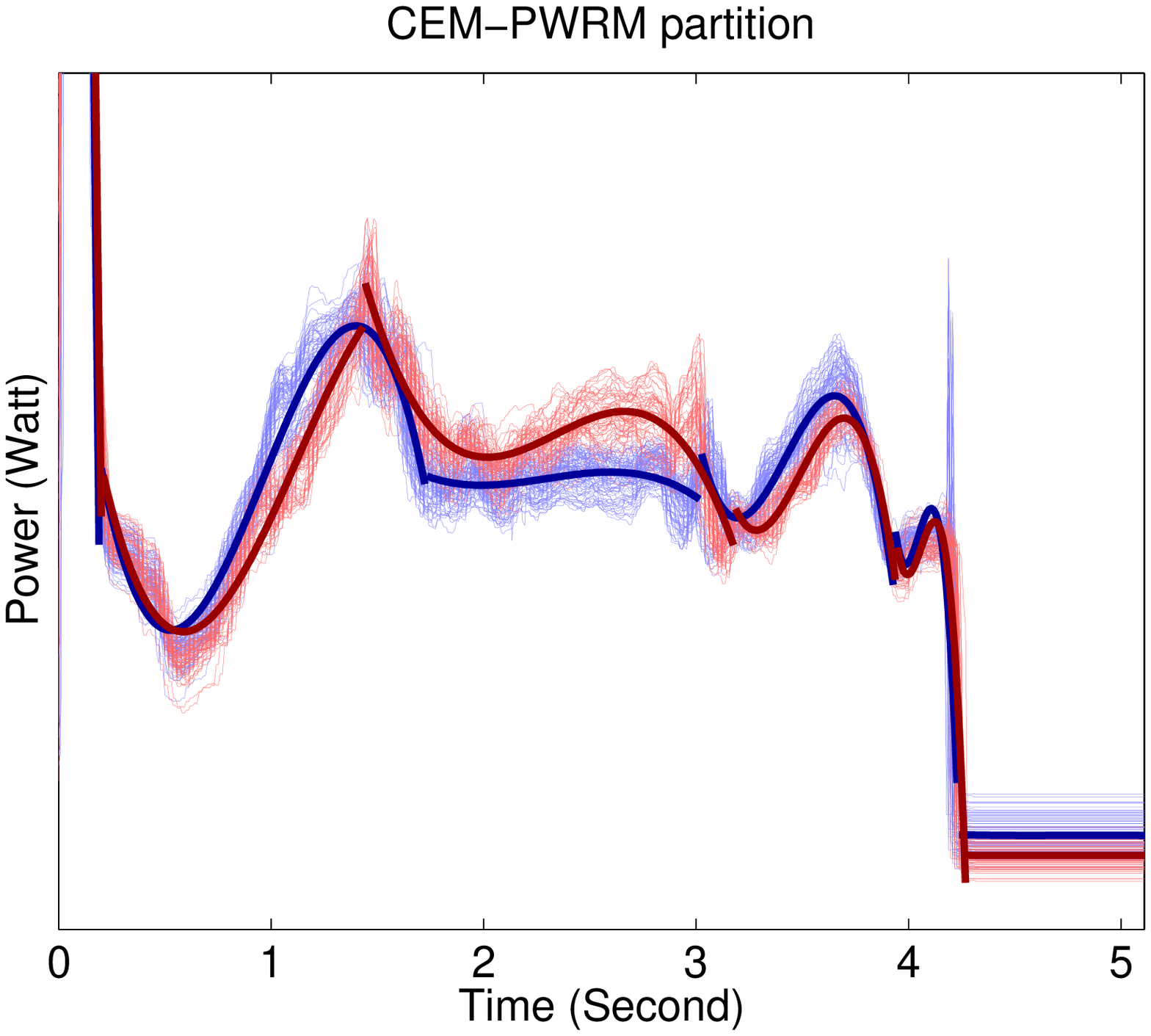} 
\includegraphics[width = 4.8cm]{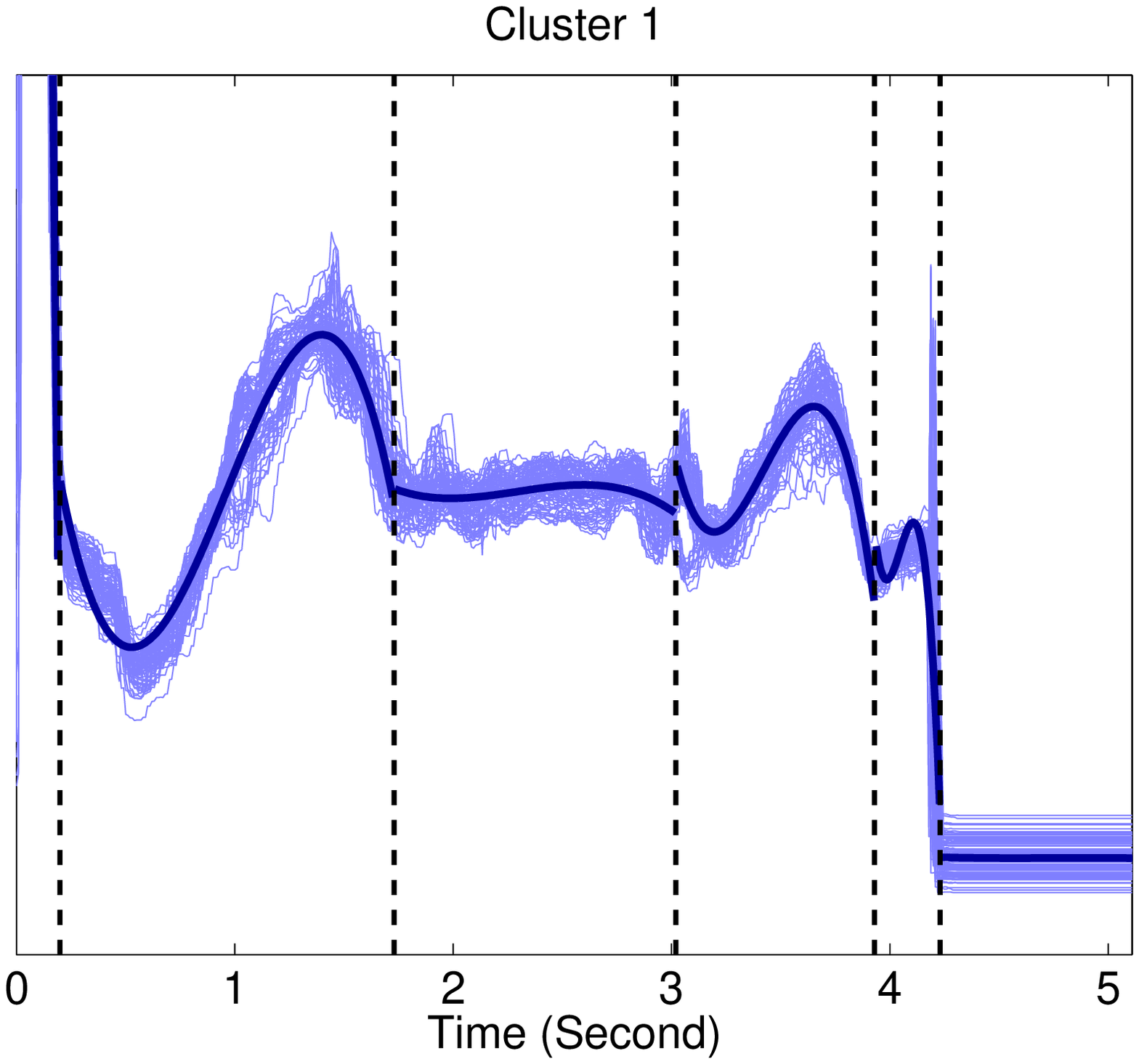}
\includegraphics[width = 4.8cm]{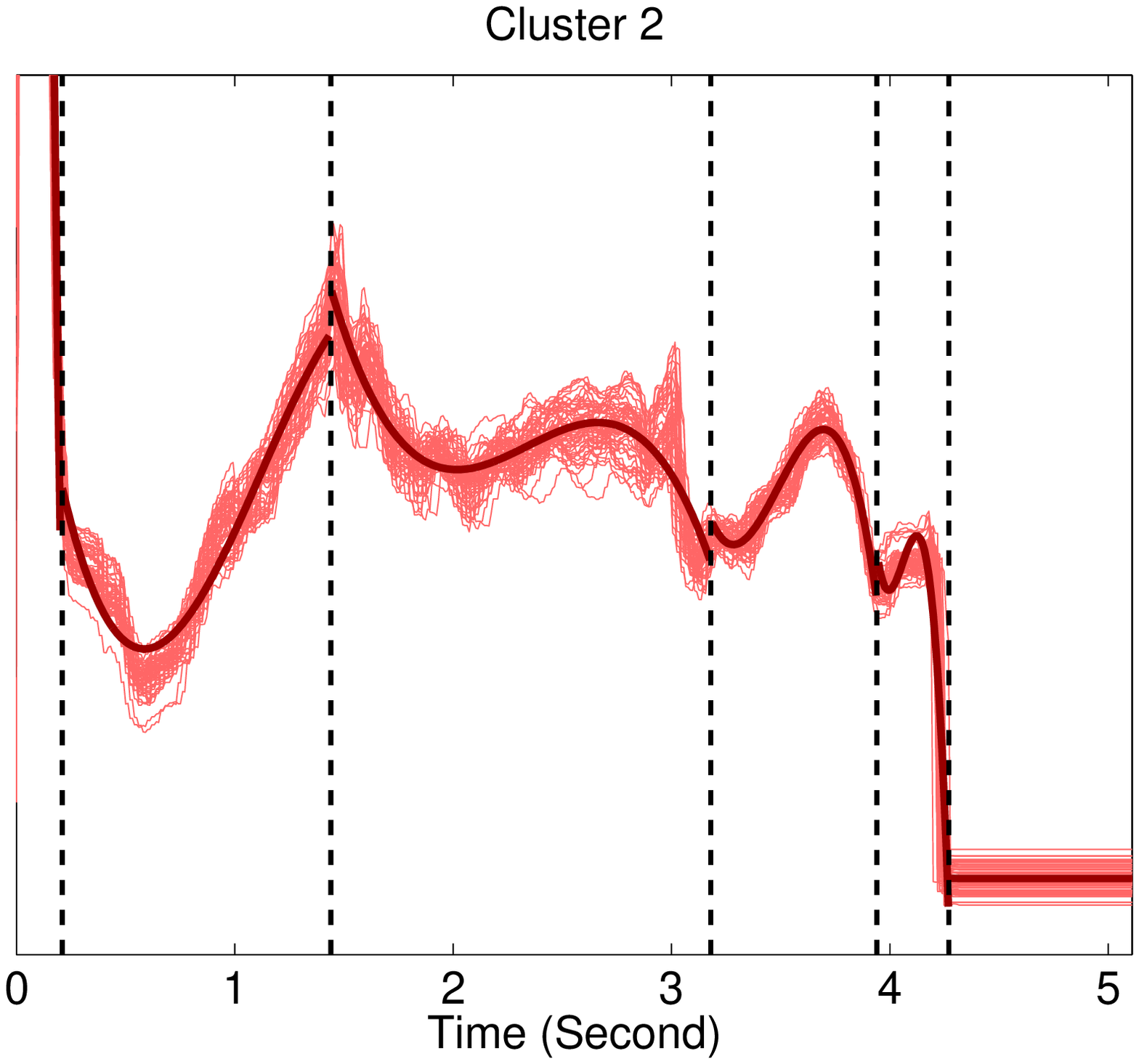}
\caption{\label{fig. clustering results for real data}Clustering results and the corresponding cluster prototypes obtained with EM-GMM, EM-PRM, EM-PSRM, and the corresponding cluster segmentations obtained with Kmeans-like and CEM-PWRM.}
\end{figure}We can see that the standard GMM clustering fails as it does not take into account the temporal aspect of the data, the obtained clusterings are not different and the mean curves are computed as an overall mean curves so that the obtained results are not very convincing.  The results provided by the PRM and  PSRM models are not convincing with regard to both the clustering and the approximation. However, the PWRM model clearly provides better results, since the cluster prototypes are more concordant with the real shape of the curves and, especially the proposed CEM-PWRM obtains to informative clusters. 
Indeed, it can be observed that for the CEM-PWRM approach, the curves of the first cluster (middle) and the second one (right) do not have the same characteristics  since their shapes are clearly different. Therefore they may correspond to two different states of the switch mechanism. 
In particular, for the curves belonging to the first cluster (middle), it can be observed that something happened at around 4.2 seconds of the switch operation. According to the experts, this can be attributed to a default in the measurement process, rather than a default of the switch itself. The device used for measuring the power would have been used slightly differently for this set of curves. 
%
%
%
Since the true class labels are unknown, we consider the results of intra-class inertia which find more significant for these data compared to the inter-class inertia of extensions. The values of inertia corresponding to the results shown in Figure \ref{fig. clustering results for real data} are given in Table \ref{table. inertia results for real data}. 
\begin{table}[H]
\centering
{\footnotesize  
\begin{tabular}{cccccc}
\hline
 EM-GMM & EM-PRM 	& EM-PSRM 	& $K$-means-like &  CEM-PWRM\\ 
 721.46 &  738.31  & 734.33 & 704.64 & 703.18 \\
\hline 
\hline
\end{tabular}}
\caption{{\small Intra-cluster inertia for the switch curves.}
}
\label{table. inertia results for real data}
\end{table}The intra-class results confirms that the piecewise regression mixture model has an advantage for giving  homogeneous and well approximated clusters from curves of regime changes. 

\subsubsection{Tecator data}
 The Tecator data\footnote{Tecator data are available at \url{http://lib.stat.cmu.edu/datasets/tecator}.} 
 consist of near infrared (NIR) absorbance spectra of 240 meat samples.
 The NIR spectra are recorded on a Tecator Infratec food and feed Analyzer working in the wavelength range $850-1050$ nm.  
The full Tecator data set contains $n=240$ spectra with $m=100$ for each spectrum, and is presented in Figure \ref{fig. tecator data}.
\begin{figure}[H] 
\centering 
\includegraphics[width=7.5cm]{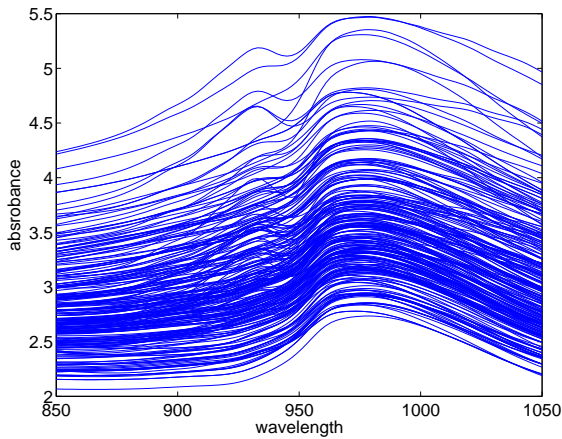} 
\caption{\label{fig. tecator data}Tecator curves.}
\end{figure} 
This data set has been considered in \cite{hebrailEtal:2010} and in our experiment we consider the same setting, that the data set is summarized with six clusters ($K=6$), each cluster being composed of five linear regimes (segments) ($R=5, p=1$).
 %
 
 Figure \ref{fig. CEM-PWRM clustering Tecator} shows the clustering and segmentation results obtained by the proposed CEM-PWRM algorithm. 
 \pagebreak
\begin{figure}[htbp] 
\centering
\includegraphics[width=14cm]{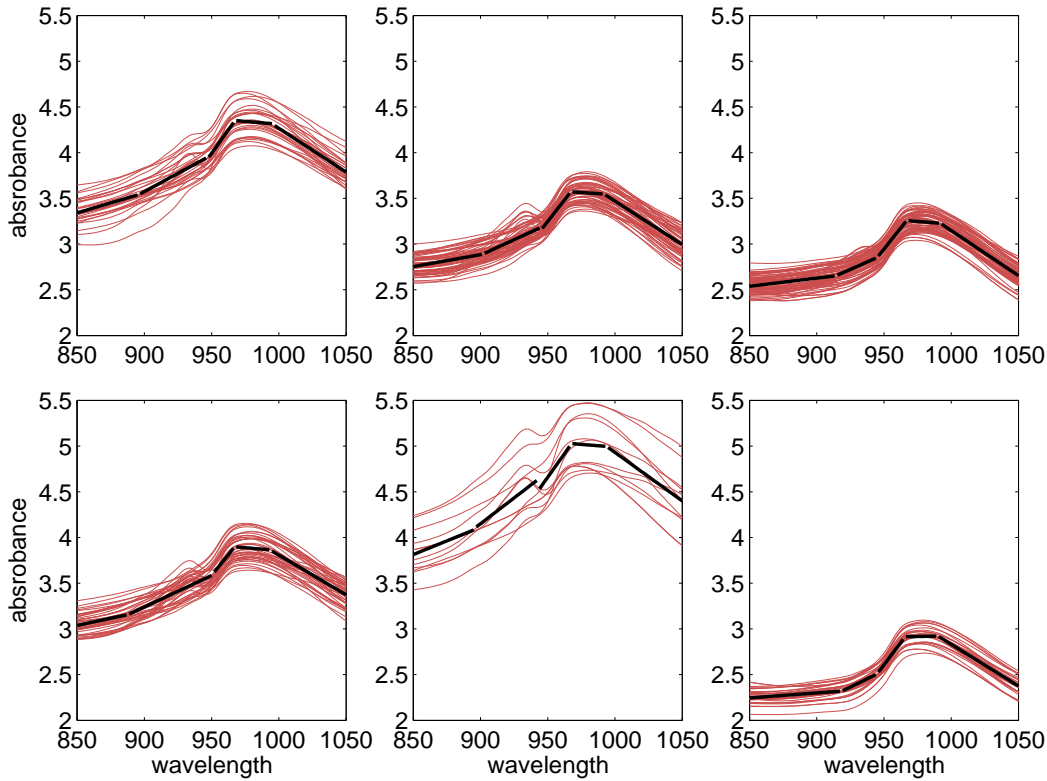}
\caption{\label{fig. CEM-PWRM clustering Tecator}Clusters and the corresponding piecewise linear prototypes for each cluster obtained with the CEM-PWRM algorithm for the full Tecator data set.}
\end{figure}One can see that the retrieved clusters are informative in the sense that the shapes of the clusters are clearly different, and  the piecewise approximation is in concordance with the shape of each cluster.
On the other hand, it can also be observed that this result is very close to the one obtained by \cite{hebrailEtal:2010} bu using the $K$-means-like approach. This not surprising and confirms that our proposed CEM-PWRM algorithm is a probabilistic alternative for the $K$-means-like approach.

\subsubsection{Topex/Poseidon satellite data}

The Topex/Poseidon radar satellite data\footnote{Satellite data are available at \url{http://www.lsp.ups-tlse.fr/staph/npfda/npfda-datasets.html}.} were registered by the satellite Topex/Poseidon around an area of 25 kilometers upon the Amazon River. The data  contain $n=472$ waveforms of the measured echoes, sampled at $m=70$ (number of echoes). 
The curves of this data set are shown in Figure \ref{fig. satellite data}.
\begin{figure}[htbp] 
\centering 
\includegraphics[width=7.5cm]{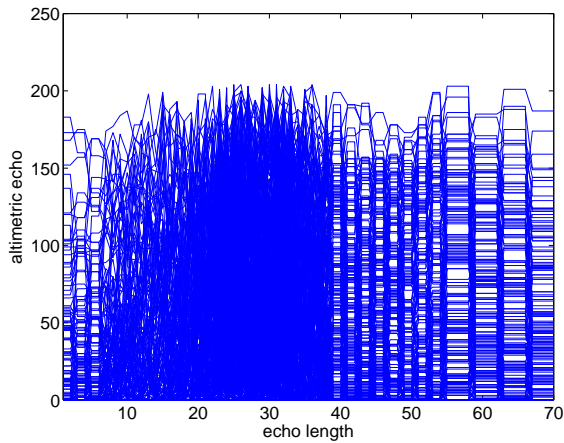}
\caption{\label{fig. satellite data}Topex/Poseidon satellite curves.}
\end{figure} 
We considered the same number of clusters (twenty) and a piecewise linear approximation of four segments per cluster as used in \cite{hebrailEtal:2010}. 
We note that, in our approach, we directly apply the proposed CEM-PWRM algorithm to raw the satellite data without a preprocessing step. However, in \cite{hebrailEtal:2010}, the authors used a two-fold scheme. They first perform a topographic clustering step using the Self Organizing Map (SOM), and then apply their $K$-means-like approach to the results of the SOM.

Figure \ref{fig. CEM-PWRM clustering Satellite}  shows the clustering and segmentation results obtained with the proposed CEM-PWRM algorithm for the satellite data set. 
First, it can be observed that the provided clusters are clearly informative and reflect the general behavior of the hidden structure of this data set.  The structure is indeed more clear with the mean curves of the clusters (prototypes) than with the raw curves. The piecewise approximation thus helps to better understand the structure of each cluster of curves from the obtained partition, and to more easily infer the general behavior of the data set.
On the other hand, one can also see that this result is similar to the one found in \cite{hebrailEtal:2010}, most of the profiles are present in the two results. The slight difference can be attributed to the fact that the result in \cite{hebrailEtal:2010} is provided from a two-stage scheme which includes and additional pre-clustering step using the SOM, rather by directly applying the piecewise regression model to the raw data.
\begin{figure}[htbp] 
\centering
\includegraphics[height=15cm, width=14cm]{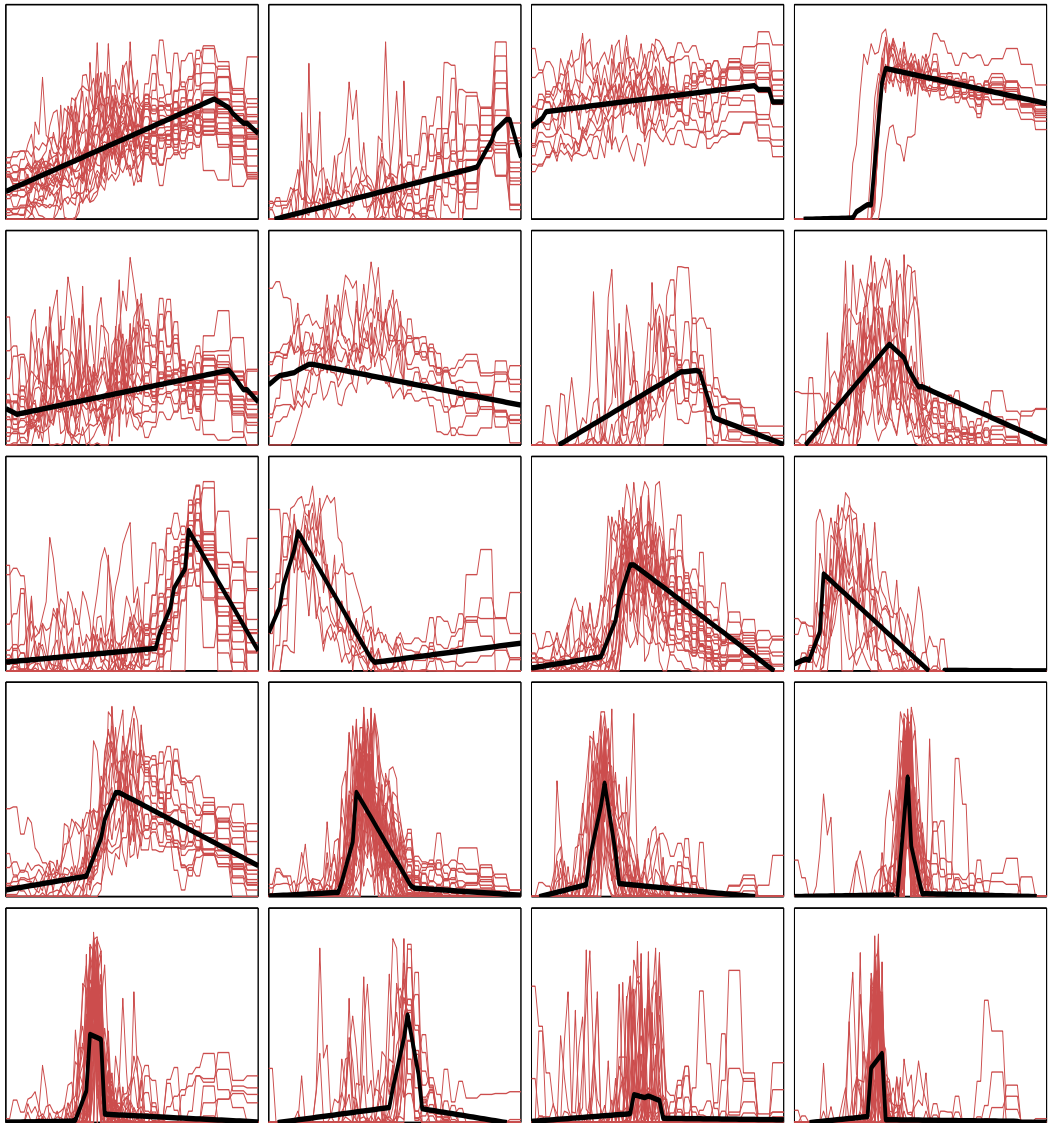}
\caption{\label{fig. CEM-PWRM clustering Satellite} Clusters and the corresponding piecewise linear prototypes for each cluster obtained with the CEM-PWRM algorithm for the satellite data set.}
\end{figure}
\section{Conclusion and discussion}
In this paper, we introduced a new probabilistic approach for simultaneous clustering and optimal segmentation  of curves with regime changes. The proposed approach is a piecewise polynomial regression mixture (PWRM).
We provided two algorithms to learn the model. The first (EM-PWRM) consists of using the EM algorithm to maximize the observed data log-likelihood and the latter (CEM-PWRM) is a CEM algorithm to maximize the complete-data log-likelihood. We showed that the CEM-PWRM algorithm is a general probabilistic-based version the $K$-means-like algorithm of \cite{hebrailEtal:2010} .
We conducted experiments on both simulated curves and real data sets to evaluate the proposed approach
and compare it to alternatives, including the regression mixture, the spline regression mixtures and the standard GMM for multivariate data.
 The obtained results demonstrated the benefit of the proposed approach in terms of both curve clustering and piecewise approximation of the regimes of each cluster. In particular, the comparisons with  the $K$-means-like algorithm approach confirm that the proposed CEM-PWRM is a general probabilistic alternative.

We note that in some practical situations involving continuous functions the proposed piecewise regression mixture, in its current formulation, may lead to discontinuities between segments for the piecewise approximation.  This can be easily avoided by slightly modifying the algorithm by adding an interpolation step as performed in \cite{hebrailEtal:2010}.
We also note that in this work we are interested in piecewise regimes which dot not overlap; only the clusters can overlap. However, one way to address the regime overlap is to augment the number of regimes in the proposed approach so that a regime that overlaps (for example it occurs in two different time ranges) can be treated as two regimes. Theses two reconstructed non-overlapping regimes would have very close characteristics so that as to correspond to a single overlapping regime.




\bibliographystyle{plainnat}

\bibliography{references}  

\begin{thebibliography}{38}
\providecommand{\natexlab}[1]{#1}
\providecommand{\url}[1]{\texttt{#1}}
\expandafter\ifx\csname urlstyle\endcsname\relax
  \providecommand{\doi}[1]{doi: #1}\else
  \providecommand{\doi}{doi: \begingroup \urlstyle{rm}\Url}\fi

\bibitem[Banfield and Raftery(1993)]{banfield_and_raftery_93}
J.~D. Banfield and A.~E. Raftery.
\newblock Model-based gaussian and non-gaussian clustering.
\newblock \emph{Biometrics}, 49\penalty0 (3):\penalty0 803--821, 1993.

\bibitem[Bellman(1961)]{bellman}
R.~Bellman.
\newblock On the approximation of curves by line segments using dynamic
  programming.
\newblock \emph{Communications of the Association for Computing Machinery
  (CACM)}, 4\penalty0 (6):\penalty0 284, 1961.

\bibitem[Biernacki et~al.(2000)Biernacki, Celeux, and Govaert]{ICL}
C.~Biernacki, G.~Celeux, and G~Govaert.
\newblock Assessing a mixture model for clustering with the integrated
  completed likelihood.
\newblock \emph{IEEE PAMI}, 22\penalty0 (7):\penalty0 719--725, 2000.

\bibitem[Brailovsky and Kempner(1992)]{brailovsky}
V.~L. Brailovsky and Y.~Kempner.
\newblock Application of piecewise regression to detecting internal structure
  of signal.
\newblock \emph{Pattern recognition}, 25\penalty0 (11):\penalty0 1361--1370,
  1992.

\bibitem[Celeux and Govaert(1992)]{celeuxetgovaert92_CEM}
G.~Celeux and G.~Govaert.
\newblock A classification {EM} algorithm for clustering and two stochastic
  versions.
\newblock \emph{Computational Statistics and Data Analysis}, 14:\penalty0
  315--332, 1992.

\bibitem[Celeux and Govaert(1993)]{celeuxetgovaert_mixture_classif_93}
G.~Celeux and G.~Govaert.
\newblock Comparison of the mixture and the classification maximum likelihood
  in cluster analysis.
\newblock \emph{Journal of Statistical Computation and Simulation},
  47:\penalty0 127--146, 1993.

\bibitem[Chamroukhi(2010)]{chamroukhi_PhD_2010}
F.~Chamroukhi.
\newblock \emph{Hidden process regression for curve modeling, classification
  and tracking}.
\newblock Ph.{D}. thesis, Universit\'e de Technologie de Compi\`egne, 2010.

\bibitem[Chamroukhi et~al.(2009{\natexlab{a}})Chamroukhi, Sam\'e, Govaert, and
  Aknin]{chamroukhiIJCNN2009}
F.~Chamroukhi, A.~Sam\'e, G.~Govaert, and P.~Aknin.
\newblock A regression model with a hidden logistic process for feature
  extraction from time series.
\newblock In \emph{International Joint Conference on Neural Networks (IJCNN)},
  2009{\natexlab{a}}.

\bibitem[Chamroukhi et~al.(2009{\natexlab{b}})Chamroukhi, Sam\'{e}, Govaert,
  and Aknin]{chamroukhi_et_al_NN2009}
F.~Chamroukhi, A.~Sam\'{e}, G.~Govaert, and P.~Aknin.
\newblock Time series modeling by a regression approach based on a latent
  process.
\newblock \emph{Neural Networks}, 22\penalty0 (5-6):\penalty0 593--602,
  2009{\natexlab{b}}.

\bibitem[Chamroukhi et~al.(2010)Chamroukhi, Sam\'{e}, Govaert, and
  Aknin]{chamroukhi_et_al_neurocomputing2010}
F.~Chamroukhi, A.~Sam\'{e}, G.~Govaert, and P.~Aknin.
\newblock A hidden process regression model for functional data description.
  application to curve discrimination.
\newblock \emph{Neurocomputing}, 73\penalty0 (7-9):\penalty0 1210--1221, March
  2010.

\bibitem[Chamroukhi et~al.(2011)Chamroukhi, Sam\'e, Aknin, and
  Govaert]{chamroukhi_ijcnn_2011}
F.~Chamroukhi, A.~Sam\'e, P.~Aknin, and G.~Govaert.
\newblock Model-based clustering with hidden markov model regression for time
  series with regime changes.
\newblock In \emph{Proceedings of IJCNN}, pages 2814--2821, 2011.

\bibitem[Chamroukhi et~al.(2013)Chamroukhi, Herv{\'e}, and
  Sam{\'e}]{chamroukhi_fmda_neucomp2013}
Faicel Chamroukhi, Glotin Herv{\'e}, and Allou Sam{\'e}.
\newblock Model-based functional mixture discriminant analysis with hidden
  process regression for curve classification.
\newblock \emph{Neurocomputing}, 112:\penalty0 153--163, 2013.

\bibitem[Dempster et~al.(1977)Dempster, Laird, and Rubin]{dlr}
A.~P. Dempster, N.~M. Laird, and D.~B. Rubin.
\newblock Maximum likelihood from incomplete data via the {EM} algorithm.
\newblock \emph{JRSS, B}, 39(1):\penalty0 1--38, 1977.

\bibitem[Fearnhead and Liu(2007)]{Fearnhead2007}
P.~Fearnhead and Z.~Liu.
\newblock {O}nline {I}nference for {M}ultiple {C}hangepoint {P}roblems.
\newblock \emph{Journal of the Royal Statistical Society, Series B},
  69:\penalty0 589--605, 2007.

\bibitem[Fearnhead(2006)]{Fearnhead2006}
Paul Fearnhead.
\newblock Exact and efficient {B}ayesian inference for multiple changepoint
  problems.
\newblock \emph{Statistics and Computing}, 16:\penalty0 203--213, 2006.

\bibitem[Ferrari-Trecate and Muselli(2002)]{ferrari1}
G.~Ferrari-Trecate and M.~Muselli.
\newblock A new learning method for piecewise linear regression.
\newblock In \emph{International Conference on Artificial Neural Networks
  (ICANN)}, pages 28--30, 2002.

\bibitem[Fraley and Raftery(2002)]{Fraley2002_model-basedclustering}
C.~Fraley and A.~E. Raftery.
\newblock Model-based clustering, discriminant analysis, and density
  estimation.
\newblock \emph{Journal of the American Statistical Association}, 97:\penalty0
  611--631, 2002.

\bibitem[Gaffney(2004)]{Gaffneythesis}
S.~J. Gaffney.
\newblock \emph{Probabilistic Curve-Aligned Clustering and Prediction with
  Regression Mixture Models}.
\newblock PhD thesis, University of California, Irvine, 2004.

\bibitem[Gaffney and Smyth(2004)]{gaffneyANDsmythNIPS2004}
S.~J. Gaffney and P.~Smyth.
\newblock Joint probabilistic curve clustering and alignment.
\newblock In \emph{In Advances in NIPS}, 2004.

\bibitem[Gaffney and Smyth(1999)]{Gaffney99trajectoryclustering}
Scott Gaffney and Padhraic Smyth.
\newblock Trajectory clustering with mixtures of regression models.
\newblock In \emph{Proceedings of the fifth ACM SIGKDD international conference
  on Knowledge discovery and data mining}, pages 63--72, 1999.

\bibitem[Gui and Li(2003)]{Gui_FMDA}
J.~Gui and H.~Li.
\newblock Mixture functional discriminant analysis for gene function
  classification based on time course gene expression data.
\newblock In \emph{Proc. Joint Stat. Meeting (Biometric Section)}, 2003.

\bibitem[H{\'e}brail et~al.(2010)H{\'e}brail, Hugueney, Lechevallier, and
  Rossi]{hebrailEtal:2010}
G.~H{\'e}brail, B.~Hugueney, Y.~Lechevallier, and F.~Rossi.
\newblock Exploratory analysis of functional data via clustering and optimal
  segmentation.
\newblock \emph{Neurocomputing}, 73\penalty0 (7-9):\penalty0 1125--1141, March
  2010.

\bibitem[Hugueney et~al.(2009)Hugueney, H\'ebrail, Lechevallier, and
  Rossi]{HugueneyEtAl:ESANN2009}
B.~Hugueney, G.~H\'ebrail, Y.~Lechevallier, and F.~Rossi.
\newblock Simultaneous clustering and segmentation for functional data.
\newblock In \emph{ESANN}, pages 281--286, 2009.

\bibitem[James and Sugar(2003)]{garetjamesJASA2003}
G.~M. James and C.~Sugar.
\newblock Clustering for sparsely sampled functional data.
\newblock \emph{Journal of the American Statistical Association}, 98\penalty0
  (462):\penalty0 397--408, 2003.

\bibitem[Liu and Yang(2009)]{liuANDyangFunctionalDataClustering}
X.~Liu and M.C.K. Yang.
\newblock Simultaneous curve registration and clustering for functional data.
\newblock \emph{Computational Statistics and Data Analysis}, 53\penalty0
  (4):\penalty0 1361--1376, 2009.

\bibitem[McGee and Carleton(1970)]{McGee}
V.~E. McGee and W.~T. Carleton.
\newblock Piecewise regression.
\newblock \emph{Journal of the American Statistical Association}, 65:\penalty0
  1109--1124, 1970.

\bibitem[McLachlan and Krishnan(1997)]{mclachlanEM}
G.~J. McLachlan and T.~Krishnan.
\newblock \emph{The EM algorithm and extensions}.
\newblock New York: Wiley, 1997.

\bibitem[McLachlan and Peel.(2000)]{mclachlanFiniteMixtureModels}
G.~J. McLachlan and D.~Peel.
\newblock \emph{Finite mixture models}.
\newblock Wiley, 2000.

\bibitem[McLachlan and Basford(1988)]{mclachlan_basford88}
G.J. McLachlan and K.E. Basford.
\newblock \emph{Mixture Models: Inference and Applications to Clustering}.
\newblock Marcel Dekker, New York, 1988.

\bibitem[Picard et~al.(2007)Picard, Robin, Lebarbier, and
  Daudin]{picardetal2007}
F.~Picard, S.~Robin, E.~Lebarbier, and J.~J. Daudin.
\newblock A segmentation/clustering model for the analysis of array {CGH} data.
\newblock \emph{Biometrics}, 63\penalty0 (3):\penalty0 758--766, September
  2007.

\bibitem[Ramsay and Silverman(2005)]{ramsayandsilvermanFDA2005}
J.~O. Ramsay and B.~W. Silverman.
\newblock \emph{Functional Data Analysis}.
\newblock Springer Series in Statistics. Springer, June 2005.

\bibitem[Sam{\'e} et~al.(2011)Sam{\'e}, Chamroukhi, Govaert, and
  Aknin]{same_chamroukhi_Adac}
Allou Sam{\'e}, Faicel Chamroukhi, G{\'e}rard Govaert, and Patrice Aknin.
\newblock Model-based clustering and segmentation of time series with changes
  in regime.
\newblock \emph{Advances in Data Analysis and Classification}, pages 1--21,
  2011.
\newblock ISSN 1862-5347.

\bibitem[Schwarz(1978)]{BIC}
G.~Schwarz.
\newblock Estimating the dimension of a model.
\newblock \emph{Annals of Statistics}, 6:\penalty0 461--464, 1978.

\bibitem[Shi and Wang(2008)]{ShiW08}
J.~Q. Shi and B.~Wang.
\newblock Curve prediction and clustering with mixtures of gaussian process
  functional regression models.
\newblock \emph{Statistics and Computing}, 18\penalty0 (3):\penalty0 267--283,
  2008.

\bibitem[Smyth(1996)]{Smyth96}
P.~Smyth.
\newblock Clustering sequences with hidden markov models.
\newblock In \emph{Advances in Neural Information Processing Systems 9, NIPS},
  pages 648--654, 1996.

\bibitem[Stone(1961)]{stone}
H.~Stone.
\newblock Approximation of curves by line segments.
\newblock \emph{Mathematics of Computation}, 15\penalty0 (73):\penalty0 40--47,
  1961.

\bibitem[Titterington et~al.(1985)Titterington, Smith, and
  Makov]{titteringtonBookMixtures}
D.~Titterington, A.~Smith, and U.~Makov.
\newblock \emph{Statistical Analysis of Finite Mixture Distributions}.
\newblock John Wiley \& Sons, 1985.

\bibitem[Xiong and Yeung(2004)]{XiongY04}
Yimin Xiong and Dit-Yan Yeung.
\newblock Time series clustering with {ARMA} mixtures.
\newblock \emph{Pattern Recognition}, 37\penalty0 (8):\penalty0 1675--1689,
  2004.

\end{thebibliography}
\end{document}